\documentclass[aip,graphicx,reprint,pop]{revtex4-2}

\usepackage{graphicx}
\usepackage{bm}
\usepackage{etoolbox} 
\usepackage{subcaption}
\usepackage{physics} 
\usepackage{placeins} 

\makeatletter
\def\@email#1#2{%
 \endgroup
 \patchcmd{\titleblock@produce}
  {\frontmatter@RRAPformat}
  {\frontmatter@RRAPformat{\produce@RRAP{*#1\href{mailto:#2}{#2}}}\frontmatter@RRAPformat}
  {}{}
}%
\makeatother

\begin{document}

\title[]{Influence of collisions on trapped-electron modes in tokamaks and low-shear stellarators}

\author{M.C.L. Morren}
    \email{m.c.l.morren@tue.nl}

\author{J.H.E. Proll}

\author{J. van Dijk}
\affiliation{Department of Applied Physics and Science Education, Eindhoven University of Technology, 5600MB Eindhoven, The Netherlands}

\author{M.J. Pueschel}
\affiliation{Department of Applied Physics and Science Education, Eindhoven University of Technology, 5600MB Eindhoven, The Netherlands}
\affiliation{Dutch Institute for Fundamental Energy Research, 5612 AJ Eindhoven, the Netherlands}

\date{\today}

\begin{abstract}
The influence of collisions on the growth rate of trapped-electron modes (TEMs) in core plasmas is assessed through both analytical linear gyrokinetics, and linear gyrokinetic simulations. Both methods are applied to the magnetic geometry of the DIII-D tokamak, as well as the Helically Symmetric eXperiment (HSX) and Wendelstein 7-X (W7-X) stellarators, in the absence of temperature gradients. \par 
Here we analytically investigate the influence of collisions on the TEM eigenmode frequency by a perturbative approach in the response of trapped particles to the mode, using an energy-dependent Krook operator to model collisions. Whilst the resulting growth rates exceed perturbative thresholds, they reveal important qualitative dependencies: a geometry-dependent stabilisation rate occurs for all wavenumbers at high collisionality, while at low collisionality a geometry-sensitive mixture of collisionless, resonantly-driven and collisionally destabilised modes are found. \par
Additionally, linear gyrokinetic simulations have been performed with a rigorous pitch-angle scattering operator for the same geometries. In the case of DIII-D and large wavenumber modes in HSX the trends predicted by analytical theory are reproduced. Dissimilarities are, however, obtained in W7-X geometry and for low wavenumber modes in HSX, which are shown to be due to a collision-induced transition to the Universal Instability (UI) as the dominant instability at marginal collisionality.
\end{abstract}

\maketitle

\section{Introduction \label{sec:intro}}
The performance of modern-day magnetic confinement fusion devices is limited by particle and heat losses as a result of turbulence. This turbulence is believed to be driven by microinstabilities in the plasma \cite{Alcuson2020,Stroth1998,Connor1994}, with the ion-temperature-gradient (ITG) mode and the trapped-electron mode (TEM) responsible for the majority of the turbulent transport in typical cases\cite{Manas2015,Shen2019,Jenko2000}. Plasma turbulence has been studied using gyrokinetics in both tokamak and stellarator geometry, where it has been found that the magnetic geometry can have a substantial influence on both the underlying microinstabilities \cite{Alcuson2020,Rewoldt2005,Proll2022,Proll2013,Nakata2014,Whelan2019,Belli2008,McKinney2019,Xanthopoulos2007} and the transport levels in the saturated state\cite{Nakata2022,Navarro2020,Belli2008,McKinney2019,Mynick2009,Proll2022}. This is particularly true for the onset of TEMs, as the details of the magnetic geometry determine the locations of the magnetic minima and thus where trapped particles reside. 
\par
For example, so-called quasi-isodynamic configurations, which are characterised by poloidally closed contours of magnetic field strength and by the second adiabatic invariant $\mathcal{J}$ being constant on flux surfaces \cite{Nuhrenberg2010,Helander2012}, achieve a trapped-electron precession in an opposite direction to their diamagnetic drift, thus avoiding the resonant instability drive of collisionless TEMs, if the second adiabatic invariant peaks on the magnetic axis, thereby constituting a maximum-$\mathcal{J}$ configuration \cite{Rosenbluth1968,Proll2012,Helander2013}. At high plasma pressure, such favourable equilibrium configurations are realised in the Wendelstein 7-X (W7-X) stellarator.
\par 
This stability of maximum-$\mathcal{J}$ configurations against TEMs holds in the absence of collisions, which also corresponds to the situation for which the majority of gyrokinetic simulations in the past have been performed \cite{Dimits2000,Nevins2006,Chowdhury2010,Xanthopoulos2007,Shen2019,Jenko2000,Proll2013,Merz2010,Proll2022,Costello2022,Shen2019,Jenko2000,Rewoldt2005,Nakata2014,Navarro2020,Rewoldt2005,Belli2008,Mynick2009,McKinney2019}. The impact of collisions on these microinstabilities is often neglected as, in typical fusion reactor scenarios, the collision frequency is typically in the low kHz range\cite{Garbet2010,Becoulet2009} whereas the oscillation frequency of the instabilities ranges from a few tens of kHz to several hundreds of kHz \cite{Austin2019,Brower1987}. Moreover, the influence of collisions is thought to be mainly stabilising as a result of their dissipative effect on the instabilities \cite{Sugama2009,Helander2021}. Studying the collisionless case would thus give an upper estimate of the linear growth rates. A notable exception to this trend occurs for microtearing modes\cite{Guttenfelder2012,Applegate2007} and resistive modes\cite{Bourdelle2012}, where collisions have a destabilising effect. However, such instabilities belong to the class of magnetic-fluctuation turbulence, which is not considered in this work. Nevertheless, especially for TEM, the role of collisions is less trivial, as collisions influence the population of trapped particles through collisional (de-)trapping. Possibly more profoundly, in the case of maximum-$\mathcal{J}$ configurations, collisions could make a resonance with the precession of trapped-electrons possible again by either affecting the propagation direction of the drift wave or upsetting the trapped-electron drift orbits, thus driving the previously stable TEM unstable. \par
In this paper, we address the influence of collisions on the TEM microinstability in the core region of general 3D large-aspect-ratio toroidal geometries. To this end, we use an analytical perturbative approach for the trapped-electron response, including the effect of de-trapping through a Krook collision operator, while assuming adiabatic behaviour for passing electrons. In the past, this approach (though neglecting the role of a decrease in trapped-electron population as a result of collisional de-trapping) has been used to assess the growth rate as a result of trapped-electron dynamics in the asymptotic limits of zero collisionality or very high collisionality \cite{Connor2006,Gang1991,Adam1976,Dominguez1992}. Here, we do not impose such restrictions on the parameter regime validity of the model, but order the collision frequency comparable to the mode frequency to obtain the growth rate dependence on the collision frequency valid over a large range of collisionalities. At the highest collisionalities, however, the validity of our model breaks down, when as a result of rapid successive (de-)trapping events the adiabaticity of passing electrons\cite{ajay2021} diminishes and exact details of the collision operator model become more important\cite{Tang1978,Pan2020,Frei2022}. In conjunction with the analytical work, numerical simulations are also performed with the \textsc{GENE} code\cite{Jenko2000} to verify these analytical results using a realistic collision operator and kinetic-electron response. \par 
To investigate differences between geometries, a set of three realistic experimental devices will be considered, consisting of the DIII-D tokamak \cite{Luxon2002}, the Helically Symmetric eXperiment (HSX) stellarator\cite{Anderson1995}, and the high-mirror configuration of the W7-X stellarator\cite{Beidler1990}. These geometries are chosen to enable highlighting of potential differences between tokamaks and stellarators with quasi-symmetry (where the magnetic field strength and particle drifts are still isomorphic to a tokamak\cite{Boozer1983}, but the magnetic shear can be substantially lower \cite{Helander2012}), or without quasi-symmetry (which have the freedom to separate the trapping wells from the bad-curvature regions \cite{Boozer1998}), such as quasi-isodynamic maximum-$\mathcal{J}$ stellarators. Strictly speaking, W7-X only achieves this maximum-$\mathcal{J}$ TEM stabilising property at high plasma $\beta$\cite{Nuhrenberg2010,Helander2014}, with $\beta$ the ratio of kinetic to magnetic pressure, whilst here we consider the vacuum magnetic field. Nevertheless, the vacuum configuration already exhibits trapping regions where the majority of deeply-trapped particles (as opposed to all trapped particles in an exact maximum-$\mathcal{J}$ configuration\cite{Helander2014}) have a favourable bounce-averaged drift, which results in significantly lower trapped-particle activity compared to tokamaks and quasi-symmetric stellarators \cite{Proll2013}. Therefore, by means of an abbreviation to this effect, we shall refer to the vacuum configuration of W7-X as being approximately maximum-$\mathcal{J}$.

This paper is organised as follows: in Section \ref{sec:gentheory} the general electrostatic gyrokinetic framework and the energy-dependent Krook collision operator model are briefly described. The perturbative method is then applied to this framework to allow for a tractable expression for the TEM growth rate. Section \ref{sec:perturbative} presents the scaling laws of the growth rate obtained from this perturbative approach in both low- and high-collisionality regimes. For comparison with the growth rates from simulations, an overview of the settings and parameters is first presented in Section \ref{sec:GENEgen}, after which the simulation results are discussed and briefly compared with analytical findings in Section \ref{sec:sims}. Lastly, a summary and short discussion is provided in Section \ref{sec:summary}. 

\section{Gyrokinetic description of trapped-electron modes \label{sec:gentheory}}
Transport-relevant plasma microinstabilities can be described by linear gyrokinetics, of which we will consider the electrostatic limit (corresponding to $\beta=0$), thereby neglecting magnetic field fluctuations. This framework, with which we will describe TEMs perturbatively, is briefly summarised below. \par 
The distribution function $f_s$ of each species $s$ is decomposed into an equilibrium (Maxwellian) $F_{Ms}$ and a fluctuating part $\delta f_s$, the latter of which can be split into an adiabatic and a non-adiabatic contribution
\begin{equation}
    f_s = F_{Ms}  + \delta f_s = F_{Ms} \textrm{$\left(1- \frac{q_s\phi}{T_s}\right)$} + g_s 
    \label{eq:ftot}
\end{equation}
with $q_s,T_s$ denoting the charge and temperature of species $s$, respectively, $\phi$ the electrostatic potential and $g_s$ the non-adiabatic part of the perturbed distribution $\delta f_s$. The evolution of the latter is then determined by the linearised gyrokinetic equation \cite{Rutherford1968,Catto1981}
\begin{equation}
   v_{\parallel} \nabla_{\parallel} \hat{g}_s - i(\omega-\omega_{ds})\hat{g}_s - \hat{C_{\bm{k}}}
     = - i \frac{q_s}{T_s} J_0 \hat{\phi} \left(\omega - \omega_{\star s}^{T}\right)F_{Ms}, 
     \label{eq:GKE}
\end{equation}
where all perturbations have been decomposed into Fourier modes perpendicular to the field lines and the ballooning transform\cite{antonsen1980,Connor1978,Dewar1998} has been applied to make the sheared toroidal magnetic field lines compatible with periodicity requirements for Fourier modes. This results in perturbations of the form $\xi = \hat{\xi}\exp(-i\omega t + i \bm{k_\perp}\cdot\bm{x})$), where the slow variation of the perturbations along the field line is given by the amplitude function $\hat{\xi}$ and the fast variation perpendicular to the field by the Eikonal, with $\xi = \left\{\phi,g_s\right\}$. In Eqn.~(\ref{eq:GKE}) $\omega = \omega_R + i \gamma$ is the complex mode frequency, with $\omega_R$ the oscillation frequency and $\gamma$ the growth rate of the instabilities, $\omega_{ds} = \bm{k_\perp} \cdot \bm{v_{ds}}$ is the magnetic drift frequency, where $\bm{v_{ds}} = \bm{e_b} \cross \left(v_{\parallel}^2 \bm{\kappa} + v_{\perp}^2 \grad \ln B / 2\right)/\Omega_{s}$ is the drift velocity, with $\Omega_s$ is the gyrofrequency of species $s$. $\bm{e_b} = \bm{B}/B$ denotes the unit vector along the magnetic field line, $\bm{\kappa} = \left(\bm{e_b} \cdot \grad\right) \bm{e_b}$ gives the magnetic-curvature vector, $i$ is the imaginary unit, and $J_0$ the zeroth-order Bessel function of the first kind with argument $k_\perp v_\perp / \Omega_s$. Furthermore, in Eqn.~(\ref{eq:GKE}) $\omega_{\star s}^T = \omega_{\star s} \left(1 + \eta_s (E/T_s - 3/2)\right)$ is the temperature-dependent diamagnetic frequency, with $\eta_s = \norm{ \bm{\nabla}\ln T_s}/\norm{\bm{\nabla} \ln n_s}$ the gradient length ratio, $E=m v^2 / 2$ is the particle energy and $\omega_{\star s} = T_s \left(\bm{k_{\perp}}\cross \bm{e_b}\right) \cdot \grad{\ln n_s}/\left(q_s B\right) $ the diamagnetic frequency. Lastly, in Eqn.~(\ref{eq:GKE}) $\hat{C_{\bm{k}}}$ denotes the corresponding Fourier-mode component of the gyro-averaged linearised collision operator \cite{Abel2008}
\begin{equation}
    \hat{C_{\bm{k}}} = \langle{\exp(i\bm{k_\perp}\cdot\bm{\rho_s})C_s\left[\hat{g}_s\exp(-i\bm{k_\perp}\cdot\bm{\rho_s})\right]\rangle}_{\bm{R}}
    \label{eq:lincol}
\end{equation}
with $\langle \cdots \rangle_{\bm{R}}$ denoting the gyroaverage and $C_s = \sum_{s'} C_{ss'}$ is the total linearised collision operator. \par 
In general, plasma collisions would be described by a Fokker-Planck collision operator as highlighted in Appendix~\ref{app:collision-formal}. While it is possible to include a realistic collision operator in the gyrokinetic framework by expanding the distribution function in a basis of Hermite-Laguerre polynomial, resulting in a hierarchy of gyromoment equations which has to solved\cite{Jorge2018,Frei2022,Frei2023}. While this method has been applied to successfully TEMs in the past\cite{Frei2023}, it was found that the necessary truncation order to reach convergence in the core region of the plasma where the influence of magnetic drifts dominates was considerable. Since the same work also found that for purely density-gradient driven TEM, which we shall consider momentarily, the TEM eigenfrequencies are fairly insensitive to the chosen model for the collision operator, which is in agreement with the numerical work from Ref.~\onlinecite{Pan2020}. Hence for the analytical work, we use an energy-dependent Krook collision operator $C_s = - \hat{\nu_s} (v_{Ts}/v)^3 g_s$ instead of a more elaborate collision operator, as also done in e.g. Refs.~\onlinecite{Rafiq2006,Rewoldt1977,Dominguez1992}, though later in the numerical work this restriction is relaxed. Despite its simplicity, the Krook operator is not overly restrictive, the form of the Krook operator represents the tendency of collisions to drive the distribution to a local equilibrium, and the energy-dependence of the collision frequency reflects roughly that of the proper Fokker-Planck operator (with the result being exact for the case of electron-ion collisions). Properly accounting for such energy dispersion in the collision rate has been found to be crucial in order to reproduce collisional effects governed by Coulomb interactions\cite{Livi1986}. And indeed, in Section~\ref{sec:GENEvsperturb} we shall see that the analytical findings obtained with this Krook operator are in close agreement with the simulation results. \par
The gyrokinetic equation Eqn.~(\ref{eq:GKE}) will have to be solved for all species $s$ in conjunction with Poisson's equation for a self-consistent solution of the electrostatic potential $\phi$ and density perturbations $\delta n_s = \int \delta f_s d^3{\bm{v_s}}$. As, under typical fusion conditions, the length scales of interest exceed the Debye length \cite{Garbet2010,Tang1978}, this reduces to an extension of the familiar quasi-neutrality condition to the perturbations, resulting in
\begin{equation}
    \sum_s n_{s} \frac{q_s^2}{T_s} \hat{\phi}(\bm{x}) = \sum_s q_s \int \hat{g}_s(\bm{R},E,\mu) J_0\left(\frac{k_\perp v_\perp}{\Omega_s}\right) \dd^3{\bm{v}},
    \label{eq:perturb-QN-condition}
\end{equation}
where $n_s$ the equilibrium density of species $s$.

\subsection{Perturbative TEM description \label{sec:perturb-machinery}} 
The preceding section considers a general description of electrostatic microinstabilities. To analyse TEMs in particular, we will consider a two-species plasma with singly charged ions and narrow down the parameter regime of the perturbations. As a result of the mass difference between ions and electrons, in order to avoid strong Landau damping on either species, the frequency of perturbations which can be driven unstable to an appreciable degree must lie in the range $k_\parallel v_{Ti} \ll \omega \ll k_\parallel v_{Te}$ \cite{Garbet2010,Helander2013}. Based on this timescale separation, the first term in Eqn.~(\ref{eq:GKE}) can be neglected for ions, resulting in an approximate solution for the non-adiabatic ion response
\begin{equation}
    \hat{g}_i \approx \frac{e}{T_i}\frac{\left(\omega - \omega_{\star i}^{T}\right)F_{Mi}}{\omega - \omega_{di} + i \nu_{i}} J_0(k_\perp \rho_i) \hat{\phi}.
\end{equation}
For the electrons, the first term on the left-hand side of Eqn.~(\ref{eq:GKE}) is dominant to lowest order, such that $\hat{g}_e$ is approximately constant along the field line. To next order in the gyrokinetic equation, the solution to $\hat{g}_e$ follows from taking the bounce-average as\cite{Dominguez1992}
\begin{equation}
    \hat{g}_e = \frac{-e}{T_e} \frac{\left(\omega-\omega_{\star e}^{T}\right)F_{Me}}{\omega - \overline{\omega_{de}}+i\nu_{e}} \overline{J_0(k_\perp \rho_e)\hat{\phi}},
    \label{eq:BAV-ge}
\end{equation}
where the bar denotes the bounce-average, defined by
\begin{equation}
    \overline{A} = \frac{\int_{l_1}^{l_2} A \dd{l}/\abs{v_\parallel} } {\int_{l_1}^{l_2} \dd{l}/\abs{v_\parallel}}.
\end{equation}
Here $l_{1,2}$ are the bounce points defined by $v_\parallel(l_{1,2})=\sqrt{2(E-\mu B(l_{1,2}))/m} = 0$, with $\mu = mv_{\perp}^2 /2 B$ the magnetic moment. As passing particles can sample the fluctuations over the complete extent of the field line, their bounce-average $\overline{\phi}\approx 0$ vanishes, such that Eqn.~(\ref{eq:BAV-ge}) effectively only describes the trapped electrons, whose bounce-averaged drift $\overline{\omega_{de}}$ results in an effective precession of the trapped particle orbits within the flux surface, and a radial drift across the flux surface, where the latter should be nearly vanishing as well to allow for good confinement of trapped particles\cite{Helander2014}. Inserting these solutions into the quasi-neutrality condition Eqn.~(\ref{eq:perturb-QN-condition}), yields a simplified description of TEMs per
\begin{eqnarray}
    (1+\tau)\hat{\phi} \approx  && \int \frac{\omega - 
    \omega_{\star i}^{T}}{\omega-\omega_{di}+i\nu_{i}}\frac{F_{Mi}}{n_0}J_0(k_\perp \rho_i)^2 \hat{\phi}\dd^3{\bm{v_i}}  \nonumber \\
    && + \tau \int \frac{\omega - 
    \omega_{\star e}^{T}}{\omega-\overline{\omega_{de}}+i\nu_{e}}\frac{F_{Me}}{n_0} \overline{\hat{\phi}}\dd^3{\bm{v_e}}
    \label{eq:QN-TEM}
\end{eqnarray}
where quasi-neutrality of the equilibrium $n_e \approx n_i = n_0$ has been used, $J_0 \approx 1$ was set for the electrons as TEMs occur on similar spatial scales as the ITG mode ($k_\perp \rho_i \sim 1$) \cite{Garbet2010,Merz2010}, and $\tau = T_i/T_e$ has been introduced.\par
The main difficulty in solving Eqn.~(\ref{eq:QN-TEM}) is that the trapped-electron response depends on the bounce-average of the mode structure, whereas the mode structure itself remains unknown. As only trapped electrons contribute to the second integral of Eqn.~(\ref{eq:QN-TEM}), that integral would be proportionally smaller to the first integral of Eqn.~(\ref{eq:QN-TEM}), describing the kinetic response of all ions, by the trapped-particle fraction, barring the effects of precession resonances. As the trapped-particle fraction is typically small\footnotemark[1234]\footnotetext[1234]{At least in (conventional) tokamaks where the trapped-particle fraction scales with the (small) inverse aspect ratio they are, though for spherical tokamaks such approximations are bound to fail.Likewise, in stellarators, this fraction depends on the helical-ripple instead, and can be substantially larger than in an equivalent tokamak with the same aspect ratio \cite{Guttenfelder2008,Beidler2011}.}, a perturbative approach is used to deal with the trapped-electron term, as considered in e.g. Refs.~\onlinecite{Connor2006,Gang1991,Adam1976,Dominguez1992}. Identifying Eqn.~(\ref{eq:QN-TEM}) as a dispersion relation $D(\omega,k_\perp,\hat{\phi})=0$, the lowest-order mode frequency $\omega_0$ will be determined from Eqn.~(\ref{eq:QN-TEM}) by first neglecting all trapping effects, resulting in a dispersion relation $D_0(\omega_0,k_\perp,\hat{\phi})=0$. Subsequently, the shift in frequency $\delta \omega$ is determined by accounting for the trapped-electron dynamics via 
\begin{equation}
    \pdv{D_0(\omega_0,k_\perp,\hat{\phi})}{\omega} \delta \omega = D_1 (\omega_0,k_\perp,\hat{\phi})
    \label{eq:perturbative-main}
\end{equation}
with $D_1$ given by the last term in Eqn.~(\ref{eq:QN-TEM}), and it has been assumed that $\delta \omega/\omega_0 \ll 1$ because of the small trapped-particle response, warranting an expansion of Eqn.~(\ref{eq:QN-TEM}) around $\omega_0$. It should then be confirmed a posteriori whether resonance effects $\omega_0 \sim \overline{\omega}_{de}$ remain sufficiently mild for the trapped-electron response to be considered as perturbative correction. \par

For simplicity, the role of collisions on ions will be neglected altogether, as the ions are too slow to be affected by (de-)trapping, and the ion-electron collision frequency is smaller than the electron-ion collision frequency (which scatter most electron momentum and therefore have largest contribution to electron (de-)trapping) by the mass ratio\cite{Helander2002,Hirshman1976}. Consequently, any collisional effect on the growth rate would rather be dominated by collisions on the electrons. The role of the ion magnetic drift is also neglected, as TEMs are not driven by ion curvature. Furthermore, the Finite Larmor Radius (FLR) effect from retaining $J_0(k_\perp \rho_i)$ in the dispersion relation, has a larger influence on the mode frequency $\omega_0$ than the ion drift would if it were taken into account\cite{Hahm1991}. Including the ion drift and collisions would only result in a small correction to $\omega_0$ and a weak damping respectively, as shown in Appendix~\ref{app:correctionsw0}. \par
\begin{figure}[!ht]
    \centering
    \includegraphics[width=0.7\linewidth]{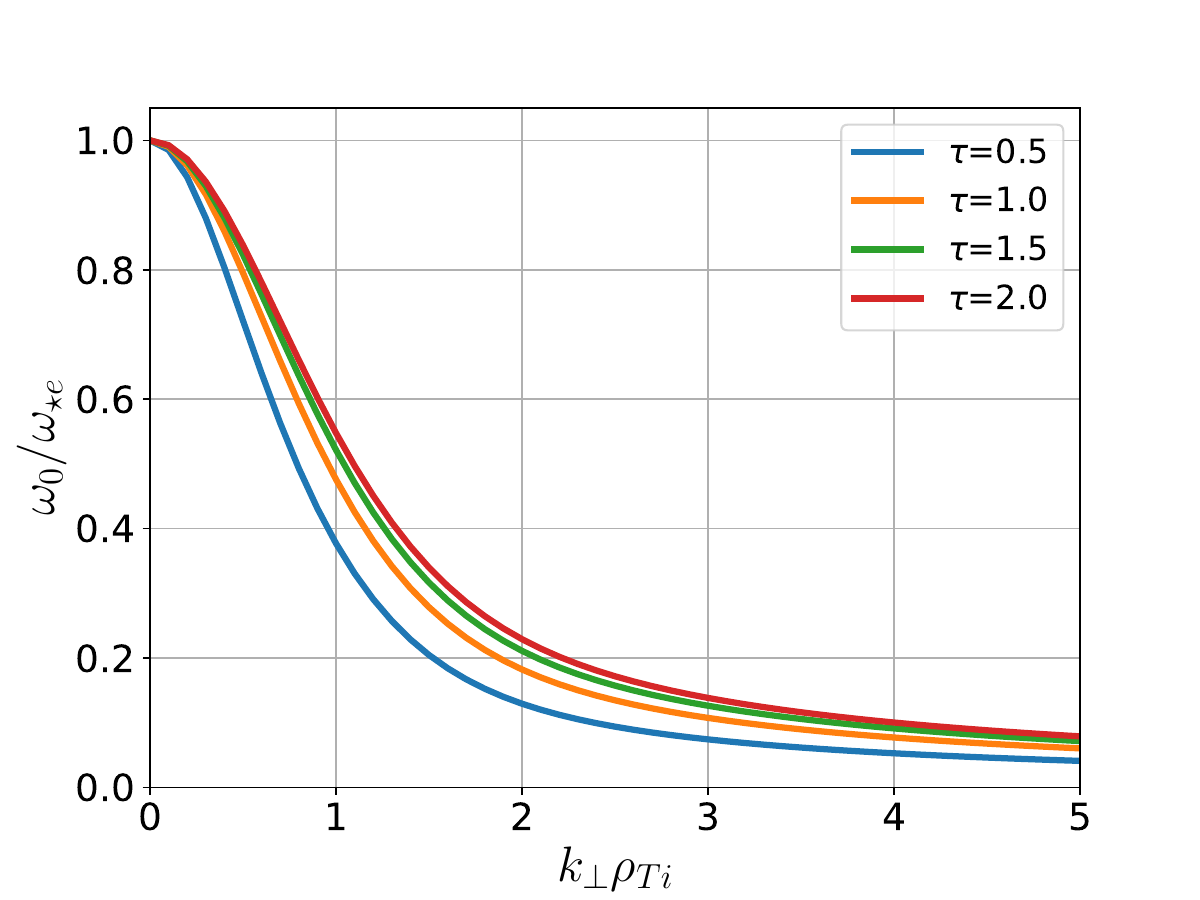}
    \caption{The lowest-order mode frequency of the perturbative approach $\omega_0$ as a function of perpendicular wave number $k_\perp \rho_{Ti}$ for various temperature ratios $\tau$, in the absence of temperature gradients ($\eta_i = \eta_e = 0$). The FLR effect has a significant influence on the allowed mode frequencies, while $\omega_0$ is rather insensitive to the temperature ratio.}
    \label{fig:omegazero}
\end{figure}

 
Consequently, the lowest-order dispersion relation simplifies to 
$$D_0 = (1+\tau) \hat{\phi} - \int \left(1 - 
\frac{\omega_{\star i}^{T}}{\omega}\right)\frac{F_{Mi}}{n_0}J_0(k_\perp \rho_i)^2 \hat{\phi}\dd^3{\bm{v_i}}.$$ 
As a last simplification, to focus solely on TEMs, only a density gradient will be considered, such that $\eta_i = \eta_e = 0$. The ion integral then becomes straightforward to evaluate, resulting in a lowest-order mode frequency 
\begin{equation}
    \frac{\omega_0}{\omega_{\star e}} =  \frac{\tau \Gamma_0(\zeta) }{1+\tau -\Gamma_0(\zeta)}
    \label{eq:omzero}
\end{equation}
which is independent of the electrostatic potential. In Eqn.~\ref{eq:omzero} we have introduced the functions $\Gamma_n(\zeta) = I_n(\zeta)\exp(-\zeta)$ with $I_n$ the $n$th-order modified Bessel function and $\zeta = (k_\perp \rho_{Ti})^2 / 2 = k_\perp^2 T_i / m_i \Omega_i^2 $, with $\rho_{Ti}$ the thermal ion gyroradius. This solution is plotted for various temperature ratios $\tau$ in Fig.~\ref{fig:omegazero}, which shows that, regardless of the the temperature ratio, the mode corresponds to a stable electrostatic drift wave in the electron diamagnetic direction, making a resonance with the precession frequency of the trapped-electrons possible in the next step. 

When proceeding to the next order in the trapped-particle fraction by including the trapped-electron response in the dispersion relation, again the difficulty arises that the dispersion relation becomes non-local. Nevertheless, the frequency shift $\delta \omega$ can still be extracted be multiplying Eqn.~(\ref{eq:perturbative-main}) with $\hat{\phi}^{*}$ and integrating along the field line in ballooning space, resulting in 
\begin{eqnarray}
    && \frac{\delta \omega}{\omega_0} =  \left(\frac{2}{\sqrt{\pi}} \frac{\omega_0-\omega_{\star e}}{\Gamma_0(\zeta)} \frac{\omega_0}{\omega_{\star e}} \bigg/ \oint \abs{\hat{\phi}}^2 \frac{\dd{l}}{B}\right) \int_{1/B_{max}}^{1/B_{min}}  \dd{\lambda} \nonumber \\ 
    && \times \sum_{w(\lambda)} \ \int\limits_{x_{cut[w]}}^{\infty} \dd{x}  \frac{x^2\exp(-x^2)}{\omega_0 - \overline{\omega_{de}}_{[w]}+i\nu_{e}} \abs{\overline{\hat{\phi}}_{[w]}}^2 L_{eff[w]}
    \label{eq:perturbfreq}
\end{eqnarray}
where pitch-angle velocity coordinates $x = v/v_{Te}, \lambda = \mu / E, \sigma = v_\parallel/\abs{v_\parallel}$ are used for the electrons, the order of integration over field-line arc length and pitch angle have been interchanged, and the bounce-averages are distinguished between different trapping wells $w$ by
\begin{eqnarray}
    \overline{A}_{[w]} = && \int_{l_{1[w]}}^{l_{2[w]}} \frac{A\dd{l}}{\sqrt{1-\lambda B}}  \bigg/  \int_{l_{1[w]}}^{l_{2[w]}} \frac{\dd{l}}{\sqrt{1-\lambda B}} \nonumber \\
    && = \int_{l_{1[w]}}^{l_{2[w]}} \frac{A \dd{l}}{\sqrt{1-\lambda B}}\bigg/L_{eff[w]}
    \label{eq:bounceav-def}
\end{eqnarray}
with $L_{eff[w]}$ the effective arc length along the well. In Eqn.~\ref{eq:perturbfreq} the lower bound on the electron energy integral is $x_{cut[w]} = \sqrt[4]{\nu^{ei}_{90}/\omega_{be[w]}^T}$ with $\nu^{ei}_{90}$ the $90^\circ$ scattering frequency and $\omega_{be[w]}^T = \pi v_{Te}/L_{eff[w]}$ the bounce frequency of a thermal electron, which accounts for the reduction in trapped-particle fraction by de-trapping.
By summing over the different trapping wells along the field line, all populations of trapped-particles at a given pitch-angle $\lambda$ are accounted for. Nevertheless, it is no longer possible to further analytically simplify Eqn.~(\ref{eq:perturbfreq}) without making assumptions on the relative magnitude of the collision term as done in the various publications cited before. As such assumptions typically undermine the energy dependence of the precession frequency $\overline{\omega_{de}}\propto x^2$ and collision frequency $\nu_e \propto x^{-3}$, they are not generally valid over the full integration. We thus propose to directly solve Eqn.~(\ref{eq:perturbfreq}) numerically instead, which also allows us to account for more general magnetic geometries beyond the simplified low-aspect-ratio limit of a circular tokamak, for which the bounce-averages can still be obtained analytically. 

\section{Evaluation of perturbative growth rates \label{sec:perturbative}}
From the perturbative paradigm of Section~\ref{sec:perturb-machinery} it follows that the lowest-order mode frequency $\omega_0$ is purely real-valued, and any growth rate thus has to be caused by the trapped-electron response to these modes, and can accordingly be obtained from the perturbative frequency shift $\gamma=\Im[\delta \omega]$. This perturbative frequency shift Eqn.~(\ref{eq:perturbfreq}) depends strongly on the magnetic geometry through the arrangement of magnetic wells $w$, the precession drift $\overline{\omega_{de}}$ and, indirectly, the bounce-average Eqn.~(\ref{eq:bounceav-def}), whereas the potential perturbation $\phi$ only enters through the bounce-average and an overall normalisation factor. Therefore, to focus on the inherent differences caused by the magnetic geometry, we make use of the so-called flute-mode approximation\cite{Connor2006,Adam1976} $\hat{\phi}\approx \phi_0$, which makes the frequency shift independent of the mode structure\footnote{Even if this approximation were not made, an arbitrary $\phi$ perturbation would only result in an overall shape-factor correction to $\delta \omega$ since Eqn.~(\ref{eq:perturbfreq}) is insensitive to the overall amplitude of the perturbation.}. \par 
With this approximation we solve Eqn.~(\ref{eq:perturbfreq}) in flux-tube geometry\cite{Beer1998} for the DIII-D tokamak and the HSX and W7-X stellarators to highlight differences to the growth rate caused by the magnetic geometry. The GIST (Geometric Interface for Stellarators and Tokamaks) code\cite{Xanthopoulos2009} was used to generate the geometrical parameters of the flux tubes from the respective magnetic equilibria of the devices. As mentioned, for W7-X we consider the (vacuum) high-mirror configuration (unless otherwise explicitly stated), whereas for HSX the standard quasi-helically symmetric configuration is used.
The flux tubes are all chosen to span a single poloidal turn and at a normalised toroidal flux label of $s = \psi / \psi_a = 0.5$, with $\psi_a$ the toroidal magnetic flux at the plasma edge. In choosing these flux-tube parameters identical, the rotational transform $\iota$ and (global) magnetic shear $\hat{s}$ differ between the geometries and are equal to $\iota = 0.39, 1.06, 0.91$ and $\hat{s} = 1.58, -0.045, -0.13$ for DIII-D, HSX, and W7-X respectively. For both stellarators the so-called ``bean'' flux-tube is used -- it is the flux tube that crosses the bean-shaped poloidal cross-section at the outboard midplane \cite{Proll2012}. \par 
For compatibility with later gyrokinetic flux-tube simulations, rather than varying the perpendicular wavenumber, we vary the bi-normal wavenumber $k_y$ instead, which relates to the perpendicular wavenumber through\cite{Faber2018} $ k_\perp = \sqrt{g^{xx} k_x^2 + 2g^{xy}k_x k_y + g^{yy} k_y^2} $ with $k_x$ the ``radial'' wavenumber. Here $g^{ij}$ denote the elements of the metric tensor for the flux-tube coordinates $(x,y,z)$, with $z$ the field-aligned coordinate taken to be the poloidal angle, and $x,y$ are the two perpendicular coordinates of the flux-tube. These are, respectively, related to the radial and bi-normal Clebsch coordinates $\psi,\alpha$ for the magnetic field $\bm{B} = \grad{\psi} \cross \grad{\alpha}$ by $x=a(\sqrt{s}-\sqrt{s_0}), \ y = a \sqrt{s_0} (\alpha-\alpha_0)$, where a $0$ subscript refers to quantities evaluated at the central magnetic field line, and $a$ is the minor radius\cite{Mynick2009}. As the metric varies along the field line, the perpendicular wavenumber $k_\perp$ will not be constant throughout the flux-tube for a given perpendicular wavevector $\bm{k_\perp} = k_x \grad{x} + k_y \grad{y}$. For consistency with Eqn.~(\ref{eq:perturbfreq}) we therefore consider the the average perpendicular wavenumber\cite{Xie2020}
\begin{equation}
    \langle k_\perp \rho_{Ti} \rangle \approx \int \sqrt{g^{yy}k_y^2 \rho_{Ti}^2} \abs{\phi}^2 \frac{\dd{l}}{B} \bigg/ \int \abs{\phi}^2 \frac{\dd{l}}{B}
    \label{eq:ky-kperp}
\end{equation}
where the radial wavenumber has been approximated as $k_x \approx 0$. This is motivated by the observation that the most unstable mode in simulations of tokamaks and bean flux tube of stellarators are strongly radially localised at the outboard side \cite{Proll2013}. In low-shear devices, including many stellarators, however, modes can become very broad in ballooning space \cite{Faber2015,Faber2018}, thus adding substantial contributions at finite $k_x$. However, since it is not possible to assess the most unstable radial wavenumber a priori and we focus on bean flux tubes for both stellarators, it is reasonable to consider only the bi-normal wavenumber $k_y$. Meanwhile we use the code package  developed in Ref.~\onlinecite{Mackenbach2023} for evaluating the bounce-averaging integrals.

\begin{figure*}[!ht]
 \centering
    \begin{subfigure}{.40\linewidth}
        \centering
        \includegraphics[width=\linewidth,height=125pt]{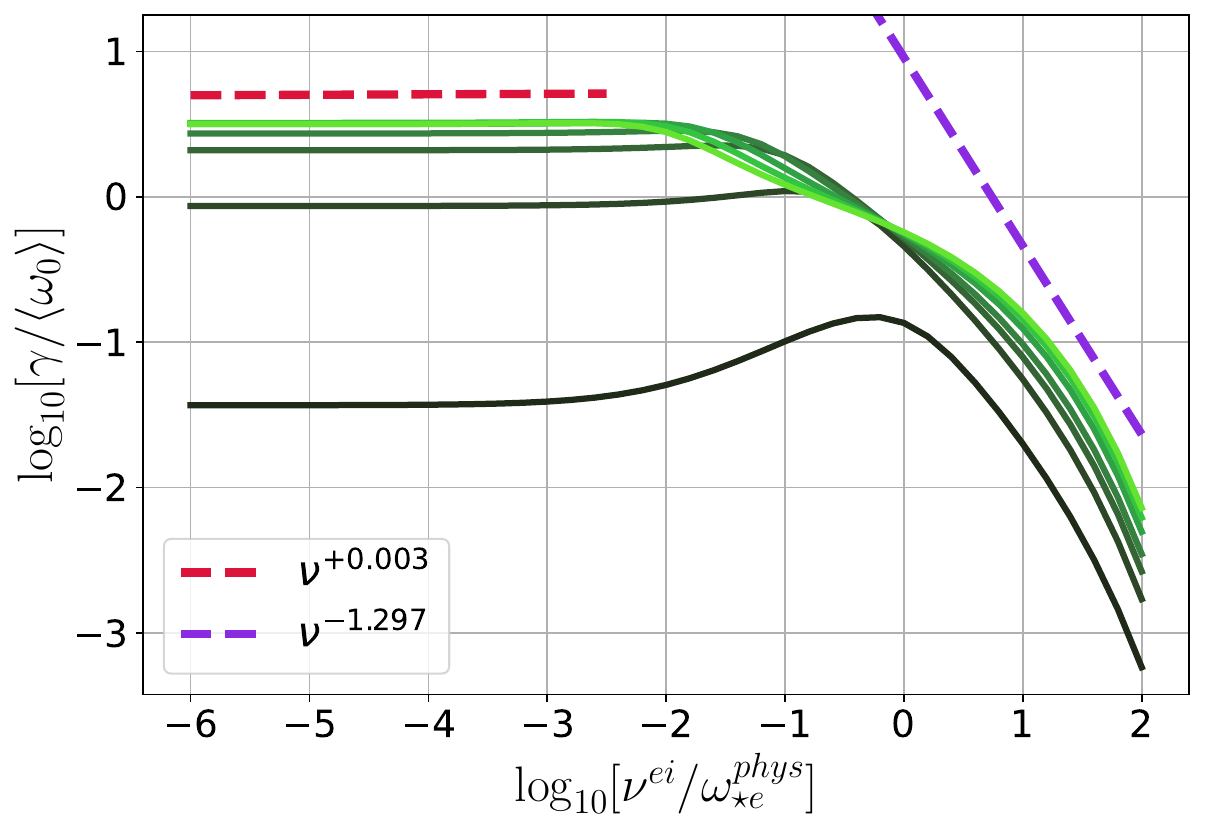}
        \caption{}      
        \label{fig:perturb-D3D-scaling}
    \end{subfigure}%
    \begin{subfigure}{.52\linewidth}
        \centering
        \includegraphics[width=\linewidth,height=125pt]{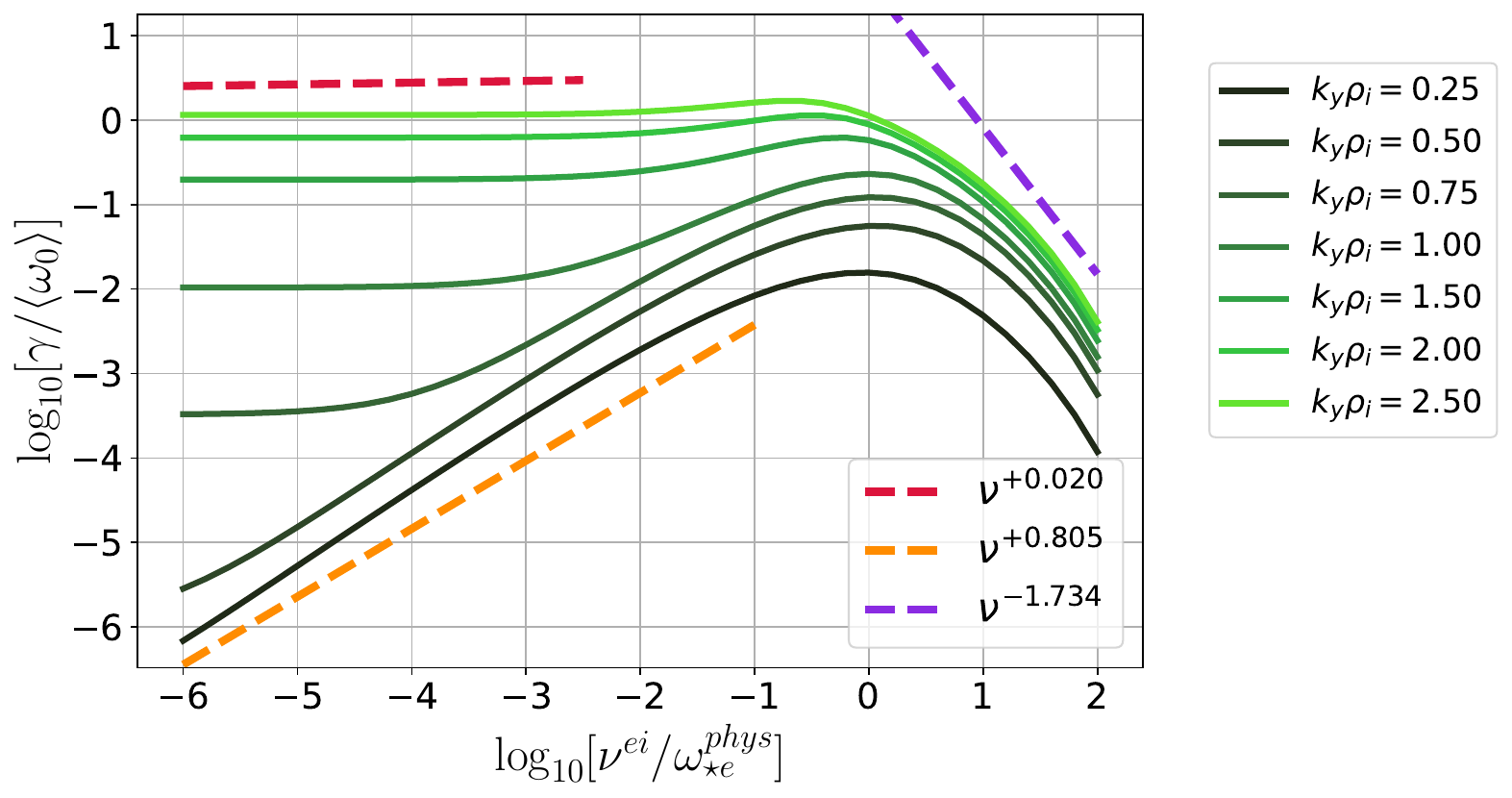}
        \caption{}        
        \label{fig:perturb-HSX-scaling}
    \end{subfigure}
\caption{\label{fig:D3D-HSX_deltagam} Perturbative growth rate as obtained by numerically integrating Eqn.~(\ref{eq:perturbfreq}) using the magnetic geometries of (a) DIII-D, and (b) HSX as a function of the collision frequency for several wavenumbers, in case of a density gradient of $a/L_n = 3$. The scaling laws for the (low collisionality) collisionless, destabilising and (high collisionality) stabilising regimes are, respectively, indicated by the red, orange and purple dotted lines. Note the absence of a destabilising regime in the DIII-D geometry.}
\end{figure*}

\begin{figure*}[!t]
 \centering
    \begin{subfigure}{.40\linewidth}
        \centering
        \includegraphics[width=\linewidth,height=125pt]{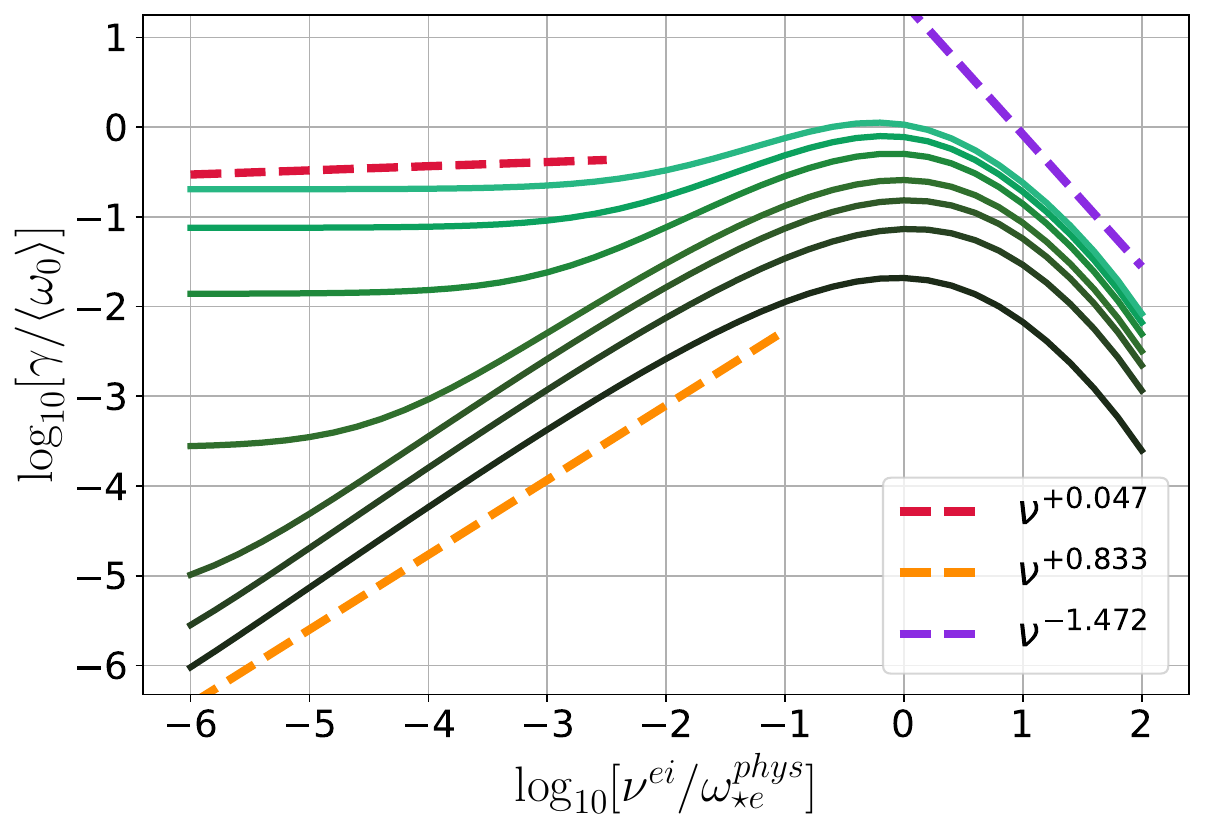}
        \caption{}        
        \label{fig:perturb-W7X-vac-scaling}
    \end{subfigure}
    \begin{subfigure}{.52\linewidth}
        \centering
        \includegraphics[width=\linewidth,height=125pt]{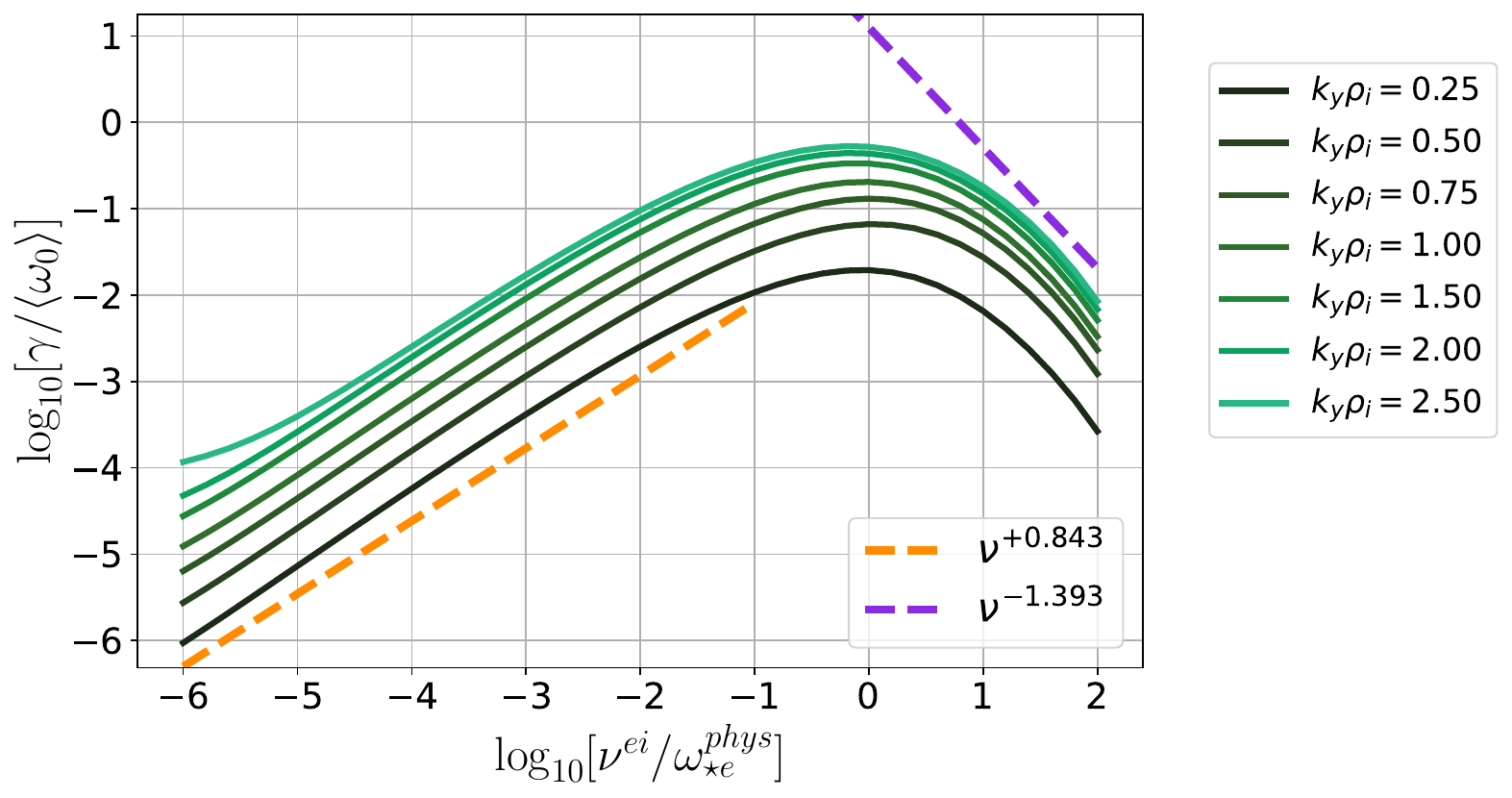}
        \caption{}
       \label{fig:perturb-W7X-beta-scaling}
    \end{subfigure}
\caption{\label{fig:W7X_deltagam} Perturbative growth rate Eqn.~(\ref{eq:perturbfreq}) for the geometry of the high-mirror configuration of W7-X at (a) vacuum, and (b) high plasma pressure of $\beta=5\%$, as a function of the collision frequency for several wavenumbers at a density gradient of $a/L_n = 3.0$. Note the absence of a collisionless regime in the high $\beta$ case, indicating stability against collisionless trapped electron modes.}
\end{figure*}

The variation of the growth rate with collision frequency for various wavenumbers obtained with these simplifications is shown in Figs.~\ref{fig:perturb-D3D-scaling},~\ref{fig:perturb-HSX-scaling},~\ref{fig:perturb-W7X-vac-scaling} for DIII-D, HSX, and W7-X respectively. Here the collision frequency is normalised to $\omega_{\star e}^{\textrm{phys}} = T_e/(e B_{ref} a \rho_{i})$, corresponding to the electron diamagnetic frequency at wavenumber $k_y \rho_i = 1$ and density gradient $a/L_n = 1$, with $\rho_i = \sqrt{T_e/m_i}/\Omega_{\textrm{ref},i}$ the ion gyroradius at sound speed and reference magnetic field strength\footnote{Note that this choice for the wavenumber normalisation introduces an additional $B_{\textrm{ref}}/B$ factor in Eqn.~(\ref{eq:ky-kperp}) over which the normalised perpendicular wavenumber $k_\perp \rho_i$ will be averaged} $B_{\textrm{ref}}$, and $L_n = -(\dd{\ln n}/\dd{x})^{-1} $ is a scale length for the density gradient. \par 
In the case of DIII-D (Fig.~\ref{fig:perturb-D3D-scaling}), a clear collisionless regime driven by a resonance with the electron precession drift can be identified at low collisionality from the near invariance of the growth over several order of magnitude in the collision frequency, which then transitions into a dissipative regime at high collisionality where the collisions stabilise the growth rate. At intermediate collisionality a transition between these regimes occurs, where only at the lowest wavenumber of $k_y \rho_i = 0.25$ a weak destabilisation occurs. \par
In both stellarator geometries, a similar stabilisation at high-collisionality is observed, although the rate of the stabilisation differs among the geometries, varying between a $\nu^{-1.3}$ and $\nu^{-1.7}$ scaling. These scalings are significantly stronger than an inverse proportionality $\gamma \propto 1/\nu$ which would be expected based on applying the ordering $\nu_e \gg \omega_0,\overline{\omega_{de}}$ to Eqn.~(\ref{eq:perturbfreq}) as commonly done in considerations for the dissipative TEM limit\cite{Dominguez1992,Gang1991,Connor2006}. The discrepancy and non-universality among the geometries of the scaling at high collisionality is due to the influence of the collisional cut-off in the energy integral of Eqn.~(\ref{eq:perturbfreq}), which is assumed negligible in aforementioned works. If this term is artificially set to zero, the growth rates in each device are found to asymptotically scale as $\gamma \propto \nu^{-1}$ at the highest collisionality.\par

In contrast, at low collisionality, the stellarator geometries support both a collisionless regime for high wavenumber modes like in the tokamak case, but a destabilised regime is observed for low wavenumber modes, where the growth rate is found to increase almost linearly proportionally with the collision frequency. Such a near linear scaling of the growth rate at low collisionality can be explained by an expansion of the velocity space integrand in Eqn.~(\ref{eq:perturbfreq}) around $\nu_e = 0$, which, in absence of a resonance with the electron precession drift, results in a small imaginary part of $\delta \omega$ purely due to collisions. Such behaviour would be expected in perfect maximum-$\mathcal{J}$ devices, which have favourable stability properties against collisionless density-gradient-driven TEMs. As the vacuum configuration for W7-X is only approximately maximum-$\mathcal{J}$, some populations of trapped particles can still be in resonance, but these will be much fewer than in HSX, where most trapped particles experience bad bounce-averaged curvature. This reduction in resonant trapped particles results in roughly one order of magnitude lower growth rates in the collisionless regime, as well as a more prominent presence of a linearly destabilised regime at low collisionality for low-wavenumber modes in W7-X compared to HSX. Furthermore, the high-wavenumber modes identified as resonant TEMs in W7-X are observed to be more prone to collisional influences by the earlier onset of a transition region, characterised by the increase in growth rates at intermediate collisionality, as compared to HSX, also indicative of the weaker resonant TEM drive characteristic of the approximate maximum-$\mathcal{J}$ feature. The significant observed differences between the stellarator devices cannot be explained based on the minor difference in $\langle k_\perp \rho_{Ti} \rangle$ introduced by the metric in Eqn.~(\ref{eq:ky-kperp}) which has only a moderate influence on the growth rate normalisation by $\langle \omega_0 \rangle$. \par

To verify the robustness of a true maximum-$\mathcal{J}$ configuration against collisionless density-gradient-driven TEMs, we also consider the perturbative growth rate Eqn.~(\ref{eq:perturbfreq}) in the magnetic geometry for a high-$\beta$ equilibrium of the W7-X high-mirror configuration, which does achieve the maximum-$\mathcal{J}$ feature to a much higher degree of approximation. The resulting growth rates are shown in Fig.~\ref{fig:perturb-W7X-beta-scaling}. There, as theory predicts, at low collisionality a collisionless regime is indeed absent and the growth rates at all wavenumbers asymptotically decay to zero at zero collisionality. However, at finite collisionality, we observe that dissipative TEMs driven unstable through de-trapping can thrive in such geometries, with growth rates initially increasing with collision frequency before ultimately being stabilised when de-trapping becomes too frequent at high collisionality.

From the results of the perturbative growth rates in Figs.~\ref{fig:D3D-HSX_deltagam} and~\ref{fig:W7X_deltagam}, it can be inferred that the perturbative method is capable of providing analytical insight into the sensitivity of TEMs to collisions in different geometries. However, in the same figures, some of the growth rates are found in the neighbourhood of $\gamma/\langle\omega_0\rangle \sim \mathcal{O}(1)$, especially in the collisionless regime of each geometry. At such large growth rates, the validity of a perturbative approach based on $\delta \omega/\omega_0 \ll 1$ breaks down. For the validity of this assumption, not just the growth rates but also the shift in propagation frequencies $\Re[\delta \omega]$ should remain small, as otherwise there is a non-negligible adjustment to the resonance $\Re[\omega]\approx \overline{\omega}_{de}$, which further limits the range of validity in parameter space. For completeness, contours of $\Im[\delta \omega],\Re[\delta \omega]$ and $\abs{\delta \omega}$ over the complete $(k_y\rho_i, \nu^{ei}/\omega_{\star e}^{\textrm{phys}})$ parameter range considered are provided in Appendix~\ref{app:contours} for the three main geometries of this work. Hence we must conclude that in most considered configurations the trapped-electron response would not generally permit a perturbative treatment. This may either be attributed to strong resonant effects (as is the case of DIII-D), a considerable fraction of trapped particles (as potentially the case for-stellarators, see\footnotemark[1234]), or for the large $k_y$ cases the sharp decrease in $\omega_0$, see Fig.~\ref{fig:omegazero}.\par 
Despite the limited parameter regime in which the perturbative approach thus provides quantitative valid answers, the scalings of the growth rates obtained from this method are at least expected to provide a qualitative measure for the influence of collision on the TEM instability, since the instability described by this approach is purely driven by the trapped electrons. Therefore, to further test these scalings and obtain accurate results for the growth rate independent of the various approximations made in Section~\ref{sec:gentheory}, gyrokinetic simulations have also been performed and will be the topic of the remainder of this paper.

\section{Simulation parameters \label{sec:GENEgen}}
Linear gyrokinetic simulations are performed with the local flux-tube version of the \textsc{GENE} code\cite{Jenko2000}. Such a local approach, however, neglects the influences of radial profile effects which play a role for devices with small $\rho^{\star}=\rho/a$ \cite{Oberparleiter2016,Candy2004,Parker1999}, neoclassical radial electric fields which are known to influence trapped-particle orbits\cite{Garcia2018,Landreman2011} and ITG turbulence saturation\cite{Sanchez2019,Fu2021}, and full-surface effects which impacts saturation scalings\cite{Navarro2020}. The latter two phenomena, however, predominantly only play an important role in non-axisymmetric geometries. For these simulations, the same flux tubes as in Section~\ref{sec:perturbative} are used, the temperature ratio is set to $\tau = 1$, and kinetic electrons are considered with a mass ratio of $m_i/m_e = 3670.48$ corresponding to an electron-deuterium plasma. At first a set of collisionless base-case simulations are performed to create a reference point for the collisional simulations, and for performing convergence test in parameter resolutions, scanning the bi-normal wavenumber over $k_y\rho_i = [0.1,0.2,\ldots,2]$. The converged resolutions are shown in Table~\ref{tab:res}. \par 
However, in the case of HSX, the resolution requirements to meet convergence for the lowest two $k_y$ values were impractically high for the purpose of investigating the change to the instabilities caused by collisions, and these wavenumbers are thus omitted. This can be attributed to simulating a flux tube consisting of only a poloidal turn, with the work of Ref.~\onlinecite{Faber2018} identifying unphysical parallel correlations along the simulation domain as a cause for poor convergence at low wavenumber. 
The same converged sets of resolutions are also used for the collisional simulations, with the exception of hyperdiffusion in velocity space \cite{Pueschel2010}, which is switched off as collisions themselves provide the necessary numerical dissipation. \par 

\begin{table}[!b]
\caption{\label{tab:res} Resolutions used in all simulation of this paper for each of the different geometries. $n_{kx}$, $n_{z},n_{v},$ and $n_{\mu}$ are the number of grid points in the radial direction, along the field line, in parallel velocity space and in the magnetic moment respectively, while $\epsilon_z$ denotes the strength of hyperdiffusion in the parallel direction according to the notation of Ref.~\onlinecite{Pueschel2010}, which has been assessed to give convergence. The strength of hyperdiffusion in parallel velocity for all geometries is set to $\epsilon_v = 0.2$ in the collisionless base-case simulations, and $0$ for the collisional simulations.}
\begin{ruledtabular}
 \begin{tabular}{c|c|c|c|c|c|}
            & $n_{kx}$ & $n_{z}$ & $n_{v}$ & $n_{\mu}$ & $\epsilon_z$ \\ \hline
         DIII-D & 21 & 64 & 48 & 16 & 0.75 \\
         HSX & 43 & 128 & 120 & 40 & 0.7 \\
         W7-X & 33 & 96 & 96 & 32 & 0.25\\
    \end{tabular}
\end{ruledtabular}
\end{table}

An additional convergence check in the velocity-space variables was performed to determine if the details of the collision operator are properly resolved. In almost all cases, both the mode structure and frequency were insensitive to resolution doubling. \par 
To focus on the role of collisions, we consider a fixed value of $a/L_n = 3$ for the density gradient, corresponding to a strong density-gradient drive, and vary the collision frequency. For the collision frequency, we scan over $\tilde{\nu} = [10^{-5}, 10^{-4}, 10^{-3}, 10^{-2}]$, where the \textsc{GENE} collision frequency is defined through
$\tilde{\nu}=2.30308\times 10^{-5} \  n[10^{19}\textrm{m\textsuperscript{-3}}] a[\textrm{m}] \ln\Lambda/T^{2}[\textrm{keV}]$, with the density and temperature measured at the central field line of the flux tube (i.e. at $s_0$). The chosen values for $\tilde{\nu}$ roughly correspond to the realistic variation of the collision frequency from the plasma core to the edge based on experimental profiles from DIII-D\cite{Grierson2018}, HSX\cite{Schmitt2014} and W7-X\cite{Beidler2021}. In the case of DIII-D, however, a numerical instability occurred at the highest collisionality for wavenumbers $k_y \rho_i = 1.9,2.0$, and these results are therefore omitted. \par 
For the linearised collision operator we only consider the pitch-angle scattering term for the test-particle part (corresponding to retaining only the first term in Eqn.~(\ref{eq:Maxwellcolop})) and a momentum- and energy conserving model for the field-particle part (see e.g. Refs.~\onlinecite{Abel2008,Sugama2009} for details and terminology) as also done in Refs.~\onlinecite{Colyer2017,Vernay2013,Manas2015} to assess the influence of collisionality on the linear growth rates in tokamaks. Only the pitch-angle scattering term of the collision operator is retained, as this term contains the relevant physics for (de-)trapping due to momentum deflection, and in the case of electron-ion collisions is the dominant contribution to the full Coulomb operator in the mass ratio. Furthermore, within the small mass ratio, the velocity dependence of the deflection rate ($\nu_{D}^{ab}$ in Eqn.~\ref{eq:coll-frequencies}) for electron-ion collisions reduces to $(v_{Te}/v)^3$, identical to the Krook operator, and thus this choice permits a fair comparison between the influence of collisional effects on the growth rate between the simulations and the earlier perturbative approach. Spot-checks with a full collision operator were performed at the highest collisionality where such differences are expected to be most prominent. Aside from an enhancement of the collisional shift in propagation frequency in the electron diamagnetic direction, no significant differences, like a transition of the dominant instability type, were observed, giving credibility that the conclusions drawn regarding (density-gradient driven) microinstability in this work will also hold for a more general collision operator. In the case of both stellarator geometries, we also assessed the sensitivity of the instabilities to the parallel extent of the simulation domain by means of spot-checking at the lowest and highest collision frequency against a flux-tube spanning two poloidal turns. The outcomes thereof are briefly discussed at the end of the relevant section of each geometry, but show that the observed trends regarding growth rate and dominant instability type remain valid. 

\section{Analysis of collisional linear gyrokinetic simulations \label{sec:sims}}

In this section, the changes to the mode frequency, growth rate, and parallel mode structure when the collision frequency is gradually increased, are studied by means of linear gyrokinetic simulations for each of the three aforementioned geometries. Note that by convention positive/negative oscillation frequencies correspond to modes propagating in the ion/electron diamagnetic direction, respectively. Furthermore, the electrostatic potential is normalised to its average value along the field line to emphasise its profile features with respect to with the geometrical features of the magnetic field. The magnetic curvature is represented by the geometric quantity $\mathcal{K}_y = \left(\bm{e_b} \cross \bm{\kappa}\right)\cdot\grad{y}$ from the GIST code \cite{Xanthopoulos2009}, which gives the contribution of the curvature component of the magnetic drift in the bi-normal direction, and for modes with $k_x \approx 0$ is thus proportional to the total local magnetic drift frequency at $\beta = 0$. Negative values of $\mathcal{K}_y$ correspond to regions of bad curvature with respect to the TEM resonance condition\cite{Helander2013,Proll2013}. Lastly, a poloidal angle of $\theta = 0$ corresponds to the outboard side of the torus, with $\theta = \pm \pi$ corresponding to the inboard side.

\subsection{DIII-D magnetic geometry}
\begin{figure*}[!ht]
 \centering
       \begin{subfigure}{.45\linewidth}
        \centering
        \includegraphics[width=\linewidth]{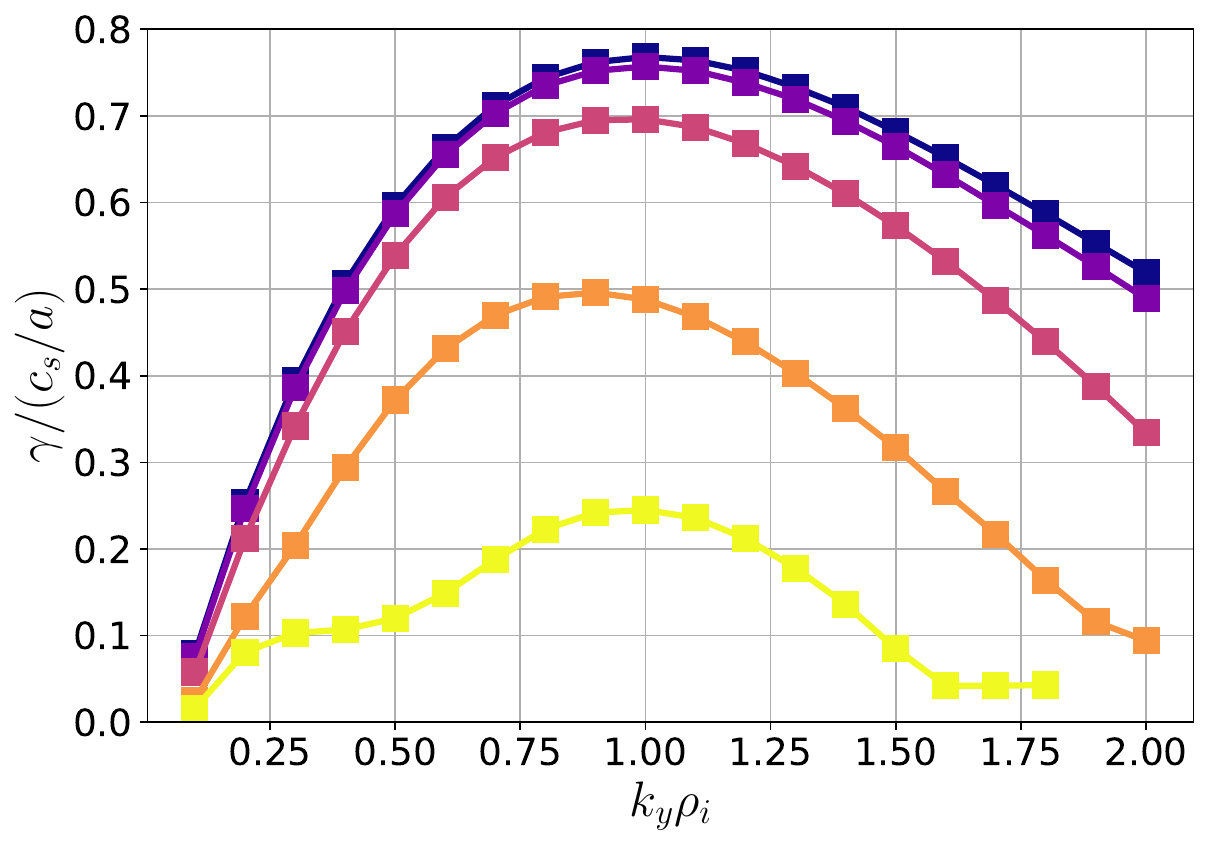}
        \caption{}        
        \label{fig:D3D-gamma}
    \end{subfigure}
    \begin{subfigure}{.45\linewidth}
        \centering
        \includegraphics[width=\linewidth]{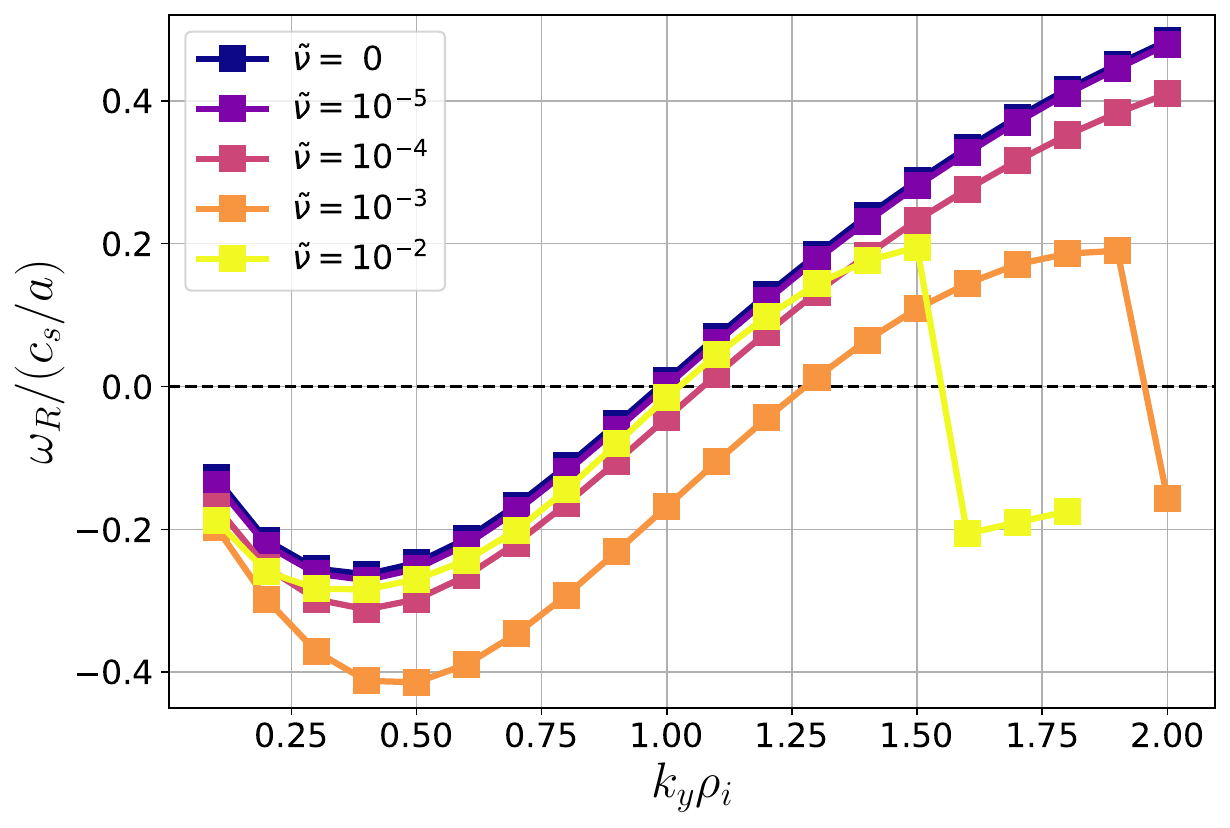}
        \caption{}      
        \label{fig:D3D-omega}
    \end{subfigure}%
\caption{\label{fig:D3D-sim} Normalised (a) growth rate and (b) oscillation frequency from linear gyrokinetic simulations in DIII-D geometry at a density gradient of $a/L_n = 3$ as the collision frequency is varied. Perceptually warmer colours indicate a higher collisionality. Note that in (a) the growth rate is decreasing monotonically with the collision frequency at all wavenumbers, indicating a purely stabilising influence of collisions.}
\end{figure*}

\begin{figure*}[!t]
 \centering
    \begin{subfigure}{.45\linewidth}
        \centering
        \includegraphics[width=\linewidth]{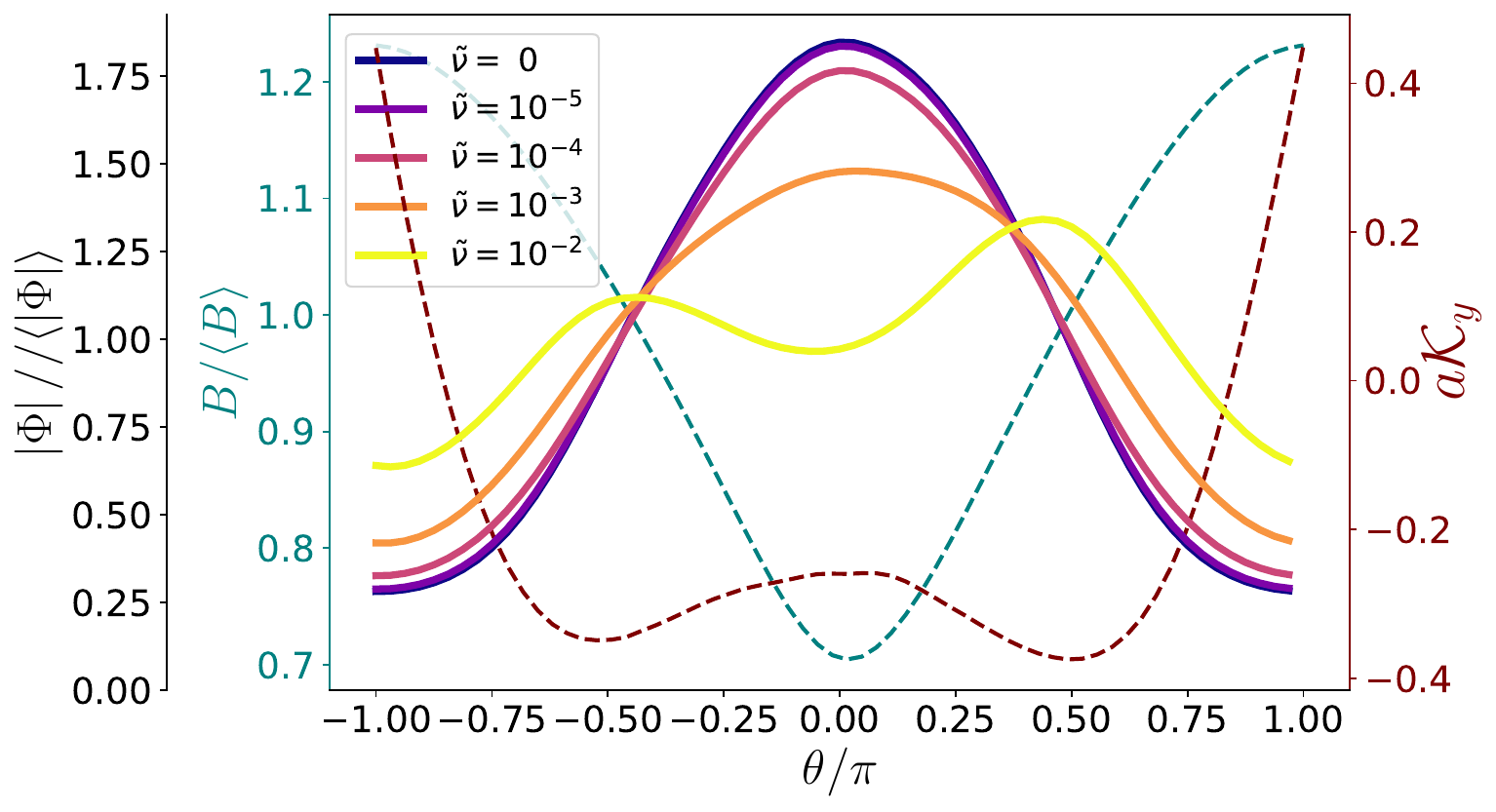}
        \caption{}      
        \label{fig:D3D-lowky}
    \end{subfigure}%
    \begin{subfigure}{.45\linewidth}
        \centering
        \includegraphics[width=\linewidth]{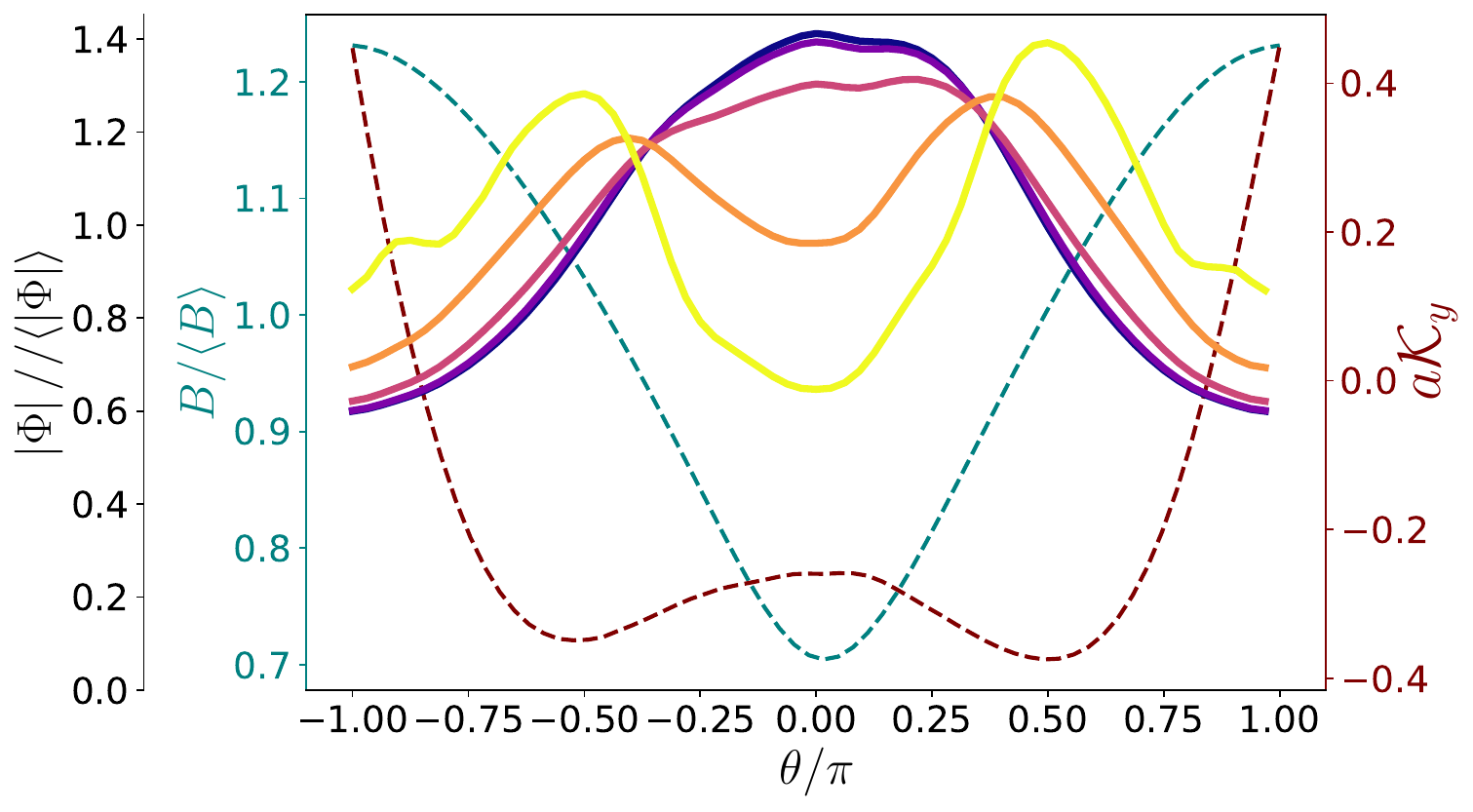}
        \caption{}        
        \label{fig:D3D-highky}
    \end{subfigure}%
\caption{\label{fig:D3D-modes} Normalised eigenmode amplitude $\Phi$ of the electrostatic potential along the field line, contrasted with the magnetic field strength and normal curvature of the DIII-D geometry (dotted) as the collision frequency is varied, for wavenumbers of (a) $k_y \rho_i = 0.3$ and (b) $k_y \rho_i = 1.5$. Perceptually warmer colours indicate a higher collisionality. The decrease in trapped-particle fraction as a result of net collisional de-trapping can be inferred from the monotonic weakening of the mode localisation at the outboard side, with the exception of $\tilde{\nu}=0.01$ where the eigenmode becomes predominantly localised in the bad curvature regions.}
\end{figure*}

The results for the (real) oscillation frequency and growth rate of the instabilities in DIII-D geometry are shown in Fig.~\ref{fig:D3D-sim}. For the collisionless case $\tilde{\nu} =0$, the mode propagates in the electron diamagnetic direction for $k_y \rho_i < 1$ corresponding to a collisionless TEM localised at the low-field side (see Fig.~\ref{fig:D3D-lowky}) where the most deeply trapped particles with bad bounce-averaged curvature reside. At larger wavenumbers, the oscillation frequency changes sign, indicating a transition to the Ubiquitous Mode\cite{Coppi1977}, which is non-resonantly driven by a combination of ion- and (trapped) electron curvature, thus resulting in a mode structure that is less strongly localised at the low-field side (see Fig.~\ref{fig:D3D-highky}). \par

As the collision frequency is increased, it is observed from Fig.~\ref{fig:D3D-sim} that the instabilities are stabilised at all wavenumbers, and the real frequency is gradually downshifted with the exception of the highest collision frequency of $\tilde{\nu}=0.01$, where a different instability regime may be occurring. The corresponding changes to the mode structure can be seen in Fig.~\ref{fig:D3D-modes}. For both low and high wavenumbers the mode structure becomes less localised at the low-field side as the collision frequency is increased, and starts to become more extended towards the region of higher magnetic field. This indicates the mode is suffering from a decreasing trapped-electron drive as a result of de-trapping and becomes more dependent on the drive from bad curvature to sustain itself. Furthermore, the behaviour of the mode structure also explains the observed deviation in the downshift trend of $\omega_R$ at the highest collision frequency of $\tilde{\nu}=0.01$, as the mode is no longer in a TEM regime at low wavenumber as opposed to all other collisionalities, but appears mostly localised in the bad-curvature regions. Since these simulations are performed in absence of temperature gradients this rules out the possibility of the ITG instability, such that the observed instability must correspond to Ubiquitous Mode-like type.

\subsection{HSX magnetic geometry}

\begin{figure*}[!ht]
	\centering
	\begin{subfigure}{.45\linewidth}
		\centering
		\includegraphics[width=\linewidth]{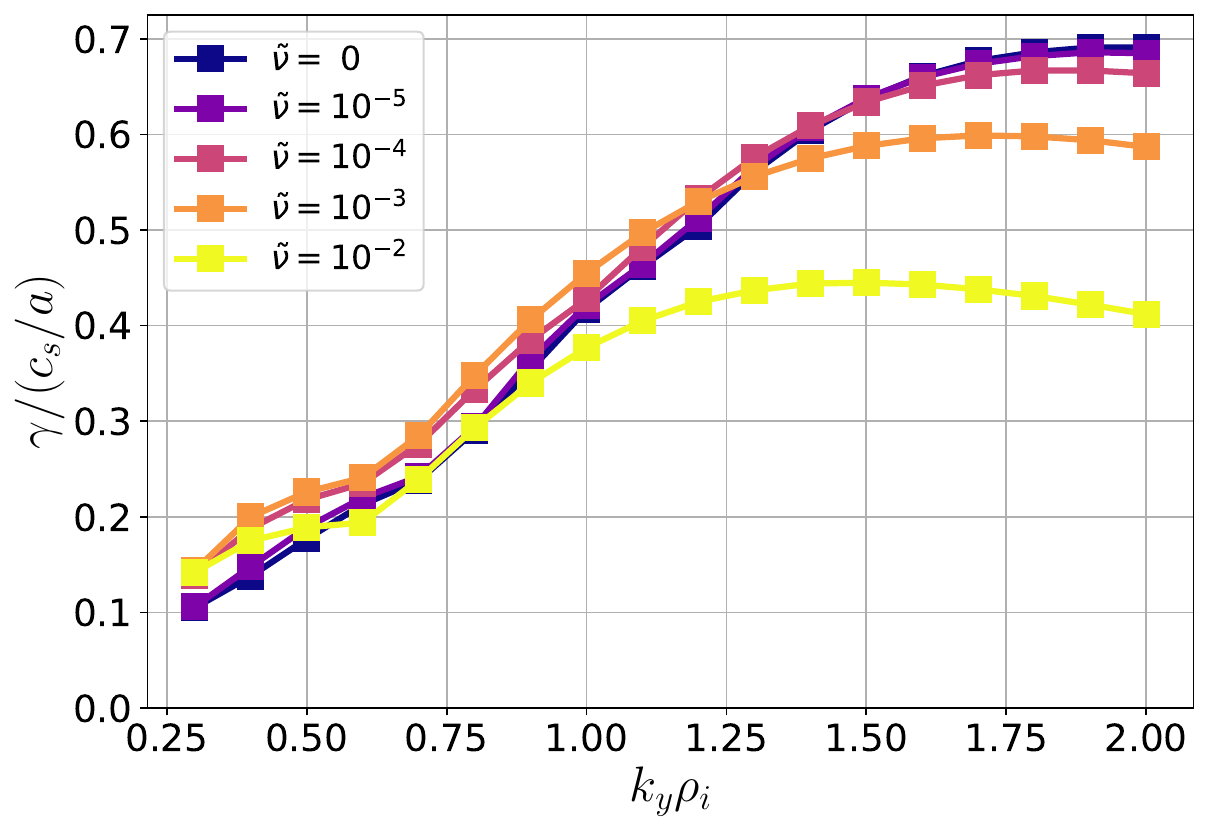}
		\caption{}        
		\label{fig:HSX-gamma}
	\end{subfigure}%
	\begin{subfigure}{.45\linewidth}
		\centering
		\includegraphics[width=\linewidth]{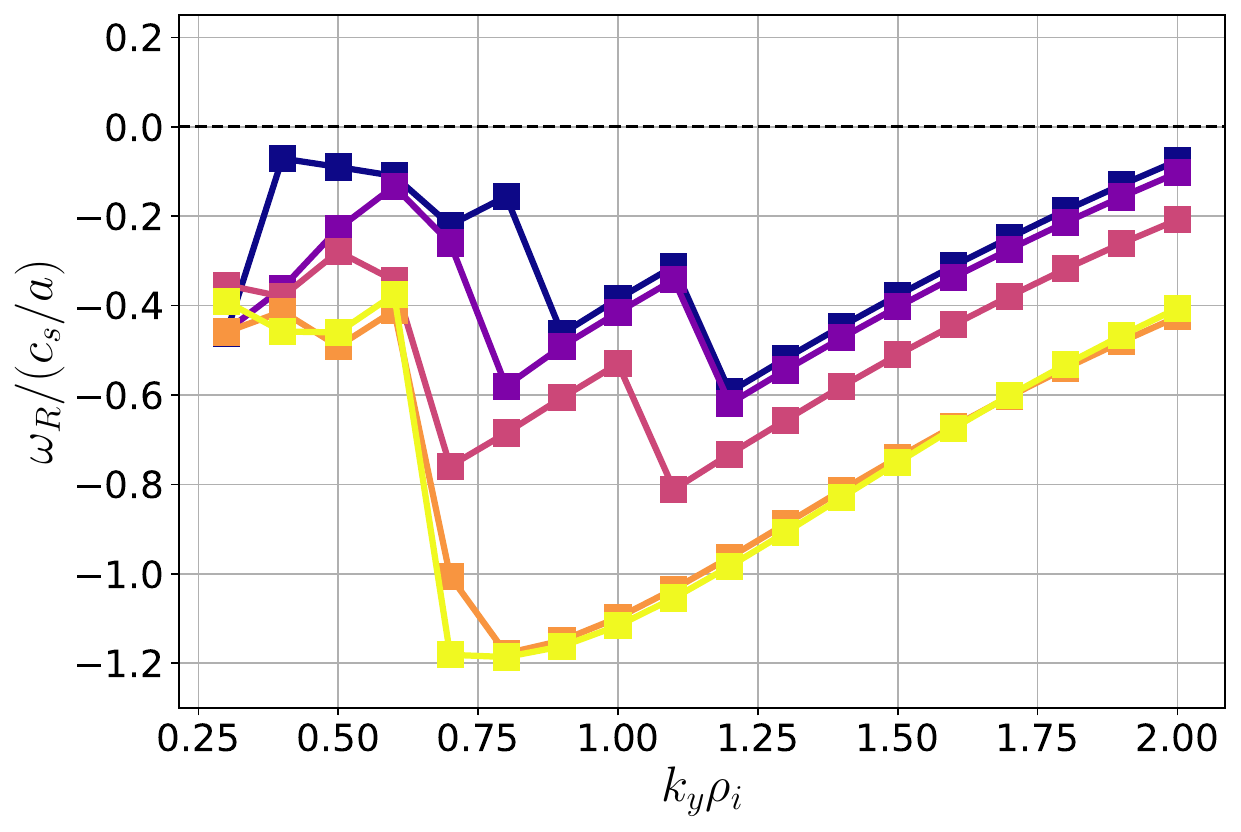}
		\caption{}      
		\label{fig:HSX-omega}
	\end{subfigure}%
	\caption{\label{fig:HSX-sim} Normalised (a) growth rate and (b) oscillation frequency from linear gyrokinetic simulations in HSX geometry at a density gradient of $a/L_n = 3$ as the collision frequency is varied. Perceptually warmer colours indicate a higher collisionality. The influence of collisions on the mode frequency is less trivial than in DIII-D geometry, with both destabilisation and stabilisation occurring at low- and high wavenumbers, respectively. The mode transitions are also altered, indicating that the influence of collisions is not identical for all branches of the dispersion relation.}
\end{figure*}

\begin{figure*}[!ht]
	\centering
	\begin{subfigure}{.45\linewidth}
		\centering
		\includegraphics[width=\linewidth]{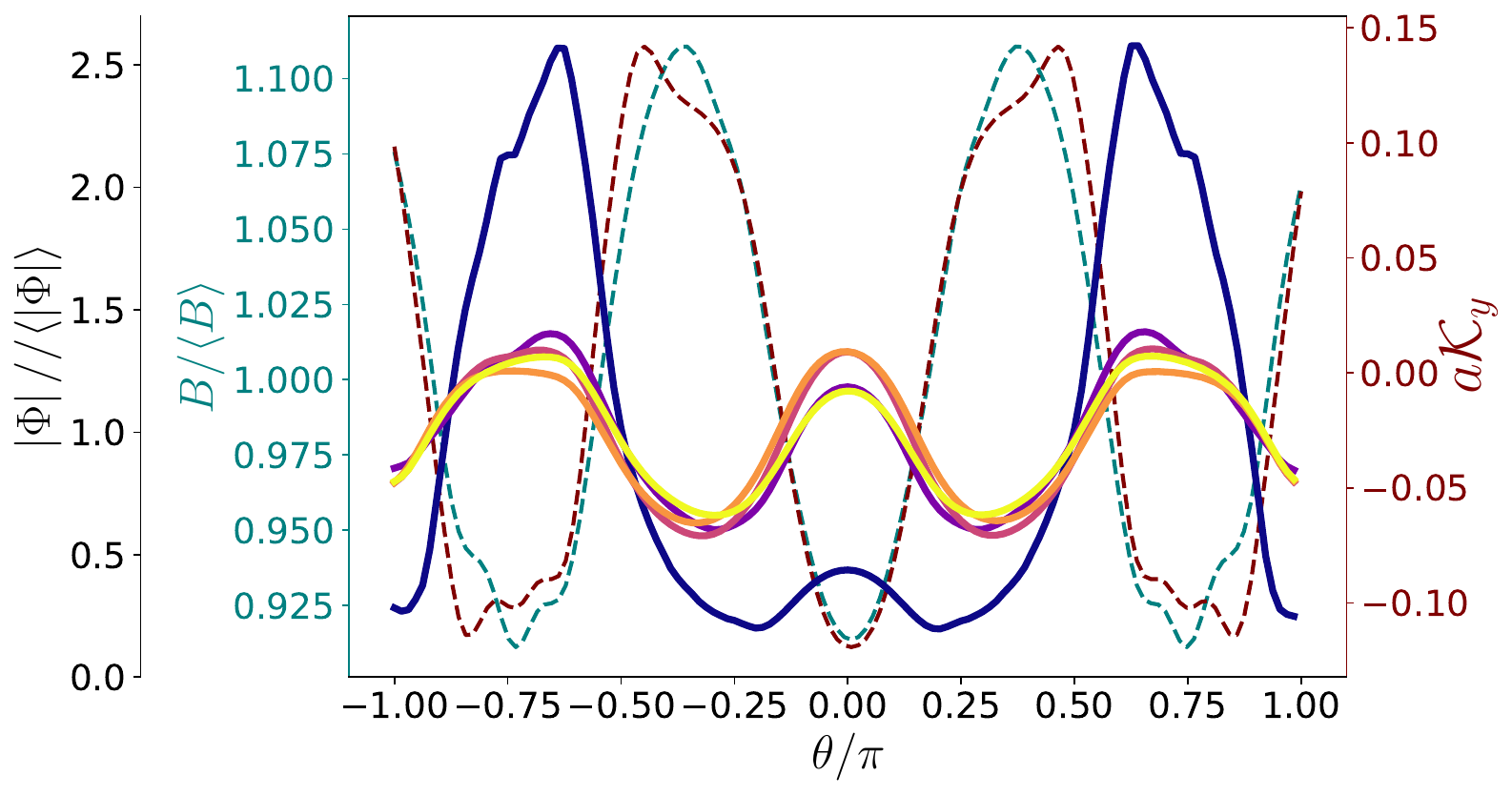}
		\caption{}   
		\label{fig:HSX-UI}
	\end{subfigure}
	\begin{subfigure}{.45\linewidth}
		\centering
		\includegraphics[width=\linewidth]{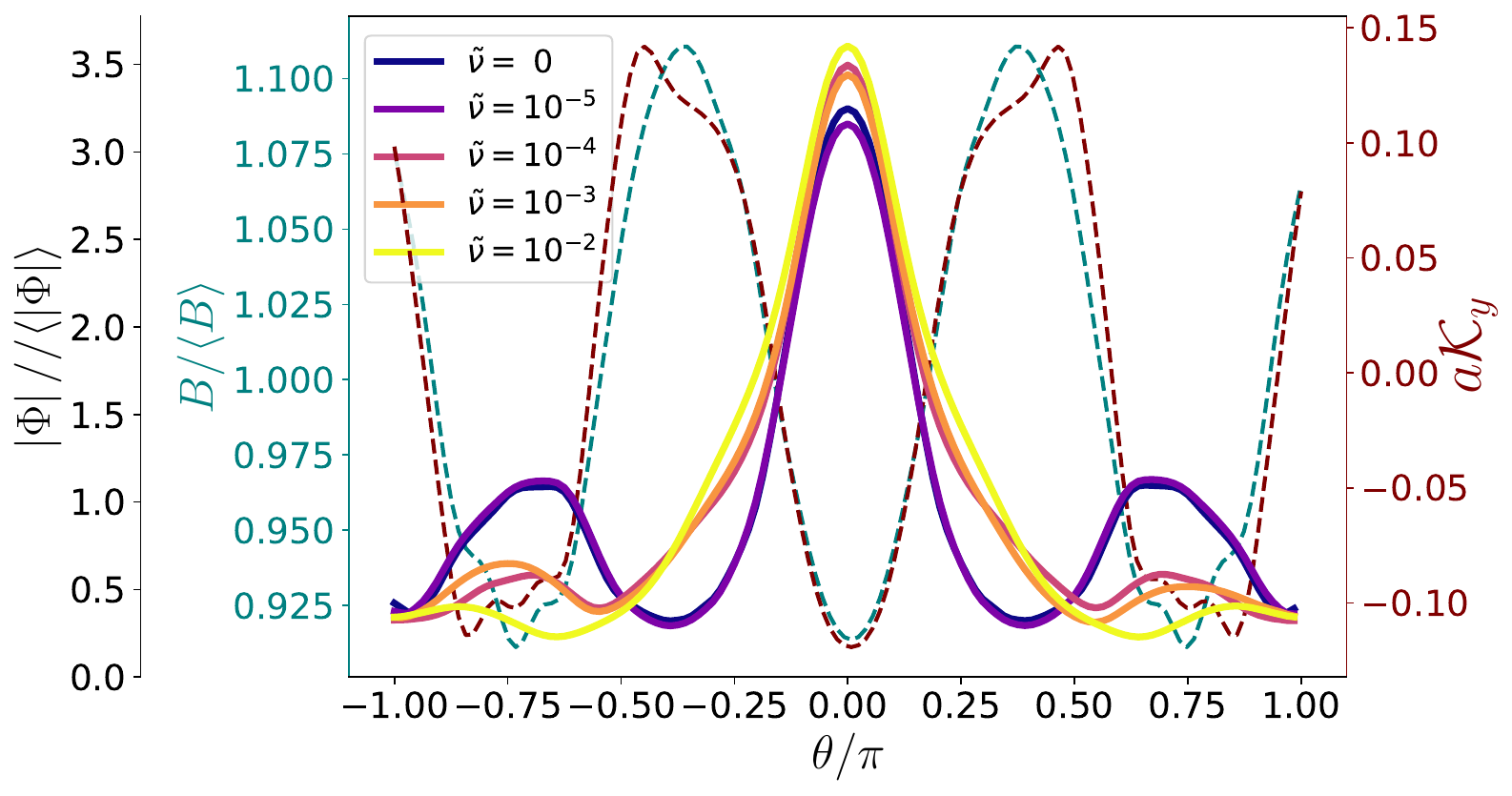}
		\caption{}       
		\label{fig:HSX-TEM}
	\end{subfigure}%
	\caption{\label{fig:HSX-modes} Normalised eigenmode amplitude $\Phi$ of the electrostatic potential along the field line, contrasted with the magnetic field strength and normal curvature of the HSX geometry (dotted) as the collision frequency is varied, for wavenumbers of (a) $k_y \rho_i = 0.4$ and (b) $k_y \rho_i = 0.7$. Perceptually warmer colours indicate a higher collisionality. The lack of a strong correlation between the collisional eigenmode localisation and geometrical features in (a) is distinctive for the UI.}
\end{figure*}

The simulated mode frequencies of instabilities in HSX geometry are shown in Fig.~\ref{fig:HSX-sim}. In contrast to the DIII-D case, several mode transitions can be observed, which are characterised by discrete jumps in $\omega_R$, in both the collisionless as well as collisional cases, with all modes propagating in the electron diamagnetic direction. As a key feature, the growth rate shows a destabilisation by the inclusion of a finite collisionality at low wavenumbers (up to about $k_y \rho_i \approx 1.2$ in the most extreme case), roughly corresponding to the wavenumber region where the dispersive behaviour of $\omega_R$ most strongly differ between the collisional and collisionless cases, while the growth rates are stabilised at high wavenumbers. Analysing the mode structure reveals that most modes in the collisionless case correspond to a TEM, with the electrostatic potential strongly localised in the wells at the inboard side for $k_y \rho_i < 0.6$, see Fig.~\ref{fig:HSX-UI}, while strongly localised at the outboard well instead for $k_y \rho_i > 0.7$, see Fig.~\ref{fig:HSX-TEM}. Since the quasi-symmetric magnetic field of HSX is similar to that of a tokamak, with the bad-curvature regions and magnetic wells overlapping, this is to be expected.\par 

From Fig.~\ref{fig:HSX-TEM} it can also be inferred that the collisionless TEM is not fully-ballooned, also being weakly localised at the inboard side wells instead of a pure localisation at the outboard side. As the wavenumber is further increased beyond $k_y \rho_i = 1.1$, corresponding to the last mode transition in Fig.~\ref{fig:HSX-omega} for the collisionless case, the electrostatic potential becomes fully localised at the outboard side. For the collisionless simulations, the only exception and thereby difference compared to DIII-D occurs at the lowest wavenumber $k_y \rho_i = 0.3$. Here a mode is observed without any preferential localisation anywhere along the field line, satisfying $\abs{\Phi} / \langle \abs{\Phi} \rangle \approx 1$. This particular mode has all the typical characteristics of the Universal Instability (UI)\cite{Landreman2015,Chowdhury2010,Helander2015} (driven by density gradient and Landau resonance of passing electrons, frequency in the electron diamagnetic direction, no strong localisation along the geometry, occurrence at low wavenumber and low magnetic shear). Despite the isomorphism of the magnetic geometries, it is thus the additional freedom arising from the much lower shear found in stellarators that allows this mode to dominate over the TEM, as the UI would be quickly stabilised by the high magnetic shear in tokamak devices. \par
For the collisional simulations we observe that at low wavenumbers $k_y \rho_i < 0.7$ the dominant instability changes from TEM to UI above $\tilde{\nu} = 10^{-5}$, as seen in Fig.~\ref{fig:HSX-UI}. This can again be explained by a decrease in the trapped-electron fraction as a result of collisional de-trapping, which requires the instabilities to increasingly rely on another driving mechanism, while the UI becomes more strongly driven due to an increase in passing particles. Meanwhile, at higher wavenumbers, the TEM remains the dominant instability, although collisions diminish the previously observed weak localisation at the inboard side for $0.7 \leq k_y \rho_i \leq 1,1$, as seen in Fig.~\ref{fig:HSX-TEM}, and the absence of mode transitions above $\tilde{\nu}=10^{-3}$ in Fig.~\ref{fig:HSX-omega} around this wavenumber region. Additionally, collisions also widen the eigenmode structure around outboard side, becoming extended into the higher-field region, as was observed for DIII-D, although the effect is more pronounced here. \par 
The simulations in the extended domain spanning two poloidal wavenumbers show excellent agreement of eigenfrequencies within 10\% in both low- and high collision frequency cases, with one case of larger discrepancy occurring at $\tilde{\nu}=10^{-5}$, where a TEM was found to be localised in wells at $\theta \approx \pm 3\pi/2$ instead. Nevertheless, the favourable transition to UI at low wavenumber remains intact, as well as the preferred localisation of the collisional TEM at the central magnetic well at $\theta = 0$. This favourable localisation is thus not an artefact of the magnetic wells at $\theta \approx \pm 3\pi/4$ being cut-off before the magnetic field strength attains a maximum, which would cause a weaker-degree of trapping due to a decrease in the mirror ratio $B_{\textrm{max}}/B_{\textrm{min}}$ (since quasi-symmetry should make all wells approximately equivalent in shape \cite{Landreman2012}). Rather, it is attributed to the better degree of overlap between the troughs of the magnetic well and the curvature profile, thus providing the most favourable condition for a TEM resonance to occur, with this higher degree of overlap also occurring for the wells at $\theta \approx \pm 3\pi/4$ in the extended domain. 

\subsection{W7-X geometry}

\begin{figure*}[!ht]
	\centering
	\begin{subfigure}{.45\linewidth}
		\centering
		\includegraphics[width=\linewidth]{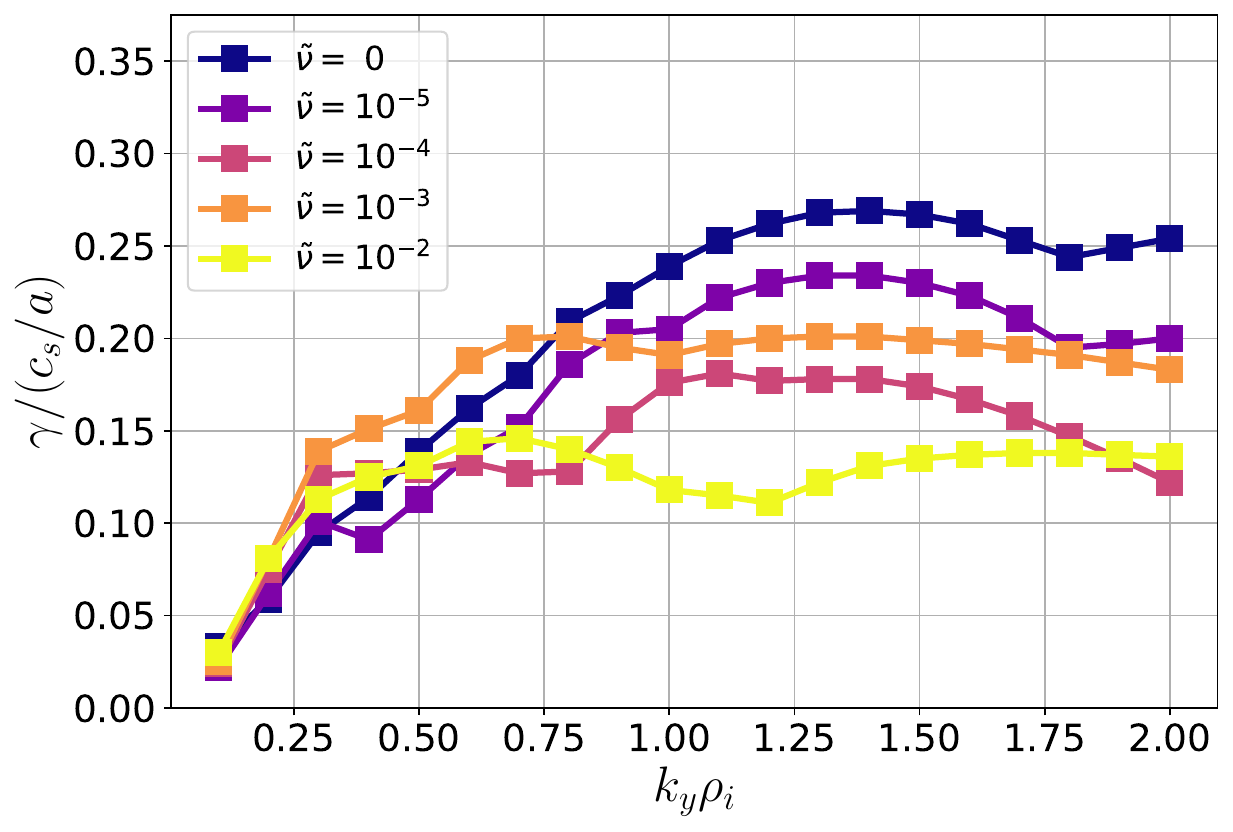}
		\caption{}        
		\label{fig:W7X-gamma}
	\end{subfigure}
	\begin{subfigure}{.45\linewidth}
		\centering
		\includegraphics[width=\linewidth]{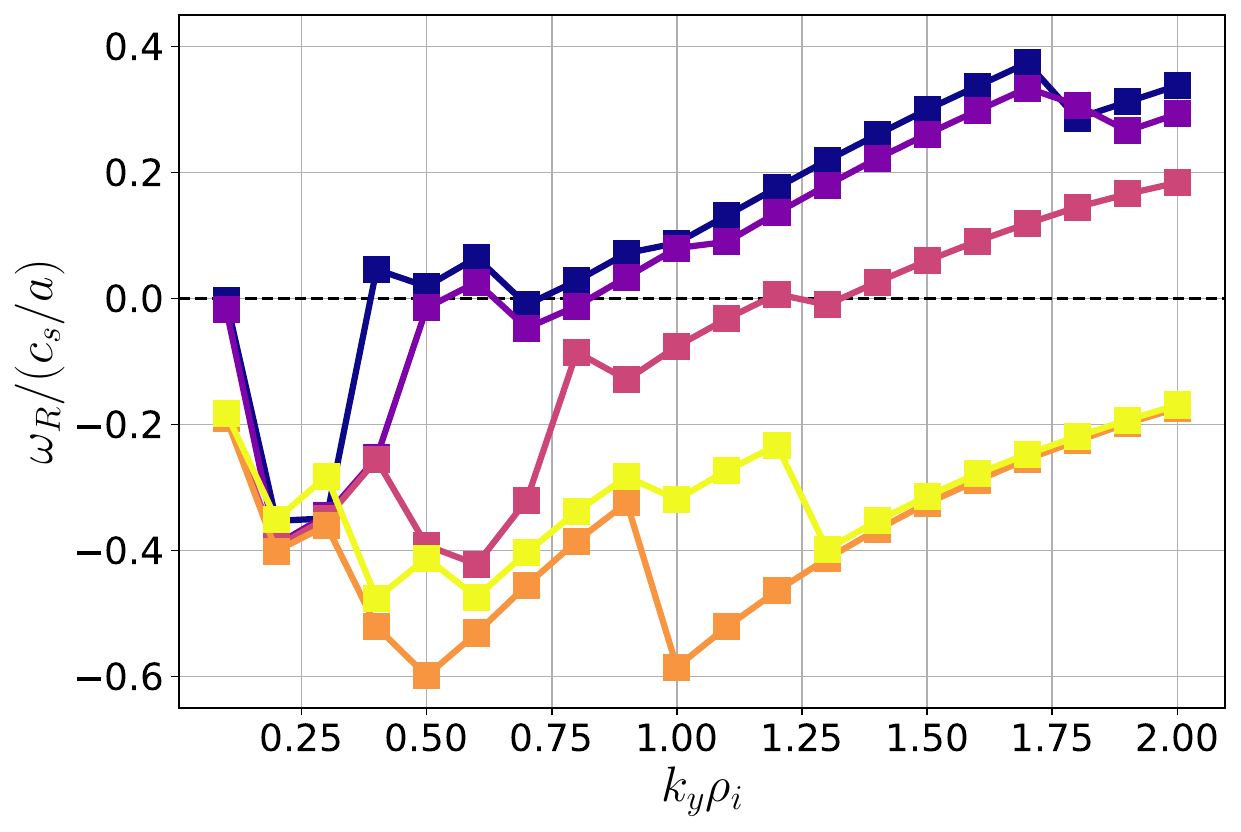}
		\caption{}      
		\label{fig:W7X-omega}
	\end{subfigure}%
	\caption{\label{fig:W7X-sim} Normalised (a) growth rate and (b) oscillation frequency from linear gyrokinetic simulations in W7-X geometry at a density gradient of $a/L_n = 3$ as the collision frequency is varied. Perceptually warmer colours indicate a higher collisionality. Similar to HSX, a collision induced destabilisation is observed for the low wavenumber modes, while the high wavenumber modes are stabilised. The transition from iTEM to UI can be identified from the reversal of the sign of $\omega_R$.}
\end{figure*}

\begin{figure*}[!ht]
 \centering
    \begin{subfigure}{.45\linewidth}
        \centering
        \includegraphics[width=\linewidth]{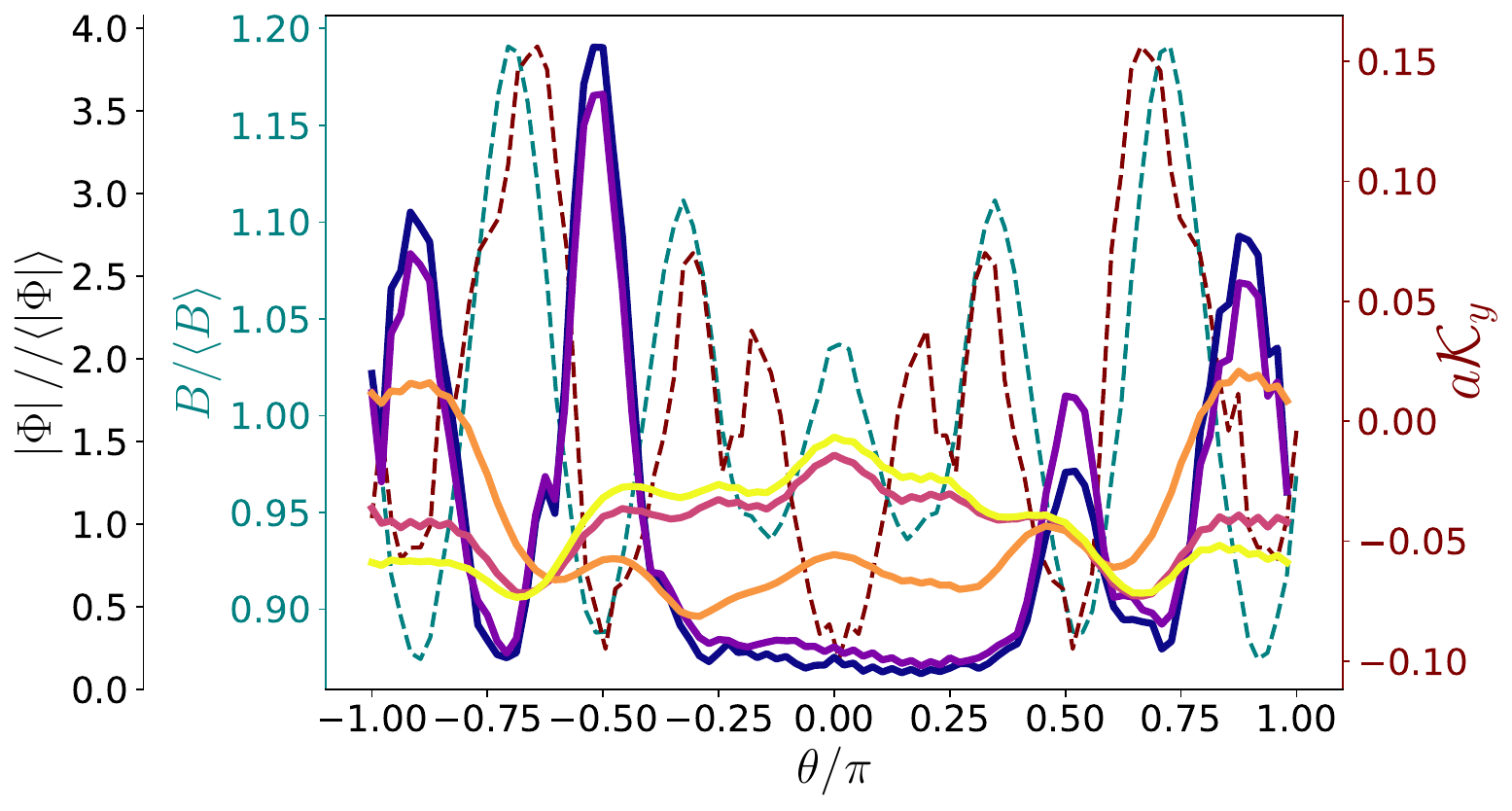}
        \caption{}      
    \end{subfigure}
    \begin{subfigure}{.45\linewidth}
        \centering
        \includegraphics[width=\linewidth]{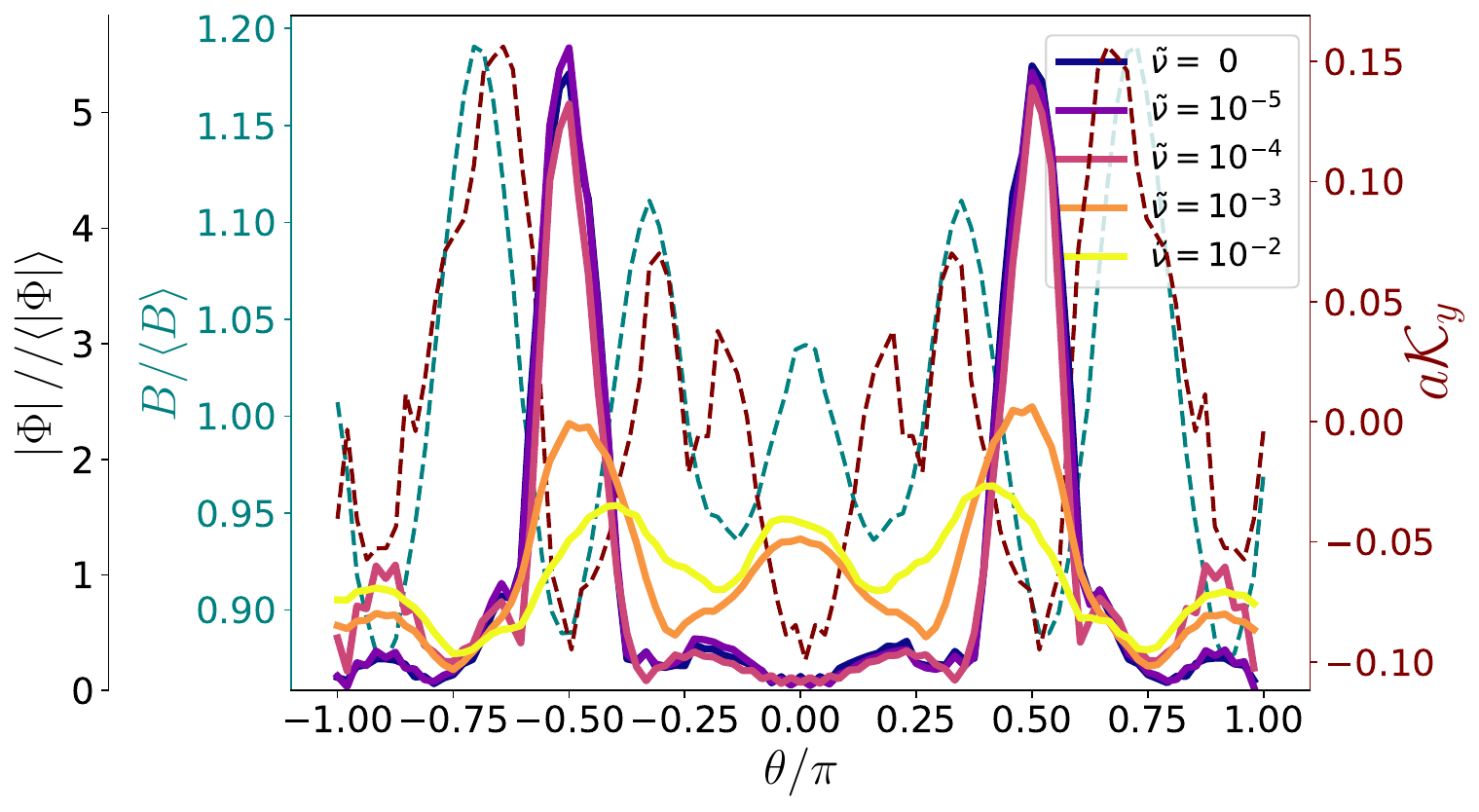}
        \caption{}        
    \end{subfigure}%
\caption{\label{fig:W7X-modes} Normalised eigenmode amplitude $\Phi$ of the electrostatic potential along the field line, contrasted with the magnetic field strength and normal curvature of the W7-X geometry (dotted) as the collision frequency is varied, for wavenumbers of (a) $k_y \rho_i = 0.5$ and (b) $k_y \rho_i = 1.5$. Perceptually warmer colours indicate a higher collisionality. The iTEMs at low collisionality are clearly distinguishable from the UIs by the contrast in the localisation of the mode structure w.r.t. the features of the magnetic geometry.}
\end{figure*}

For the high-mirror configuration of W7-X, in the collisionless case, most of the instabilities are found to propagate in the ion diamagnetic direction, see Fig.~\ref{fig:W7X-omega}, even though the mode structure is found to be localised in the magnetic wells, see Fig.~\ref{fig:W7X-modes}. These modes are identified as so-called ion-driven trapped-electron modes (iTEMs)\cite{Plunk2017}, which can be considered as an extension of the Ubiquitous Mode to general geometry, which is driven unstable by ions in regions of local bad-curvature but requiring coupling to the trapped-electrons, hence explaining the localisation of the electrostatic potential in the magnetic wells. This instability has appeared in various previous simulations of the (vacuum) high-mirror configuration\cite{Alcuson2020,Proll2013,Proll2022}, which is attributed to the approximately maximum-$\mathcal{J}$ property, as can be seen from the non-overlap between the magnetic well and bad-curvature region at the outboard side in Fig.~\ref{fig:W7X-modes}, which enhances stability against collisionless TEMs. \par 

Meanwhile, the modes propagating in the electron diamagnetic direction at $k_y \rho_i = 0.2,0.3$ are found to be instances of the UI, which have recently been rigorously proven to exist in collisionless gyrokinetic simulations of the high-mirror configuration\cite{Costello2022} up to $k_y \rho_i = 0.5$. Another exception to the iTEM occurs at $k_y \rho_i = 0.1$, where the mode is strongly localised in the magnetic well at $\theta / \pi \approx -0.5$, but propagates in the electron diamagnetic direction instead, although it is marginally close to the $\omega_R = 0$ threshold. Based on the oscillation frequency, this mode is believed to be a TEM, which can still exist as the geometry is not strictly maximum-$\mathcal{J}$ since the wells near $\theta = \pm \pi/2$ do coincide with bad-curvature regions, thus not precluding some degree of instability against TEMs. \par

We observe substantial changes to the instabilities when the collisionality is increased, with the majority of modes propagating in the electron diamagnetic direction for a collision frequency above $\tilde{\nu} = 10^{-4}$, as well as a destabilisation up to $k_y \rho_i = 0.7$, with stabilisation occurring at higher wavenumbers. \par 
Examining the corresponding changes in the mode structure of the electrostatic potential portrayed in Fig.~\ref{fig:W7X-modes} we find this mainly corresponds to collision-induced transition from iTEM to UI for both low and high wavenumbers, as characterised by a lack of strong correlation between eigenmode localisation with features from the magnetic geometry. Like in HSX, this transition is orchestrated by the collisional de-trapping mechanism reducing the trapped-electron population necessary to drive the iTEM unstable, while the lower magnetic shear facilitates favourable conditions for the UI. However, as iTEM growth rates are weaker than for conventional TEMs observed in HSX, this trend of the UI becoming the dominant mode also continues up to higher wavenumbers in W7-X. This confirms the hypothesis of Ref.~\onlinecite{Helander2015} that the UI is likely to be found in maximum-$\mathcal{J}$ configurations due to a lack of resonant instability drive from trapped-electrons, which otherwise overshadows the UI drive from passing electrons. This shift towards the UI above a collision frequency of $\nu = 10^{-4}$ also occurs for the $k_y \rho_i = 0.1$ mode previously believed to be a potential collisionless TEM. \par
\begin{figure*}[!ht]
    \centering
    \includegraphics[width=0.85\linewidth]{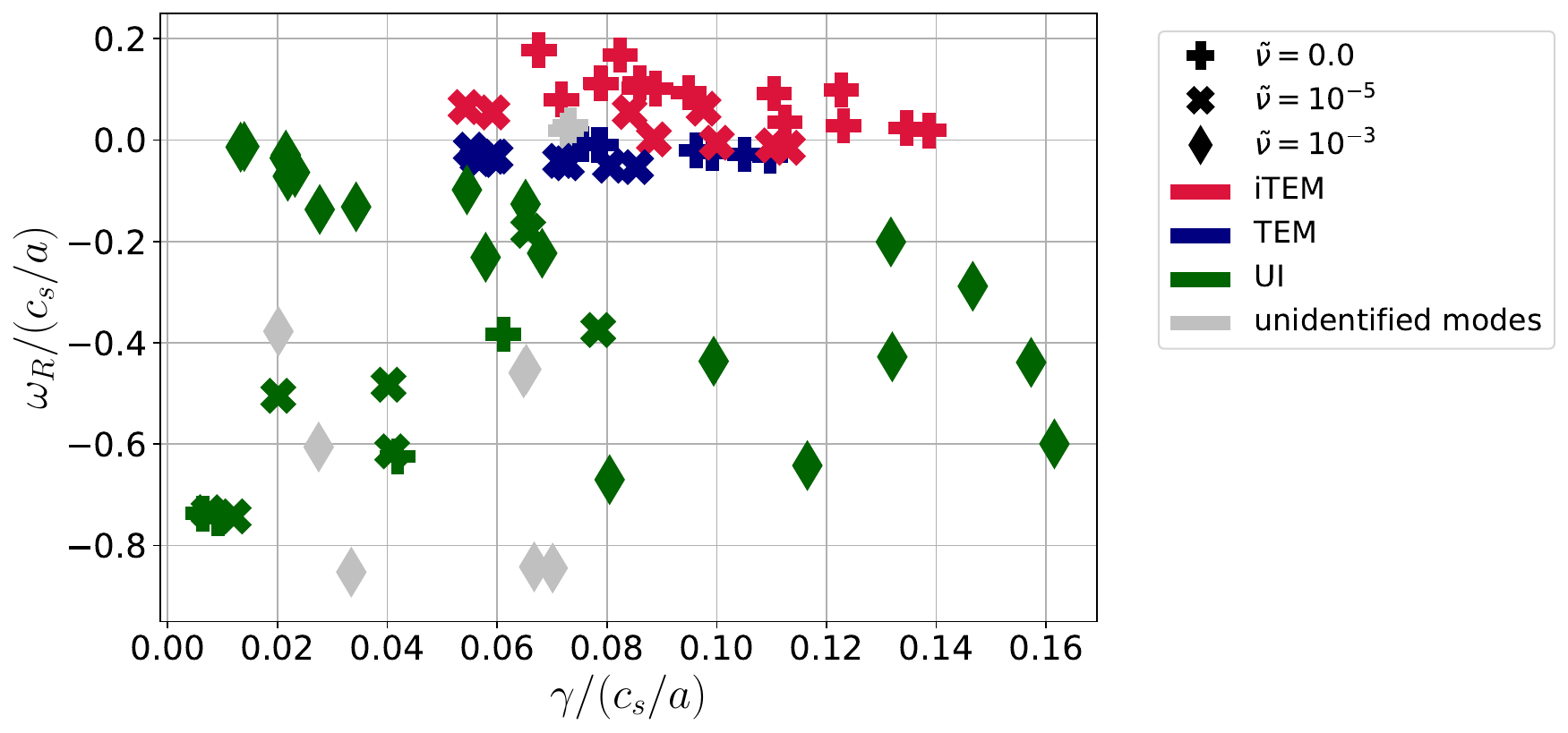}
    \caption{Eigenmode identification in the eigenspectrum of the $\sim 25$ most unstable eigenvalues for $k_y \rho_i = 0.5$ at collision frequencies of $\tilde{\nu}=[0,10^{-5},10^{-3}]$ in the high-mirror configuration of W7-X. The collision frequencies are indicated by different symbols whilst the different type of instabilities are catalogued by colour. The eigenmodes are identified based on the oscillation frequency $\omega_R$, localisation of the eigenmode in the features of the background geometry and cross-phases between perturbations in electrostatic potential and density/temperature. The handful of modes in grey could not be evidently be identified as any of the instabilities discussed before using these methods.}
    \label{fig:EV-spectrumW7Xky0.5}
\end{figure*}

The simulations performed in the longer flux-tube spanning two poloidal turns show excellent agreement of the eigenfrequencies within 10\% at the low collisionality case, with a high degree of similarity in the eigenmode localisation showing clear trapped-particle mode features. In the high collisionality case, with the exception of low wavenumbers, the propagation frequencies were found to be shifted more upwards (though only up until 30\% thus remaining within the electron diamagnetic direction) whilst the growth rates were reduced by about 20\%. These discrepancies are, like in the exceptional case of HSX, due to dissimilarities in the parallel mode structure. However, even in the elongated domain, the observed electrostatic potential did not show preferential localisation to features of the magnetic field nor the curvature, and were strongly extended along the field line. Hence, these modes are nonetheless an instance of the UI, likely being a previously subdominant variant in a cluster of UIs, whose individual growth rates are slightly affected by the extended geometry causing a change in the dominant instability to which the simulation converges (see Fig.~\ref{fig:EV-spectrumW7Xky0.5} for examples of such UI clusters in the single poloidal turn case.). Thus the generally observed trend of UI overtaking the iTEM as collisionality is increased is insensitive to the parallel extent of the simulation domain.

\begin{figure*}[!ht]
 \centering
    \begin{subfigure}{.45\linewidth}
        \centering
        \includegraphics[width=\linewidth]{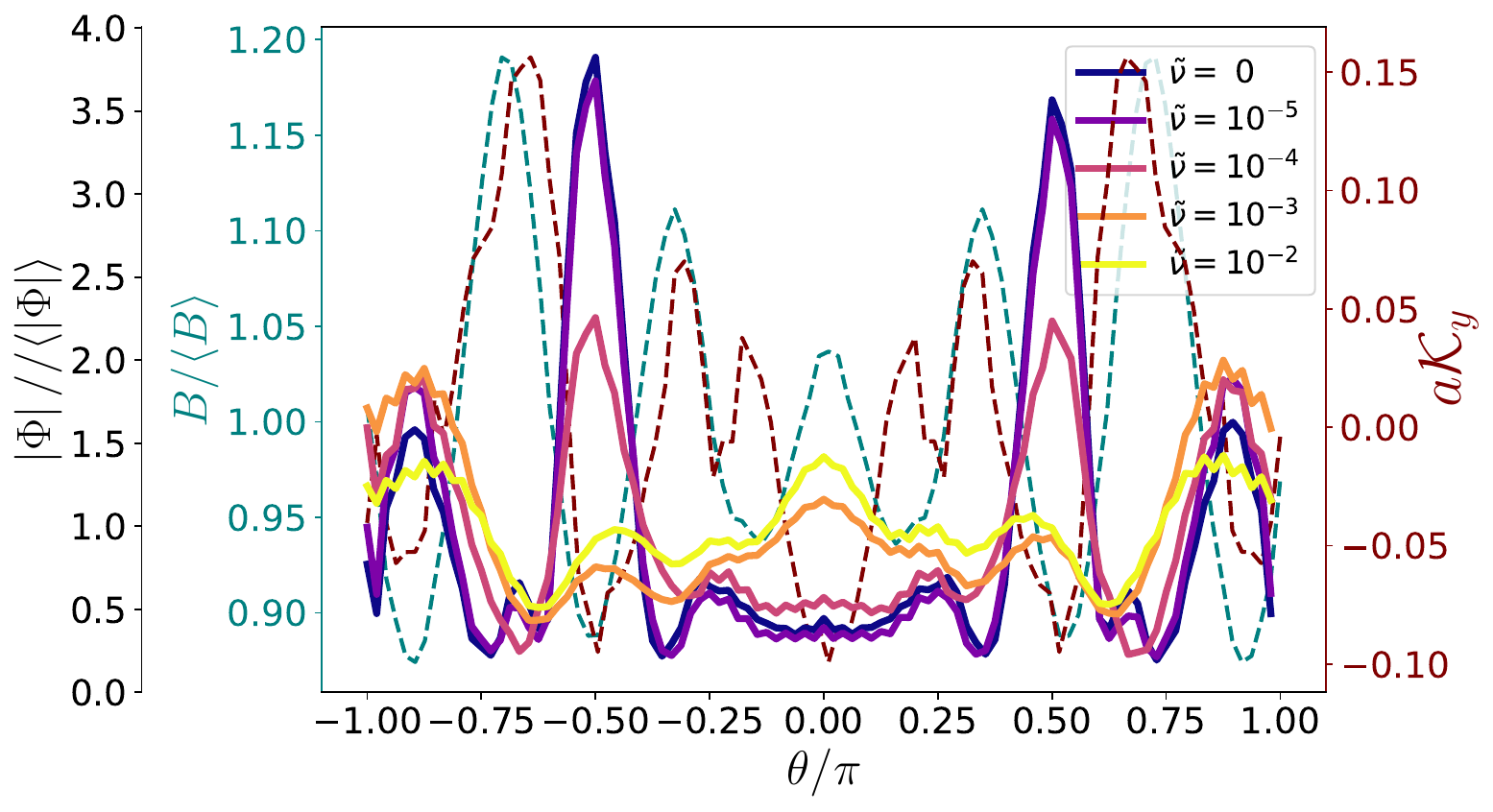}
        \caption{}   
        \label{fig:W7Xky09total}
    \end{subfigure}%
    \begin{subfigure}{.45\linewidth}
        \centering
        \includegraphics[width=\linewidth]{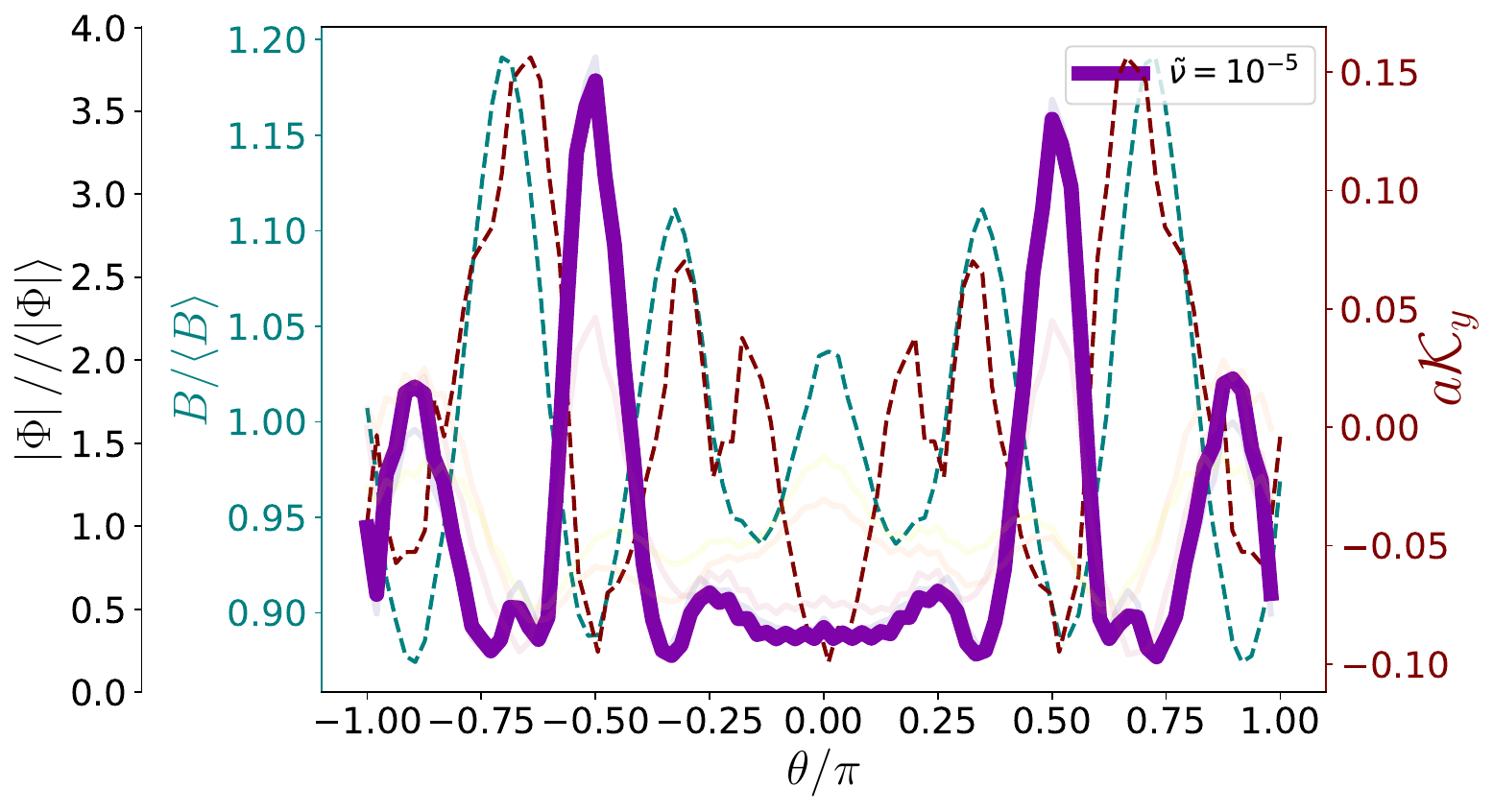}
        \caption{}       
        \label{fig:W7Xky09lowcol}
    \end{subfigure}
    \begin{subfigure}{.45\linewidth}
        \centering
        \includegraphics[width=\linewidth]{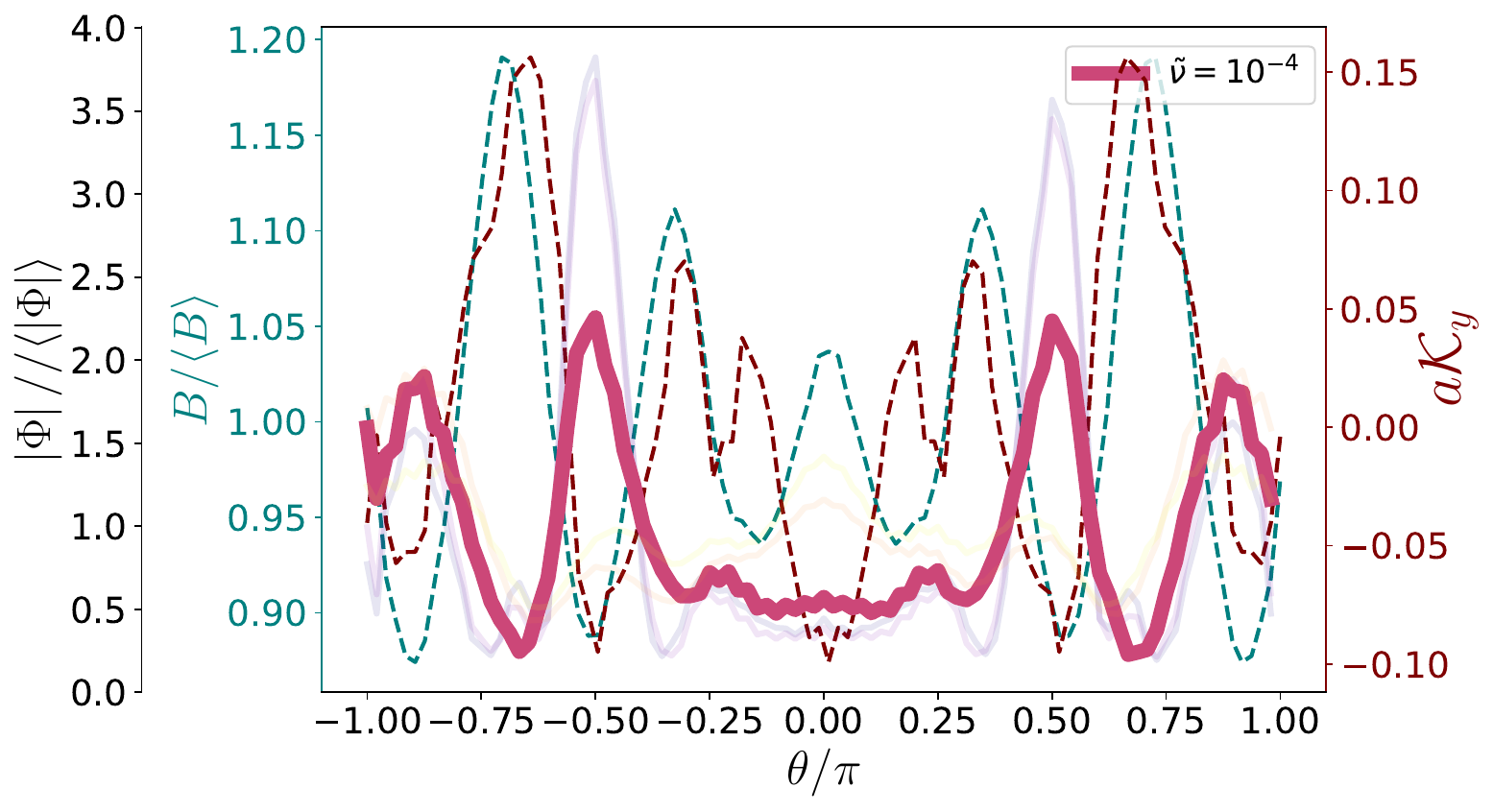}
        \caption{}       
        \label{fig:W7Xky09TEM}
    \end{subfigure}%
    \begin{subfigure}{.45\linewidth}
        \centering
        \includegraphics[width=\linewidth]{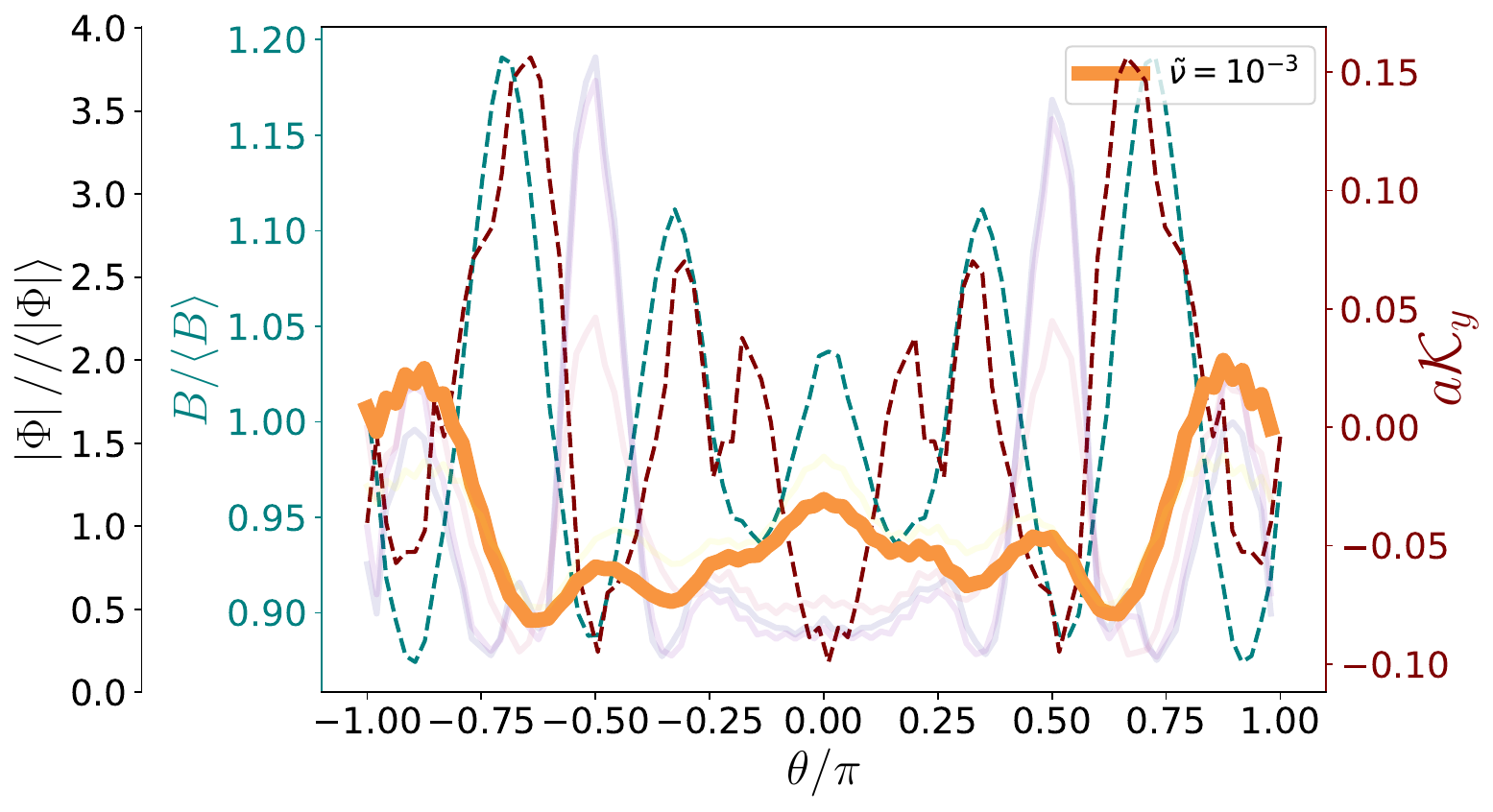}
        \caption{}       
        \label{fig:W7Xky09highcoll}
    \end{subfigure}%
\caption{\label{fig:W7X-TEMsus} Normalised eigenmode amplitude $\Phi$ of the electrostatic potential along the field line, contrasted with the magnetic field strength and normal curvature of the W7-X geometry, at wavenumber $k_y\rho_i = 0.9$ showing (a) overview of the modes at all collisionalities, (b) highlight of the mode at $\tilde{\nu}=10^{-5}$, (c) highlight of the mode at $\tilde{\nu}=10^{-4}$, and (d) highlight of the mode at $\tilde{\nu}= 10^{-3}$. The localisation differences between these highlighted mode structures, combined with the disparity in their propagation frequencies from Fig.~\ref{fig:W7X-omega}, and the energy transfer analysis identify the modes at $\tilde{\nu}=10^{-5},10^{-4},10^{-3}$ as an iTEM, TEM and UI, respectively.}
\end{figure*}

However, one notable exception from the trend with the UI becoming the dominant instability as collisionality is introduced is found in the form of the small cluster of modes around $0.8 <k_y \rho_i < 1.2$ for $\tilde{\nu}=10^{-4}$. These modes are characterised by a separate frequency band in the electron diamagnetic direction which is clearly distinguishable from the frequency band of the UI, see Fig.~\ref{fig:W7X-omega}, indicating that a mode from a different branch of the dispersion relation has become most unstable. Indeed, closely examining the changes to the mode structure at these wavelengths, as done in Fig.~\ref{fig:W7X-TEMsus}, we find that the mode structure at $\tilde{\nu}=10^{-4}$ is clearly distinguishable from the UI ($\tilde{\nu} \geq 10^{-3}$) by its localisation in the magnetic wells at the outboard side where the overlap with bad-curvature region occurs, however, the localisation is significantly less pronounced than the iTEM ($\tilde{\nu} \leq 10^{-5}$), as characterised by a lower peaking factor $\max{(\abs{\Phi}/\langle \abs{\Phi} \rangle)}$. Hence, this cluster of modes likely correspond to collisionally excited TEMs.\par

To confirm this hypothesis, we use an additional diagnostic developed in Ref.~\onlinecite{BanonNavarro2011}, measuring the power $(\partial \mathcal{E}_\phi/\partial t)_{i,e}$ that ions and electrons supply (respectively) to the electrostatic energy $\mathcal{E}_\phi$ stored in the mode, as was used in Ref.~\onlinecite{Proll2013} to identify the iTEM for the first time. Applying this diagnostic to the modes shown in Fig.~\ref{fig:W7Xky09total}, we find that the electron-to-ion power transfer ratio $\Delta \mathcal{E}_{ei} \equiv (\partial \mathcal{E}_\phi/\partial t)_{e}/(\partial \mathcal{E}_\phi/\partial t)_i$ is always negative, indicating that one species acts to stabilise the mode while the other destabilises it. Consequently, the driving species can be inferred directly from this ratio by requiring that the total supplied power $\sum_{s=i,e} (\partial \mathcal{E}_\phi/\partial t)_{s}>0$ must hold for an instability to occur. For $\tilde{\nu} \leq 10^{-5}$ the electron-to-ion power transfer ratio satisfies $-0.2 < \Delta \mathcal{E}_{ei} < 0$, thus these modes are driven by ions and hence iTEMs. In contrast, for $\tilde{\nu} \geq 10^{-4}$, the power ratio satisfies $\Delta \mathcal{E}_{ei} < -1.2$, indicating that these modes are driven unstable by the electrons, and thus the mode in the separate frequency band at $\tilde{\nu}=10^{-4}$ is indeed a TEM.

This emergence of the TEM at finite collisionality with a growth rate below that of the collisionless iTEM suggests the question of whether these instabilities are collisionally excited (and thus a dissipative TEM (DTEM)\cite{Dominguez1992}), or they already exist as a subdominant mode at zero collisionality. To investigate this, we performed a coarse-grained collisionality scan over $\tilde{\nu} = [0, 10^{-5}, 10^{-3}]$ using \textsc{GENE}'s eigenvalue solver, at wavenumber $k_y \rho_i = 0.5$. The eigenmodes are identified based on similarities in their propagation frequency, the localisation of the mode structure at specific geometry features, and, the cross-phases between perturbations of electrostatic potential $\phi$ and density/temperatures of each species. Details of this analysis can be found in Appendix~\ref{app:EV-supplement}. The resulting spectrum of identified eigenmodes is shown in Fig.~\ref{fig:EV-spectrumW7Xky0.5} ,where different symbols correspond to collision frequencies and different colours correspond to instability types. A handful of eigenmodes could not, however, uniquely be identified as either of the three instabilities discussed in this work using these methods and have been greyed out. \par
Two clear trends regarding the iTEM and UI can be observed: as the collision frequency increases the iTEM is rapidly stabilised while the UI grows increasingly unstable. Aside from the iTEM and UI, a cluster of instabilities with oscillation frequencies $\omega/(c_s/a) \sim -0.1$ is found for both $\tilde{\nu}=0$ and $10^{-5}$, which is comparable to the range where the TEMs were observed in the initial-value simulation near $k_y \rho_i \sim 1$, and these modes can indeed be identified as TEMs, whose growth rates are also decreased by introducing collisionality. This shows that the TEM observed before is not a DTEM, and collisionless TEMs do in fact exist as a subdominant instability in W7-X. As the magnetic configuration is only approximately maximum-$\mathcal{J}$ rather than strictly maximum-$\mathcal{J}$, this is to be expected.

\subsection{Comparison with perturbative approach}\label{sec:GENEvsperturb}

\begin{figure*}[!ht]
	\centering
	\begin{subfigure}{.45\linewidth}
		\centering
		\includegraphics[width=\linewidth]{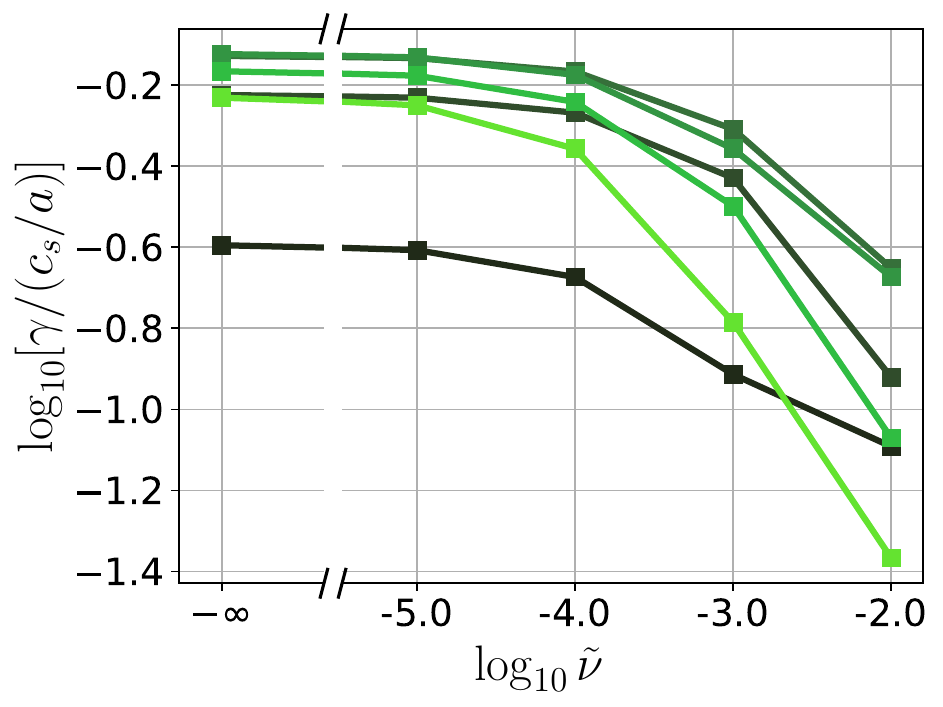}
		\caption{}
	\end{subfigure}%
	\begin{subfigure}{.45\linewidth}
		\centering
		\includegraphics[width=\linewidth]{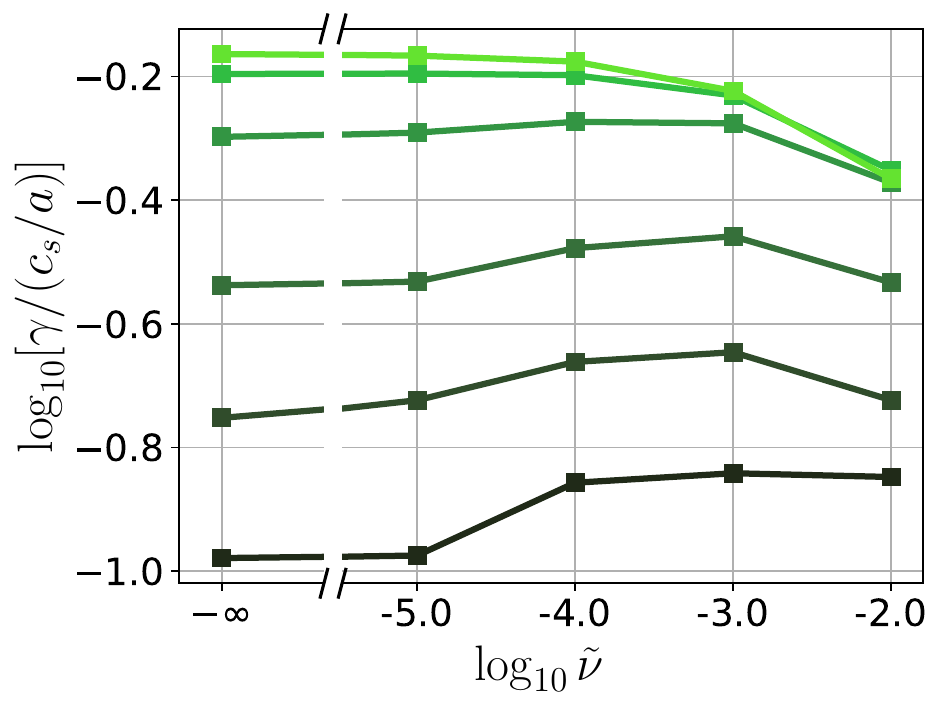}
		\caption{}
	\end{subfigure}
	\begin{subfigure}{.75\linewidth}
		\centering
		\includegraphics[width=\linewidth]{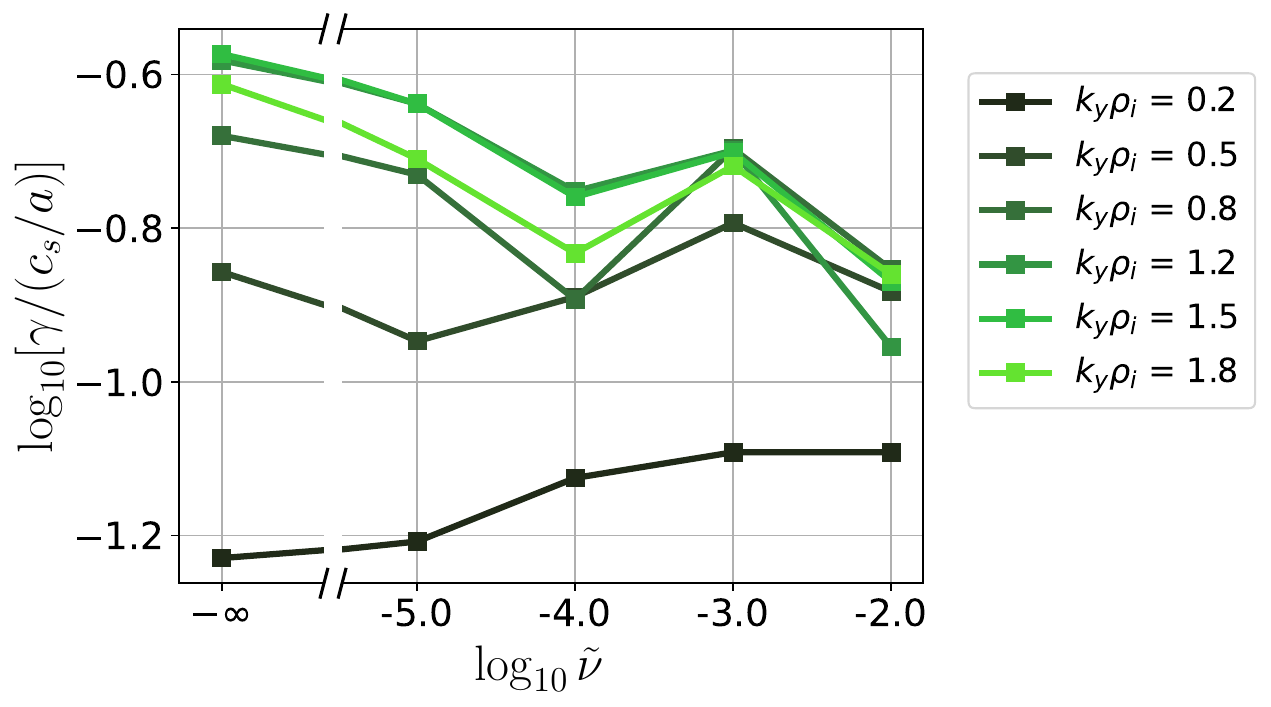}
		\caption{}
	\end{subfigure}
	\caption{Variation of normalised growth rates with \textsc{GENE} collision for a subset of the simulated wavenumbers for (a) DIII-D, (b) HSX, and (c) W7-X. Note that the collision frequency axis is broken to incorporate the reference point $\tilde{\nu}=0$ of the collisionless simulations on a logarithmic axis.}
	\label{fig:GENEgamma-loglog}
\end{figure*}

With the result from linear gyrokinetic simulations at hand we briefly return to the earlier findings of the perturbative approach. While those growth rates were found to be too large to be considered as perturbative (and hence quantitatively valid), the obtained qualitative trends may now be tested against the behaviour of the growth rates obtained from the simulations. For this purpose, we present the \textsc{GENE} growth rates from Figs.~\ref{fig:D3D-gamma},~\ref{fig:HSX-gamma},~\ref{fig:W7X-gamma} on a double-logarithmic axis versus the GENE collisionality $\tilde{\nu}$ for a handful of representative wavenumbers in Fig.~\ref{fig:GENEgamma-loglog}. This facilitates direct comparison with their perturbative counterparts from Figs.~\ref{fig:perturb-D3D-scaling},~\ref{fig:perturb-HSX-scaling},~\ref{fig:perturb-W7X-vac-scaling} respectively. Because of the relative scarcity of collisional data, detailed comparisons are not feasible, and rather we focus mostly on the asymptotic behaviour. \par 
We note that, aside from the absence of a weak destabilisation at intermediate collisionality, the DIII-D simulations match the qualitative behaviour obtained with the perturbative approach. The only observable discrepancy is the lack of a uniform scaling at all wavenumbers, which can be explained from the inherent $k_y$ dependent normalisation by $\langle \omega_0 \rangle$, whilst the \textsc{GENE} growth rates employ a fixed normalisation. Likewise, for HSX the behaviour at large wavenumber matches the trends observed in the perturbative approach. For low wavenumber however, the growth rates do not asymptotically approach zero, which makes their destabilisation with collisionality less pronounced than in the perturbative approach. In the case of W7-X, the trends in growth differ quite strongly between the simulations and perturbative approach. For the large wavenumber modes, the collisionless regime at low collisionality is absent, and those modes- instead are mostly subject to a stabilisation. However, at intermediate collisionality, modes over all considered spatial scales undergo some degree of destabilisation, before being further stabilised at the largest collisionality. Furthermore, the lowest wavenumber modes are also destabilised as the collisionality is increased, both of which are consistent with the perturbative approach. The discrepancies in W7-X, and to a lesser degree in HSX, can be attributed to the appearance of the UI in the simulations, whilst the perturbative approach only considers the TEM driving mechanism. All in all, as far as TEMs are considered, it may be concluded that the qualitative trends from the simple perturbative approach are well in agreement with the linear gyrokinetic simulations.

\section{Summary \label{sec:summary}}
In conclusion, we have investigated the role of collisions on the growth rate of electrostatic density-gradient-driven microinstabilities in core plasmas of different realistic toroidal magnetic geometries from both an analytical and numerical viewpoint, with the former focused on the TEM instability, which is commonly the dominant instability in this parameter regime. Analytically, the contribution from the trapped electrons to mode frequencies was treated perturbatively, based on small trapped-particle fraction expansion of the dispersion relation. This perturbative approach considers a non-asymptotic treatment of collisionality as the collision frequency has been ordered as $\nu/\omega \sim \nu/\overline{\omega_{de}} \sim \mathcal{O}(1)$, thereby extending previous treatments which assumed these to be either small or large. From the perturbative approach, it follows that collisions non-trivially affect the growth rate in different geometries. In particular, by taking into account the decrease in trapped-particle population through de-trapping, at high collisionality a stabilising effect stronger than the common $\gamma\propto 1/\nu$ scaling for the dissipative limit has been found. Most notably, the perturbative approach has successfully recovered the result that maximum-$\mathcal{J}$ configurations are stable against collisionless TEMs, but additionally revealed that such configurations are therefore most susceptible to destabilisation when a finite collisionality is introduced. While the growth rates obtained from this relatively simple perturbative approach exceeded its quantitative validity limits ($\gamma/\omega_0 \ll 1$), attributable to break-down of the underlying assumptions, over a considerable part of the parameter regime in ($k_y,\nu$)-space, the essential qualitative features of its predictions -- like an absolute stabilisation in the DIII-D tokamak and destabilisation at low wavenumber in both HSX and W7-X -- are reproduced by high fidelity gyrokinetic simulations. \par
These simulations, additionally provide insight into the localisation behaviour of the mode structure along the magnetic field line, which revealed that this destabilisation in stellarators is due to a change of the dominant instability to the Universal Instability, which in case of W7-X is found to exist up to much larger wavenumber than previously observed in collisionless simulations \cite{Costello2022}. However, a departure from this trend is found at an intermediate collision frequency of $\tilde{\nu}=10^{-4}$ in a narrow range of wavenumbers around $k_y \rho_i = 1$, where for the first time it has been observed that a TEM emerges as the dominant instability in the high-mirror configuration. By means of an additional investigation with an eigenvalue solver, it was found that these modes are subdominantly unstable in collisionless simulations. We also note that these simulations significantly expand upon the investigation of the influence of collisions on microinstability in the W7-X stellarator in Ref.~\onlinecite{Alcuson2020}, where only a single collision frequency was considered in the high-iota configuration, which is a slightly worse approximation to maximum-$\mathcal{J}$ than the high-mirror configuration considered in the present work.\par
For future work we consider improving the perturbative approach with a more realistic mode structure, to consistently account for additional geometry differences resulting in eigenmode localisation, which is neglected by invoking the flute-mode approximation. Additionally, we will seek for a way to include the trapped-electron response in a non-perturbative manner, to make the render the analytical growth rates also quantitatively valid. Furthermore, we will seek to characterise whether the observed range of UIs belong to the slab branch or toroidal branch of the instability. Additionally, it would be worthwhile to consider whether the (subdominant) existence of TEMs in W7-X would cease in simulations using a finite-$\beta$ equilibrium, where the maximum-$\mathcal{J}$ property is fully achieved rather than only approximately, as suggested by both theory and the perturbative approach. Such an investigation would, however, require an artificial suppression of magnetic fluctuations arising at finite-$\beta$, to keep within the domain of electrostatic instabilities discussed here for a fair comparison. Ultimately, when the linear behaviour is better understood, non-linear simulations will allow us to assess whether the general stabilisation of the linear microinstabilities also leads to reduced turbulent fluxes, like mixing-length estimates would suggest\cite{Garbet2006,Weiland2016}, and what the role of the emerging Universal Instability means for the saturation of turbulent transport.

\begin{acknowledgments}
We would like to thank P.T. Mulholland for providing assistance in working with the \textsc{GENE} code, as well as P. Costello for insightful discussions on the Universal Instability. We would like to thank R.J.J. Mackenbach for providing access to and assistance with his codes for calculating bounce-integrals and the Available Energy of trapped-electrons, which form the essential building blocks for the code used to calculate the perturbative frequency shift in this work. \par
We are grateful to B.A. Grierson, J.C. Schmitt and C. Beidler for sharing the profile data from DIII-D, HSX and W7-X, respectively, which allowed us to restrict the numerical values of the \textsc{GENE} collision frequency to experimentally reasonable values. \par
Simulations presented in this work have been performed on MARCONI supercomputer at CINECA under EUROfusion allocation. This work has been carried out within the framework of the EUROfusion Consortium, funded by the European Union via the Euratom Research and Training Programme (Grant Agreement No 101052200 — EUROfusion). Views and opinions expressed are however those of the author(s) only and do not necessarily reflect those of the European Union or the European Commission. Neither the European Union nor the European Commission can be held responsible for them. 
\end{acknowledgments}

\bibliography{paper_refs}

\appendix

\section{ \label{app:collision-formal} Rigorous description of the collision operator}
The influence of small velocity deflections on the distribution function of particles of species $a$ as a result of the large-range nature of Coulomb collisions with particles of species $b$ is given by the Fokker-Planck collision operator, as first derived in Ref.~\onlinecite{Rosenbluth1957}
\begin{equation}
    C_{ab}(f_a,f_b) = L^{ab} \pdv{}{v_k} \left[\frac{m_a}{m_b} \pdv{\varphi_b}{v_k} f_a - \pdv{\psi_b}{v_k}{v_l} \pdv{f_a}{v_l}\right],
    \label{eq:collop-general}
\end{equation}
where $L^{ab} = (q_a q_b/m_a \epsilon_0)^2 \ln \Lambda$ is a measure of the interaction strength, with $\ln \Lambda$ the Coulomb logarithm, $f_{a,b},m_{a,b},q_{a,b}$ are the distribution function, mass, and charge of species $a$ and $b$ respectively, $\epsilon_0$ is the vacuum permittivity, the $v_{k,l}$ denote velocity components with repeated indices $l,k$ implying Einstein summation, and $\varphi_b,\psi_b$ are the Rosenbluth potentials determined by
\begin{equation}
    \begin{split}
        \varphi_b(\bm{v}) =& - \frac{1}{4\pi} \int \frac{f_b(\bm{v'})}{\norm{\bm{v}-\bm{v'}}} \dd^3{\bm{v'}}, \\
        \psi_b(\bm{v}) = & - \frac{1}{8\pi} \int \norm{\bm{v}-\bm{v'}} f_b(\bm{v'})\dd^3{\bm{v'}}.
    \end{split}
    \label{eq:Rosenbluth-pots}
\end{equation}
In the limiting case that species $b$ is described by a Maxwellian distribution $F_{M}$ Eq.~(\ref{eq:collop-general}) takes the more practical form of \cite{Hirshman1976,Helander2002}
\begin{eqnarray}
    C(f_a,F_{Mb}) = \nu_{D}^{ab} \mathcal{L}\{f_a\} && + \frac{1}{v^2} \pdv{}{v} \left[v^3\left(\frac{m_a}{m_a+m_b} \nu_{s}^{ab} f_a \right. \right. \nonumber \\
    && \left.\left.+ \frac{1}{2} \nu_{\parallel} v \pdv{f_a}{v}\right)\right],
    \label{eq:Maxwellcolop}
\end{eqnarray}
where the first part describes pitch-angle scattering by the Lorentz scattering operator 
$\mathcal{L} = \frac{1}{2}\left(\frac{1}{\sin\theta}\pdv{}{\theta}\left(\sin\theta \pdv{}{\theta}\right)+\frac{1}{\sin^2\theta}\pdv[2]{}{\varsigma}\right)$
with $(v,\theta,\varsigma)$ spherical velocity variables, while the second part describes energy scattering, with the three collision frequencies for the relevant physical processes of drag, parallel- and perpendicular diffusion w.r.t. the initial velocity being, respectively,
\begin{eqnarray}
    &&\nu_{s}^{ab} =  \hat{\nu}^{ab} \left(1+\frac{m_b}{m_a}\right) \frac{T_a}{T_b} \frac{2G(v/v_{Tb})}{v/v_{Ta}}, \quad \nu_{\parallel}^{ab} =  \hat{\nu}^{ab} \frac{2G(v/v_{Tb})}{\left(v/v_{Ta}\right)^3}, \nonumber \\ 
    &&\nu_{D}^{ab} =  \hat{\nu}^{ab} \frac{\erf(v/v_{Tb})-G(v/v_{Tb})}{\left(v/v_{Ta}\right)^3}
    \label{eq:coll-frequencies}
\end{eqnarray}
where the common collision frequency $\hat{\nu}^{ab}$ differs from the practical $90^\circ$ collision frequency between the species by a factor $ 3 \sqrt{\pi}/4$, and the velocity-dependent function is given by $G(x) = \left(\erf(x)-x \dv{\erf(x)}{x}\right)/(2x^2)$, with $\erf(x)$ the error function, and $v_{Ts} = \sqrt{2T_s / m_s}$ the thermal velocity of species $s$.

\section{\label{app:correctionsw0} Influence of ion drift and collisions on the leading order mode frequency.} 

The effect of ion collisions can also be accounted for by a Krook operator, however, the energy dependence will be different than for the electrons as the velocity functions in Eqn.~\ref{eq:coll-frequencies} should be expanded in the inverse mass ratio $m_i / m_e \gg 1$ instead. Consequently, the collisions of ions with electrons is analogous to Brownian motion \cite{Helander2002}, resulting in a virtually constant and small friction frequency and a velocity dependent diffusion frequency $\nu_D = \nu_\parallel \propto \tilde{\nu} (v_{Ti}/v)^2$ which dominates at low particle energy. Hence it is the latter which gives the largest collisional response when integrated over the distribution function, as there is no lower velocity cut-off for the ions. Therefore, for the ions we choose the diffusion frequency for the Krook model such that $C_i = \tilde{\nu}(v_{Ti}/v)^2 g_i$. Furthermore, consistent with the electrostatic limit we approximate the curvature vector by $\bm{\kappa} \approx \bm{\nabla}_\perp \ln B$, as appropriate for low $\beta$ plasmas \cite{Helander2014}. Consequently the ion magnetic drift frequency is given by $\omega_{di} \approx \hat{\omega}_{di} \left((v_\perp/v_{Ti})^2 / 2 + (v_\parallel/v_{Ti})^2\right)$ with $\hat{\omega_{di}} = \bm{k_\perp} \cdot(\bm{e_b}\cross \grad{\ln B}) v_{Ti}^2/\Omega_i$ as characteristic drift frequency. As we expect both terms to have a weak influence for TEMs we expand the denominator for $\omega_{di}/\omega \sim \nu_i/\omega \ll 1$, which respectively avoid a resonance with the ion drift in bad curvature regions (which contributes to driving the toroidal ITG branch \cite{Navarro2020,Xanthopoulos2007}), and is justified based on the collision frequency ordering $\nu_i / \nu_e \propto \sqrt{m_e/m_i} \ll 1$ together with $\omega/\nu_{e} \sim \mathcal{O}(1)$. Including these effects then the leading order dispersion relation becomes
\begin{eqnarray}
    &&D_0 \approx 1+\tau - \int \left(1-\frac{\omega_{\star i}}{\omega}\right)\left[1 + \frac{\omega_{di}(\bm{v_i})}{\omega} - i \frac{\nu^{ie}(\bm{v_i})}{\omega}\right] \nonumber \\
    && \times \left(\frac{m_i}{2\pi T_{i0}}\right)^{3/2}\exp(-\frac{m_i v_i^2}{2T_{i0}})J_0\left(k_\perp \frac{v_{\perp,i}}{\Omega_i}\right)^2 \dd^3{\bm{v_i}}
    \label{eq:expanded-D0}
\end{eqnarray}

\begin{figure*}[!ht]
	\centering
	\begin{subfigure}{.45\linewidth}
		\centering
		\includegraphics[width=\linewidth]{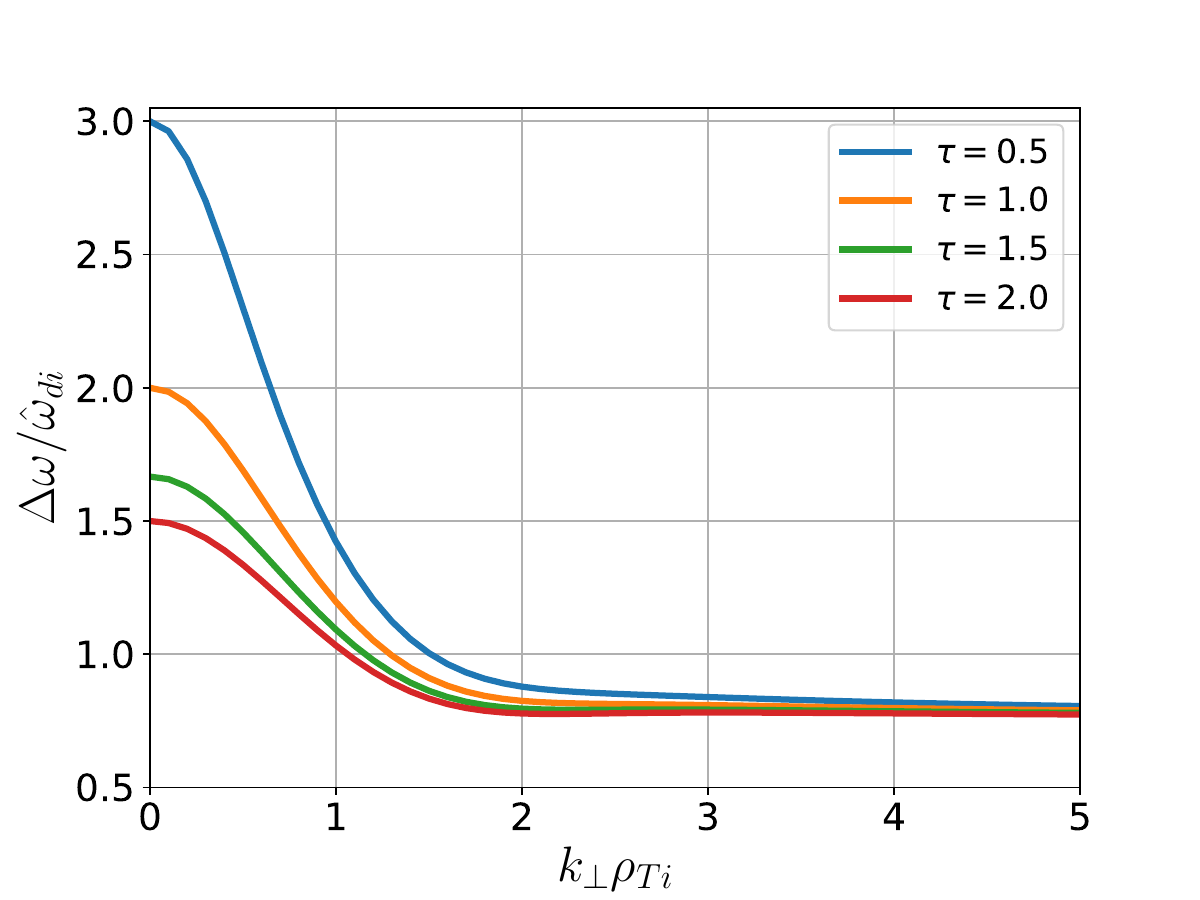}
		\caption{}      
	\end{subfigure}%
	\begin{subfigure}{.45\linewidth}
		\centering
		\includegraphics[width=\linewidth]{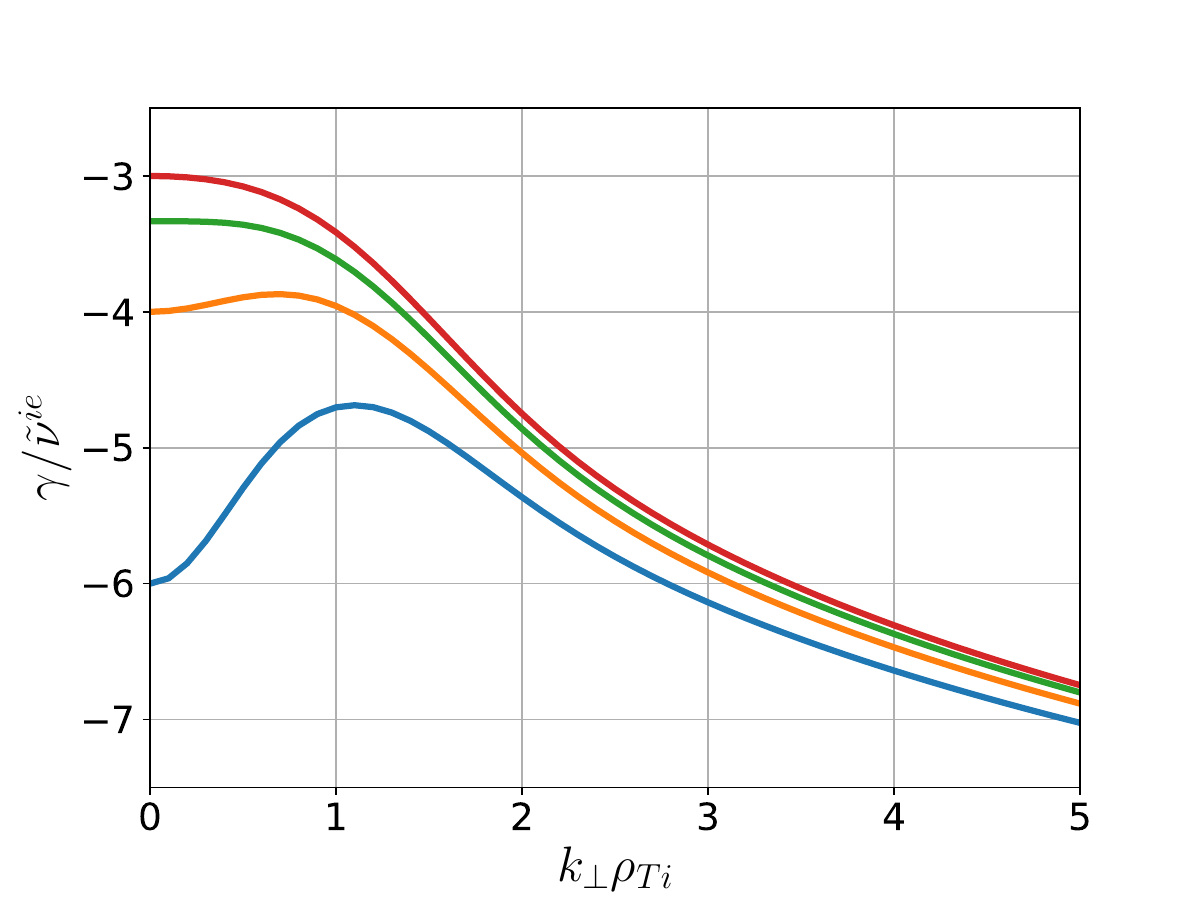}
		\caption{}        
	\end{subfigure}%
	\caption{\label{fig:om0-corrections} The corrections to the lowest-order mode frequency of the perturbative approach when finite ion drift and collisionality are accounted for, in the absence of temperature gradients ($\eta_i = \eta_e = 0$). Shown are (a) the shift in (real) frequency in the direction of the ion drift (second term in Eqn.~(\ref{eq:omega-zero-full})), and (b) the growth rate as a result of collisions (third term in Eqn.~(\ref{eq:omega-zero-full}) as a function of perpendicular wave number $k_\perp \rho_{Ti}$ for different temperature ratios $\tau=T_i/T_e$. Note that in all cases the collisions provide a weak damping.}
\end{figure*}

where we set the temperature gradient to zero. Integrating Eqn.~(\ref{eq:expanded-D0}) using cylindrical velocity coordinates $\dd^3{\bm{v_i}} \rightarrow \dd{\vartheta} v_\perp \dd{v_\perp} \dd{v_\parallel}$ leads to a quadratic dispersion relation in $\omega$ whose roots can be expanded for $\hat{\omega_{di}}/\omega_{\star e} \sim \tilde{\nu}/\omega_{\star e} \ll 1$, giving the solution for the self-consistent root of $\omega_0$ as

\begin{widetext}
    \begin{equation}
        \frac{\omega_0}{\omega_{\star e}} =  \frac{\tau \Gamma_0(\zeta) }{\left(1+\tau - \Gamma_0(\zeta)\right)} + \frac{\hat{\omega}_{di}}{\omega_{\star e}} \frac{\left(\left[1-\zeta/2\right]\Gamma_0(\zeta)+\zeta \Gamma_1(\zeta)/2  \right)\left(1+\tau\right)}{\Gamma_0(\zeta)\left(1+\tau - \Gamma_0(\zeta)\right)}
        - i \frac{\tilde{\nu}^{ie}}{\omega_{\star e}} \frac{2 \times{ }_{2} F_{2}\left[\begin{array}{cl}1 / 2 & 1 / 2 \\ 1 & 3 / 2\end{array} ;-2 \zeta\right]\left(1+\tau\right)}{\Gamma_0(\zeta)\left(1+\tau - \Gamma_0(\zeta)\right)} .
    \label{eq:omega-zero-full}
    \end{equation}
\end{widetext}

where $_{p} F_{q}$ is the generalised hypergeometric function. The first term in Eqn.~(\ref{eq:omega-zero-full}) corresponds exactly to Eqn.~(\ref{eq:omzero}) as if ion drifts and collisions were neglected, and the other two terms present respectively a shift of the propagation frequency in the direction of ion drift and a small growth rate due to ion collisions. These additional terms are plotted for various temperature ratios in Fig.~\ref{fig:om0-corrections}, showing that for practical parameters relevant for transport under fusion conditions ($\tau \approx 1, k_\perp \rho_{Ti} < 1$) the shift in real frequency will have less influence than the dispersive effect of Eqn.~(\ref{eq:omzero}) and the ion collisions provide a weak damping mechanism regardless of parameters.

\section{\label{app:contours} Contours of the frequency shift}

\begin{figure*}[!t]
 \centering
    \begin{subfigure}{.33\linewidth}
        \centering
        \includegraphics[width=\linewidth]{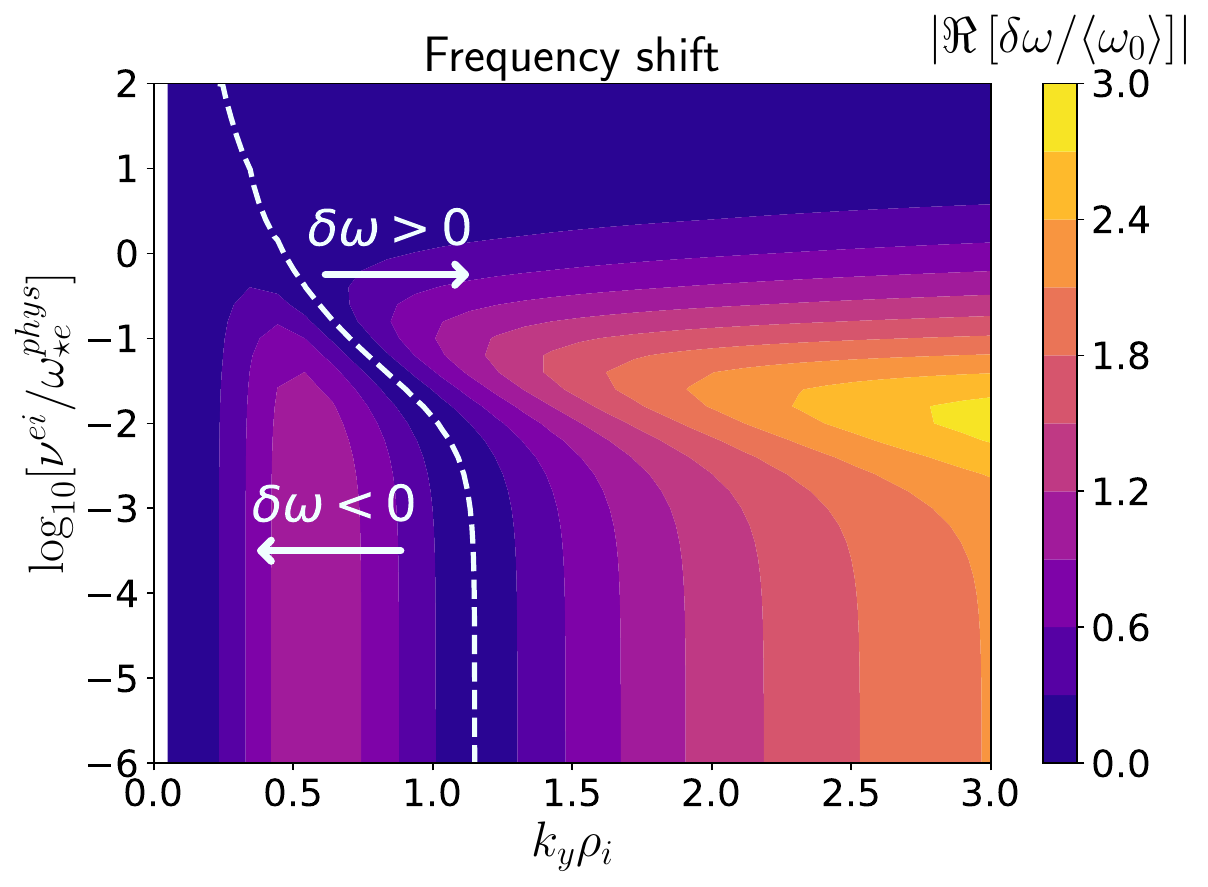}
        \caption{}      
    \end{subfigure}%
    \begin{subfigure}{.33\linewidth}
        \centering
        \includegraphics[width=\linewidth]{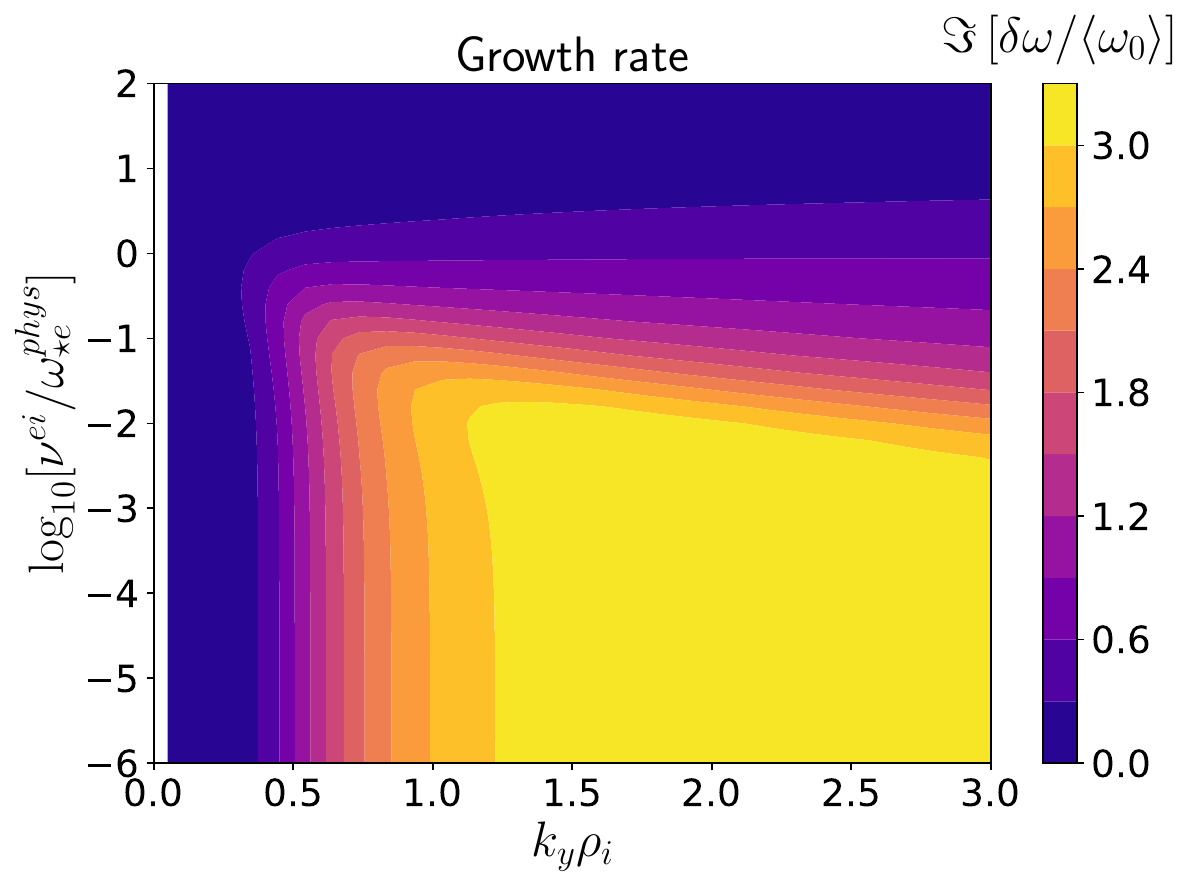}
        \caption{}        
    \end{subfigure}%
    \begin{subfigure}{.33\linewidth}
        \centering
        \includegraphics[width=\linewidth]{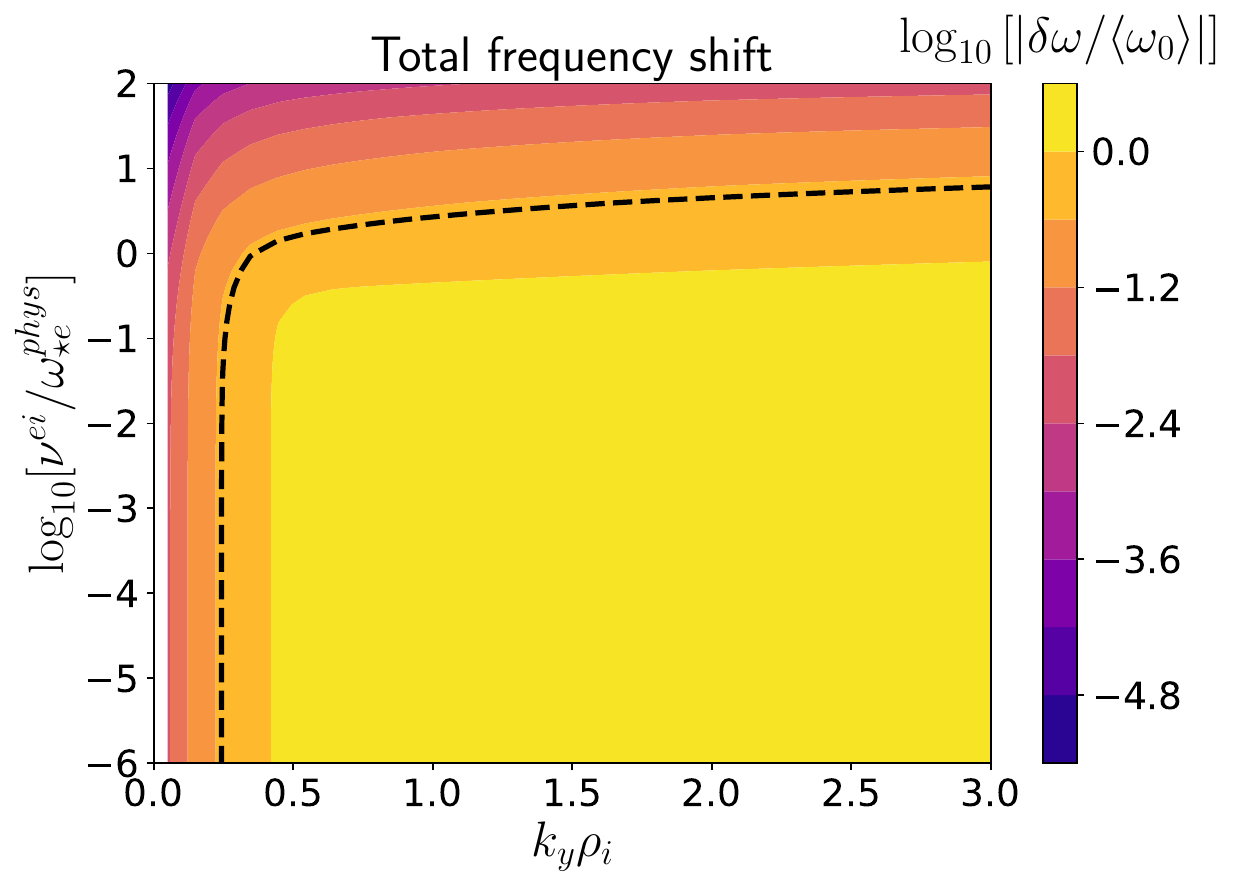}
        \caption{}        
    \end{subfigure}
\caption{\label{fig:D3Dperturbfull} The full perturbative frequency shift $\delta \omega$ as a function of collision frequency and wavenumber obtained by numerically integrating Eqn.~(\ref{eq:perturbfreq}) in DIII-D geometry, in case of a density gradient $a/L_n = 3$. Shown are (a) its real part, (b) its imaginary part, and (c) its absolute value. The dotted line in (c) indicates where $\abs{\delta \omega}/\omega_0 = 10^{-0.5} \approx 0.3$ and is a measure for the limited validity range of the approach.}
\end{figure*}

\begin{figure*}[!t]
 \centering
    \begin{subfigure}{.33\linewidth}
        \centering
        \includegraphics[width=\linewidth]{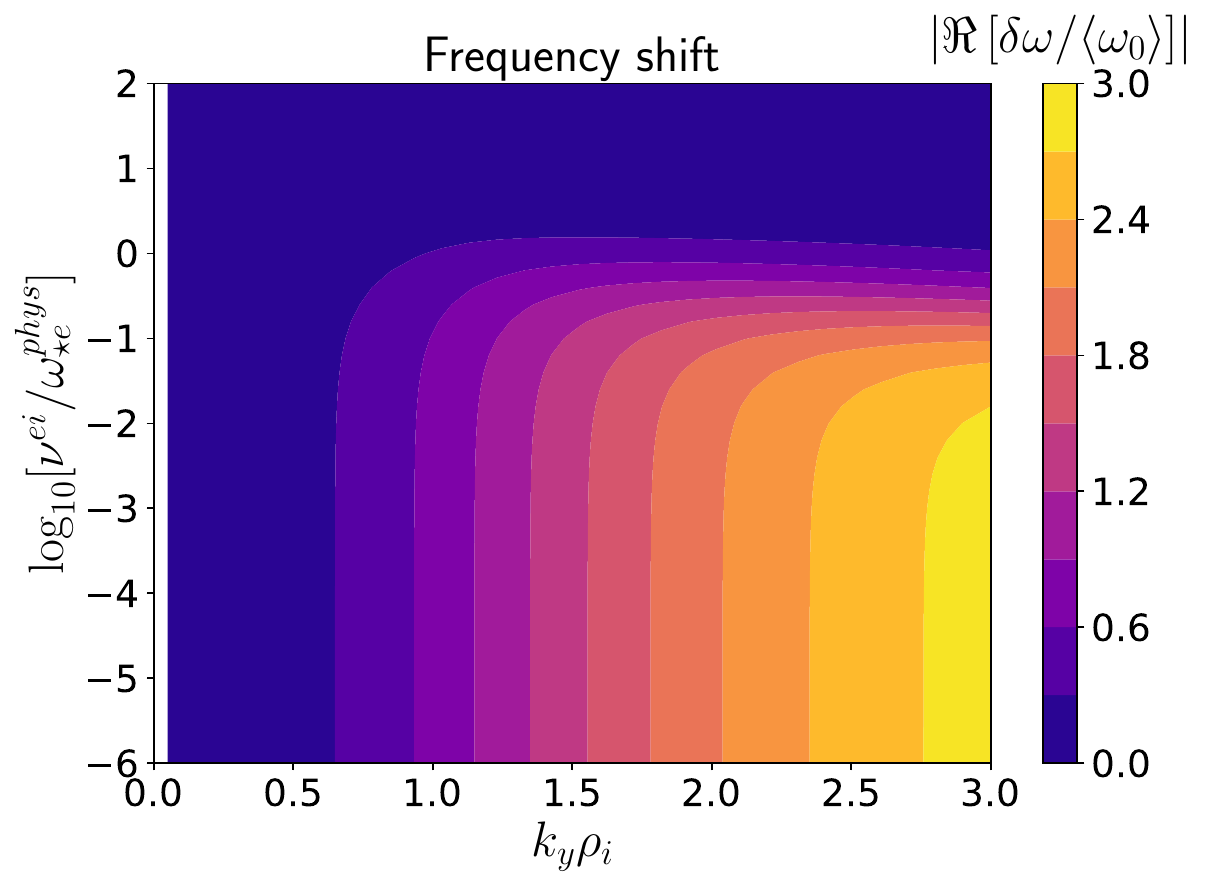}
        \caption{}      
    \end{subfigure}%
    \begin{subfigure}{.33\linewidth}
        \centering
        \includegraphics[width=\linewidth]{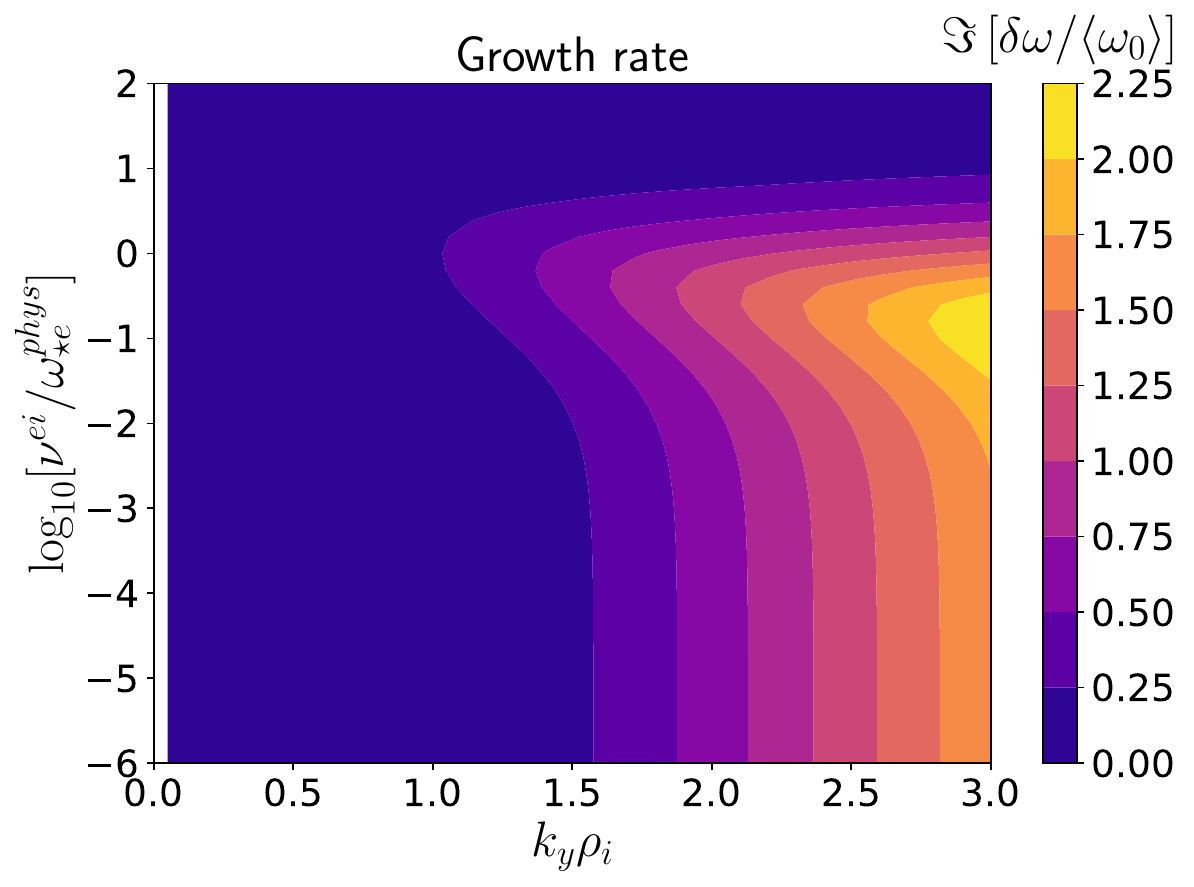}
        \caption{}        
    \end{subfigure}%
    \begin{subfigure}{.33\linewidth}
        \centering
        \includegraphics[width=\linewidth]{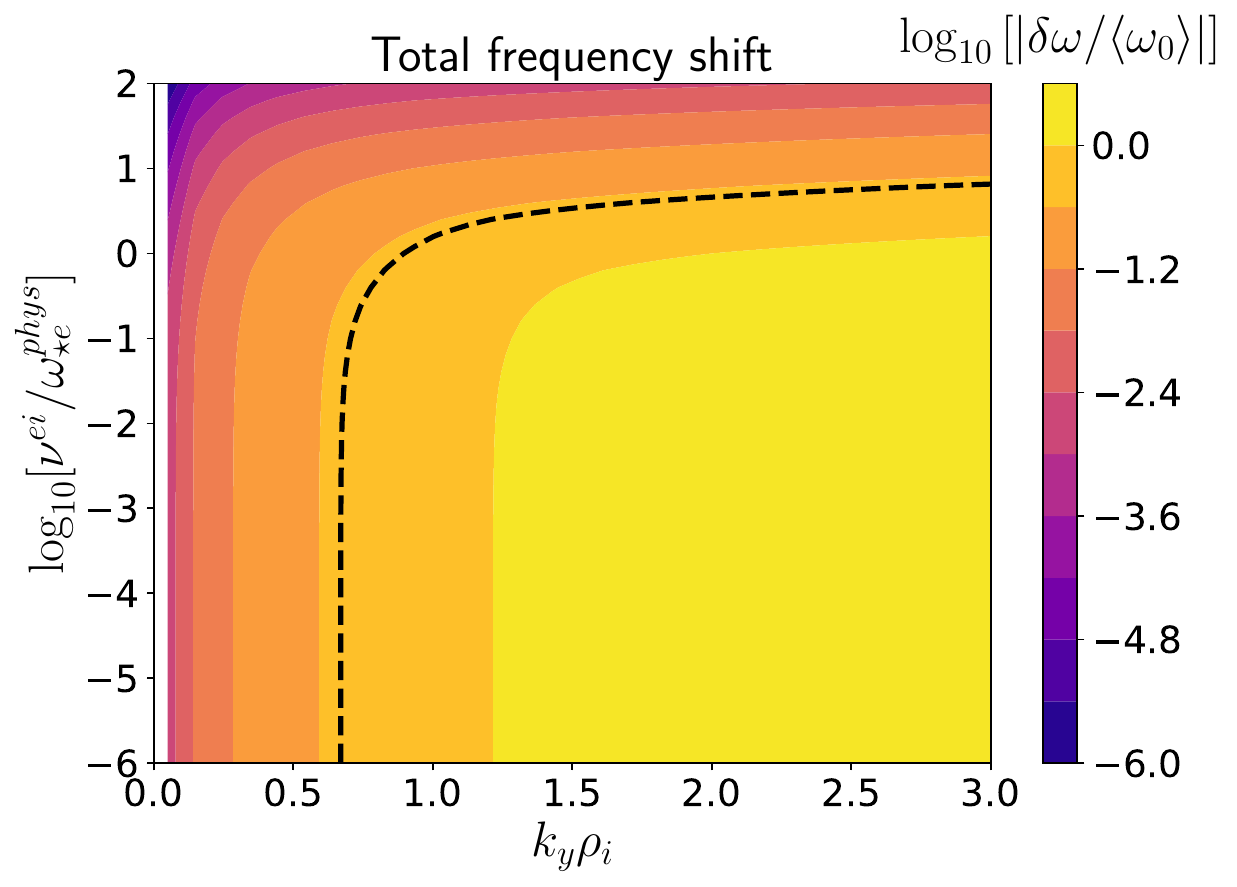}
        \caption{}        
    \end{subfigure}
\caption{\label{fig:HSXperturbfull} The full perturbative frequency shift $\delta \omega$ as a function of collision frequency and wavenumber obtained by numerically integrating Eqn.~(\ref{eq:perturbfreq}) in HSX geometry, in case of a density gradient $a/L_n = 3$. Shown are (a) its real part, (b) its imaginary part, and (c) its absolute value. The dotted line in (c) indicates where $\abs{\delta \omega}/\omega_0 = 10^{-0.5} \approx 0.3$ and is a measure for the limited validity range of the approach.}
\end{figure*}

\begin{figure*}[!ht]
 \centering
    \begin{subfigure}{.33\linewidth}
        \centering
        \includegraphics[width=\linewidth]{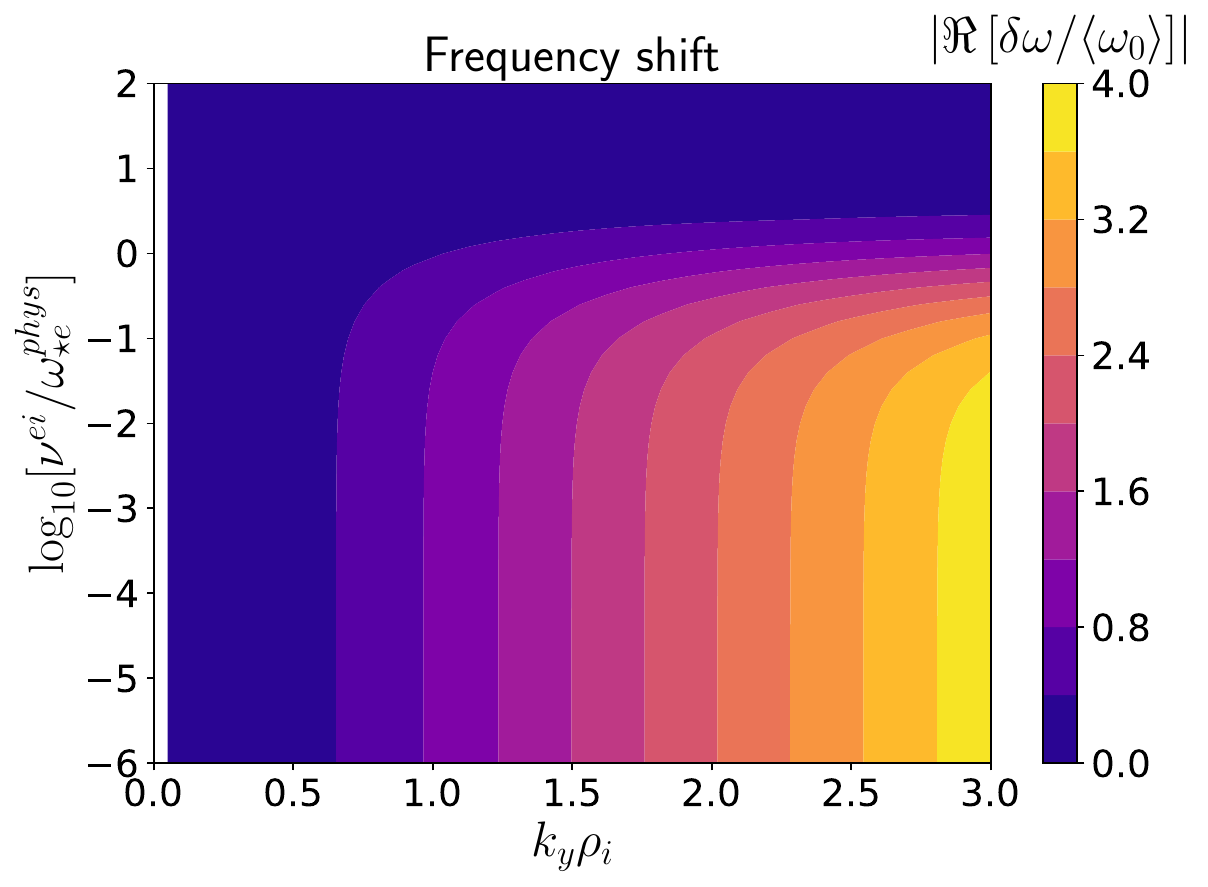}
        \caption{}      
    \end{subfigure}%
    \begin{subfigure}{.33\linewidth}
        \centering
        \includegraphics[width=\linewidth]{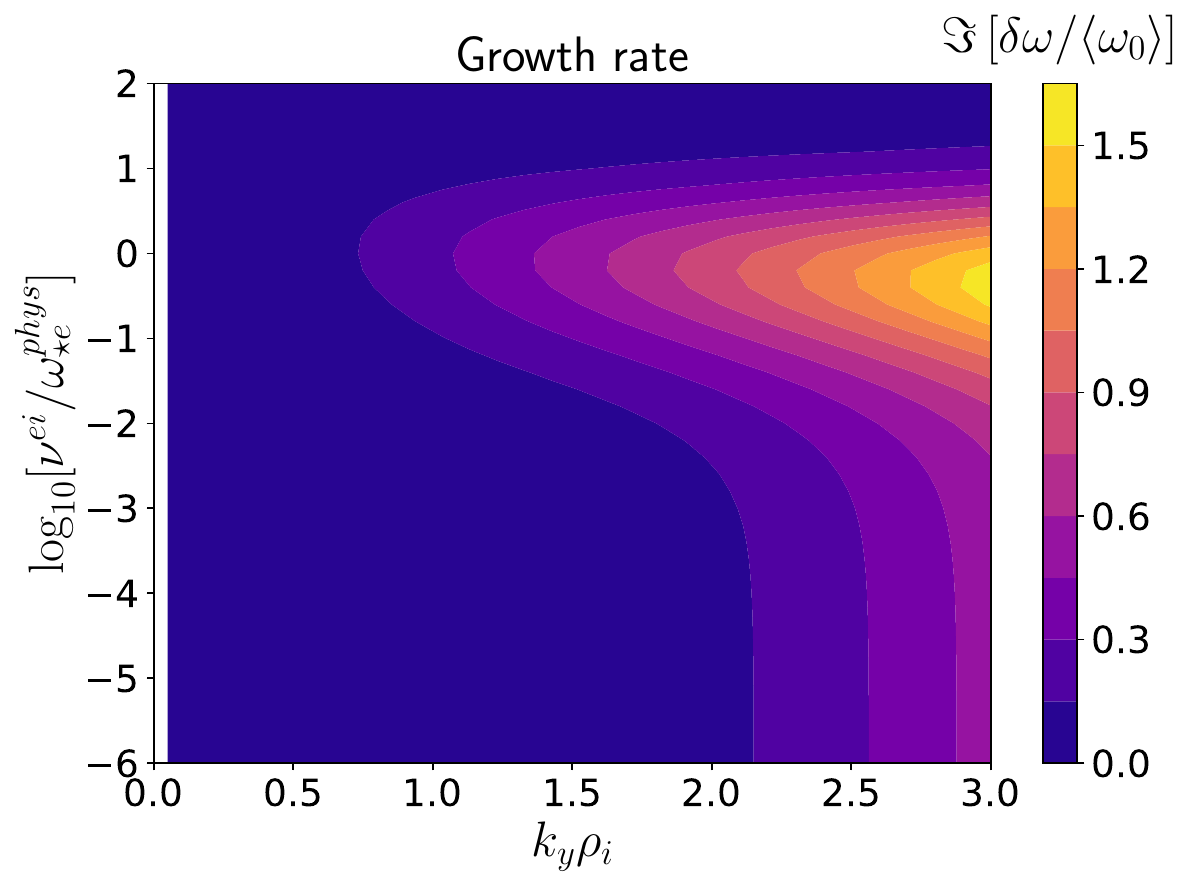}
        \caption{}        
    \end{subfigure}%
    \begin{subfigure}{.33\linewidth}
        \centering
        \includegraphics[width=\linewidth]{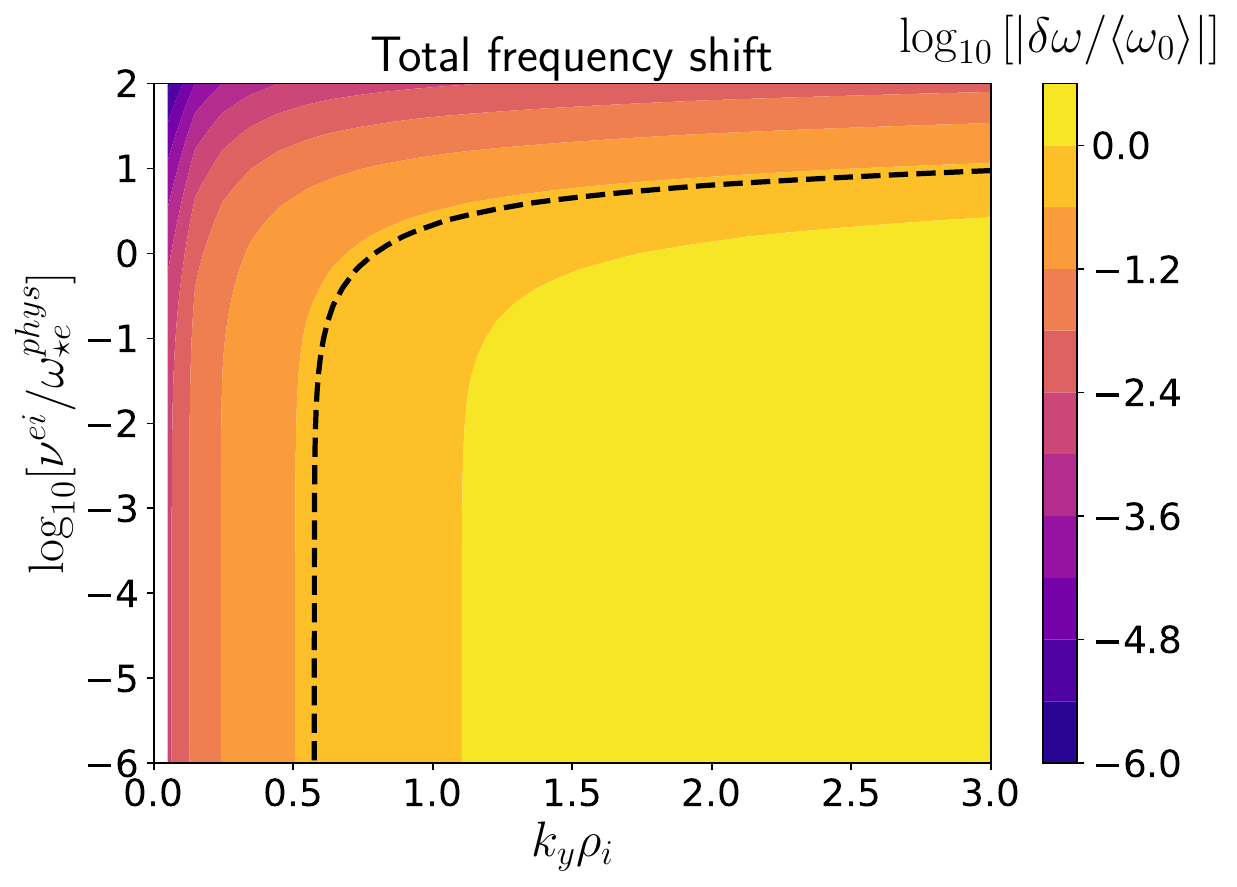}
        \caption{}        
    \end{subfigure}
\caption{\label{fig:W7Xperturbfull} The full perturbative frequency shift $\delta \omega$ as a function of collision frequency and wavenumber obtained by numerically integrating Eqn.~(\ref{eq:perturbfreq}) in (vacuum) W7-X geometry, in case of a density gradient $a/L_n = 3$. Shown are (a) its real part, (b) its imaginary part, and (c) its absolute value. The dotted line in (c) indicates where $\abs{\delta \omega}/\omega_0 = 10^{-0.5} \approx 0.3$ and is a measure for the limited validity range of the approach.}
\end{figure*}

Contours of the perturbative frequency shift Eqn.~(\ref{eq:perturbfreq}), explicitly separating its real and imaginary parts, in the $(k_y\rho_i, \nu^{ei}/\omega_{\star e}^{\textrm{phys}})$ plane for the DIII-D, HSX and W7-X geometries, are shown in Figures \ref{fig:D3Dperturbfull}, \ref{fig:HSXperturbfull} and \ref{fig:W7Xperturbfull}, respectively. Note that only in case of DIII-D, the real frequency shift changes sign while it is strictly negative for both stellarators. \par

This places strong limitations on the validity of the perturbative approach, as a frequency shift of $\Re[\delta \omega]/\langle \omega_0 \rangle < -1$ corresponds to a mode propagating in the ion diamagnetic direction instead, thus completely changing the (in)stability properties regarding a resonance with the electron precession drift. Furthermore, even if a sign reversal of the propagation frequency does not occur, too large growth rates would also invalidate a perturbative approach, as the $\delta \omega/\langle \omega_0 \rangle \ll 1$ assumption applies to both real and imaginary parts. Therefore, as an overall validity measure, we consider the absolute value of $\abs{\delta \omega/\langle \omega_0 \rangle} = \sqrt{\Re[(\delta \omega/\langle \omega_0 \rangle)]^2 + \Im[(\delta \omega/\langle \omega_0 \rangle)]^2}$ and require this to be below $30\%$. This critical level is indicated by the black dotted lines, and mostly limits the validity of the perturbative approach to low wavenumbers, since the lowest-order mode frequency Eqn.~(\ref{eq:omzero}) is a rapidly decreasing function of the perpendicular wavenumber.

\section{\label{app:EV-supplement} Eigenmode identification of eigenvalue scan.}

\begin{figure*}[!ht]
    \centering
    \includegraphics[width=0.7\linewidth]{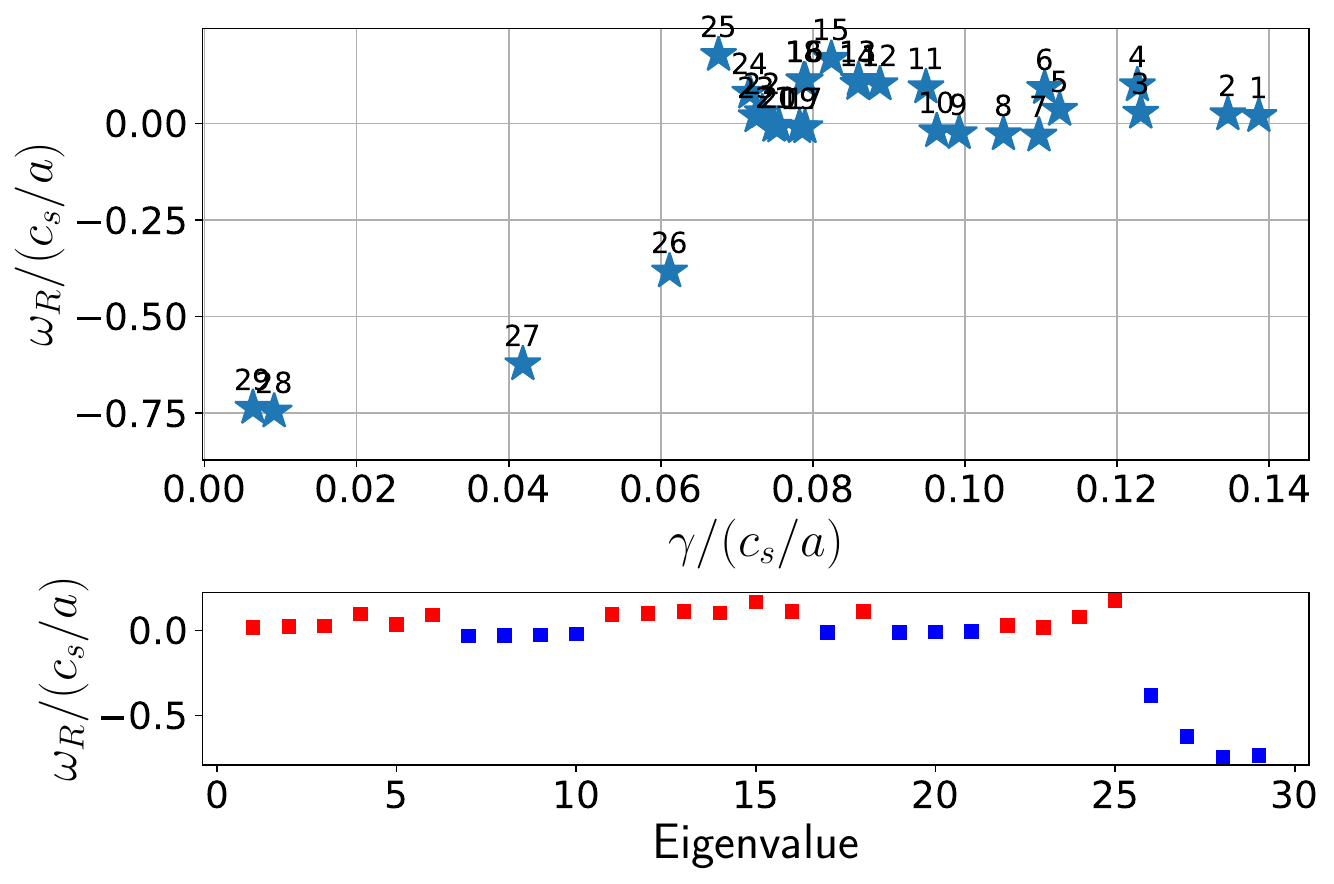}
    \caption{Eigenfrequency spectrum obtained by the eigenvalue solver at $k_y \rho = 0.5$ and $\tilde{\nu}=0$, showing at the top the eigenfrequencies in the complex plane labelled in order of highest growth rates and in the bottom a detail along the $\omega_R$ axis, with modes indicated by red/blue propagating in the ion/electron diamagnetic direction respectively.}
    \label{fig:app_nu0_spec}
\end{figure*}

From the eigenvalue simulations we obtain both the (complex) eigenmode frequencies, as well as the eigenmode structures of the electrostatic potential $\phi$. The eigenfrequencies are ordered of decreasing growth rate, and the real frequency can be used as a first indicator for which instabilities the eigenmodes correspond to, as iTEM are found at positive frequency $\omega_R$, while UI/TEM are found at negative frequency $\omega_R$. This results in a spectrum as shown in Fig.~\ref{fig:app_nu0_spec}, where the sign of $\omega_R$ has been colour labelled, for the $\tilde{\nu}=0$ case. \par 

\begin{figure*}[!ht]
    \centering
    \includegraphics[width=0.75\linewidth]{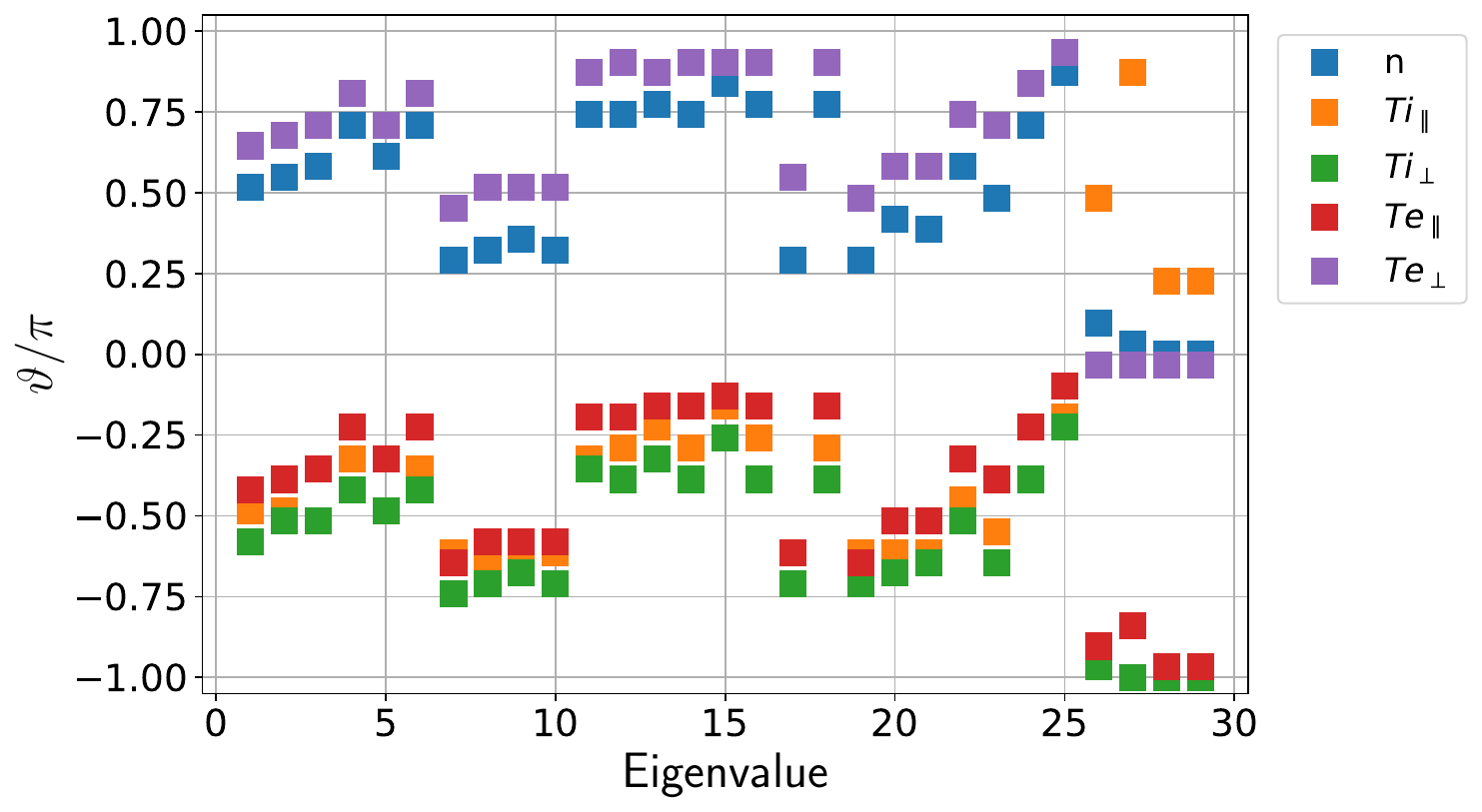}
    \caption{Cross-phases between the electrostatic potential $\phi$ and density/temperature perturbations $\delta n, \delta T_{\parallel}, \delta T_{\perp}$ of each species at $\tilde{\nu}=0$. Note that due to quasi-neutrality, the density perturbations must be equal $\delta n_e = \delta n_i$. Several clusters of eigenmodes with similar cross-phases can be identified, serving as a signature for the instability type.}
    \label{fig:app_nu0_XP}
\end{figure*}

This identification method is not absolute, however, as the introduction of collisions introduces (negative) frequency shifts, as observed in Section \ref{sec:sims}. Like in Section \ref{sec:sims}, the different type of eigenmodes are also characterised by distinct mode structures of the electrostatic potential $\phi$. The mode structures, however, are not the only spatial profiles with distinct structures between the different type of modes, so are the perturbations in density $\delta n$, and temperatures $\delta T_\parallel, \delta T_\perp$. Consequently, the different type of modes are characterised by distinct cross-phases between these fluctuations\cite{Faber2015,Costello2022,Hatch2009}. The cross-phases for the eigenspectrum highlighted in Fig.~\ref{fig:app_nu0_spec} are shown in Fig.~\ref{fig:app_nu0_XP}, where several clusters of modes with similar cross-phases can be identified.\par

\begin{figure*}[!b]
    \centering
    \begin{subfigure}{.33\linewidth}
        \centering
        \includegraphics[width = \linewidth]{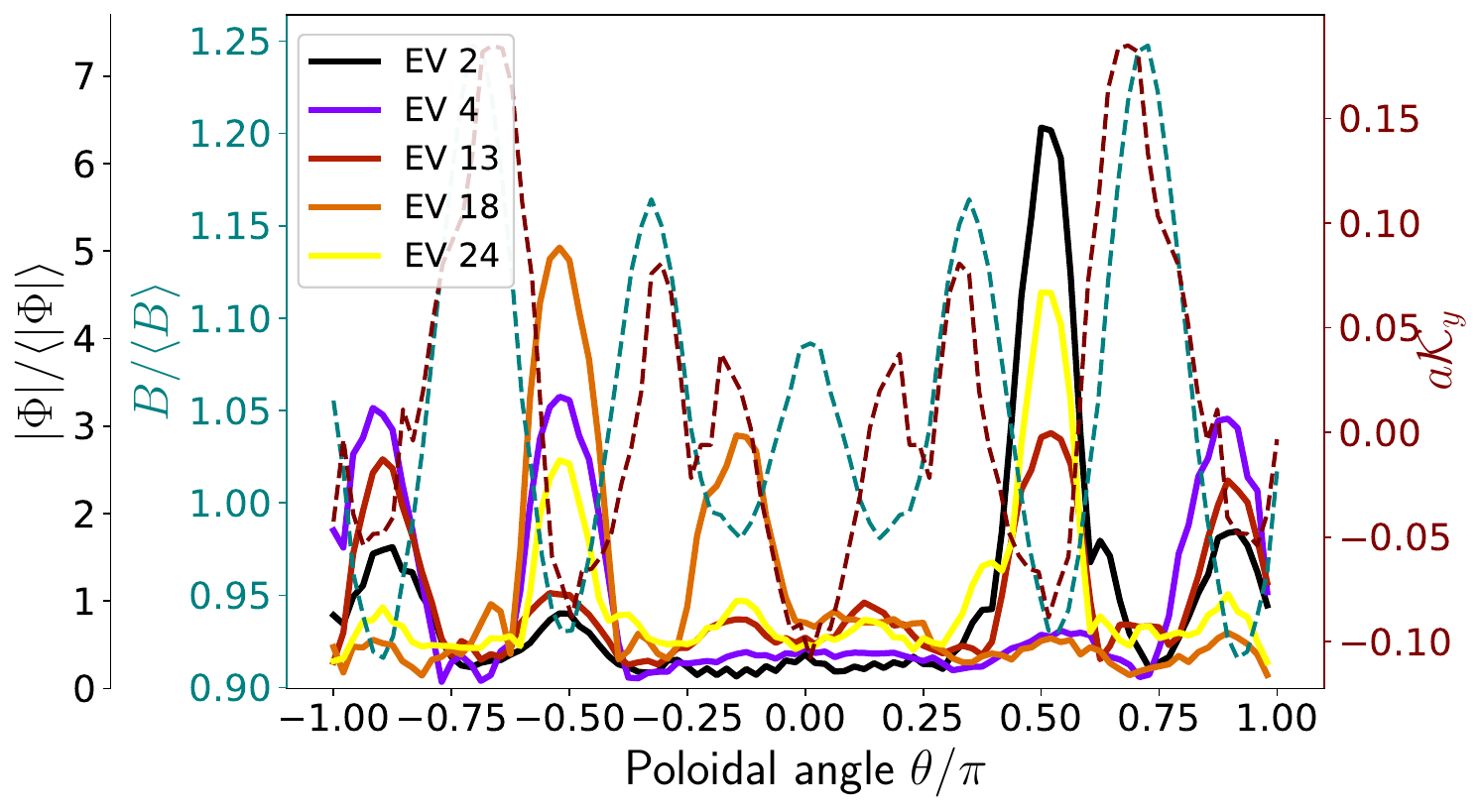}
        \caption{}
        \label{fig:app-iTEM-example}
    \end{subfigure}%
    \begin{subfigure}{.33\linewidth}
        \centering
        \includegraphics[width = \linewidth]{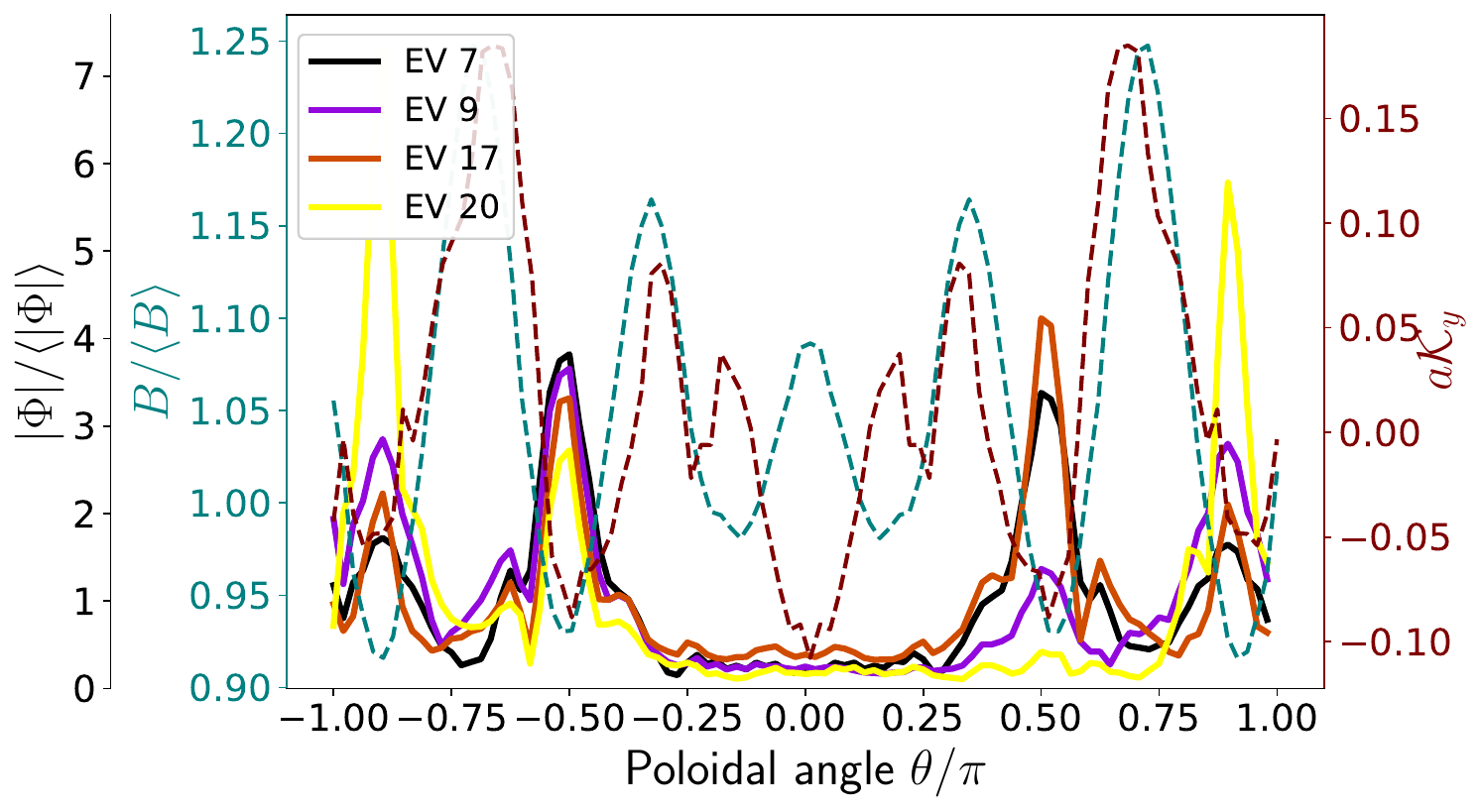}
        \caption{}
        \label{fig:app-TEM-example}
    \end{subfigure}%
    \begin{subfigure}{.33\linewidth}
        \centering
        \includegraphics[width = \linewidth]{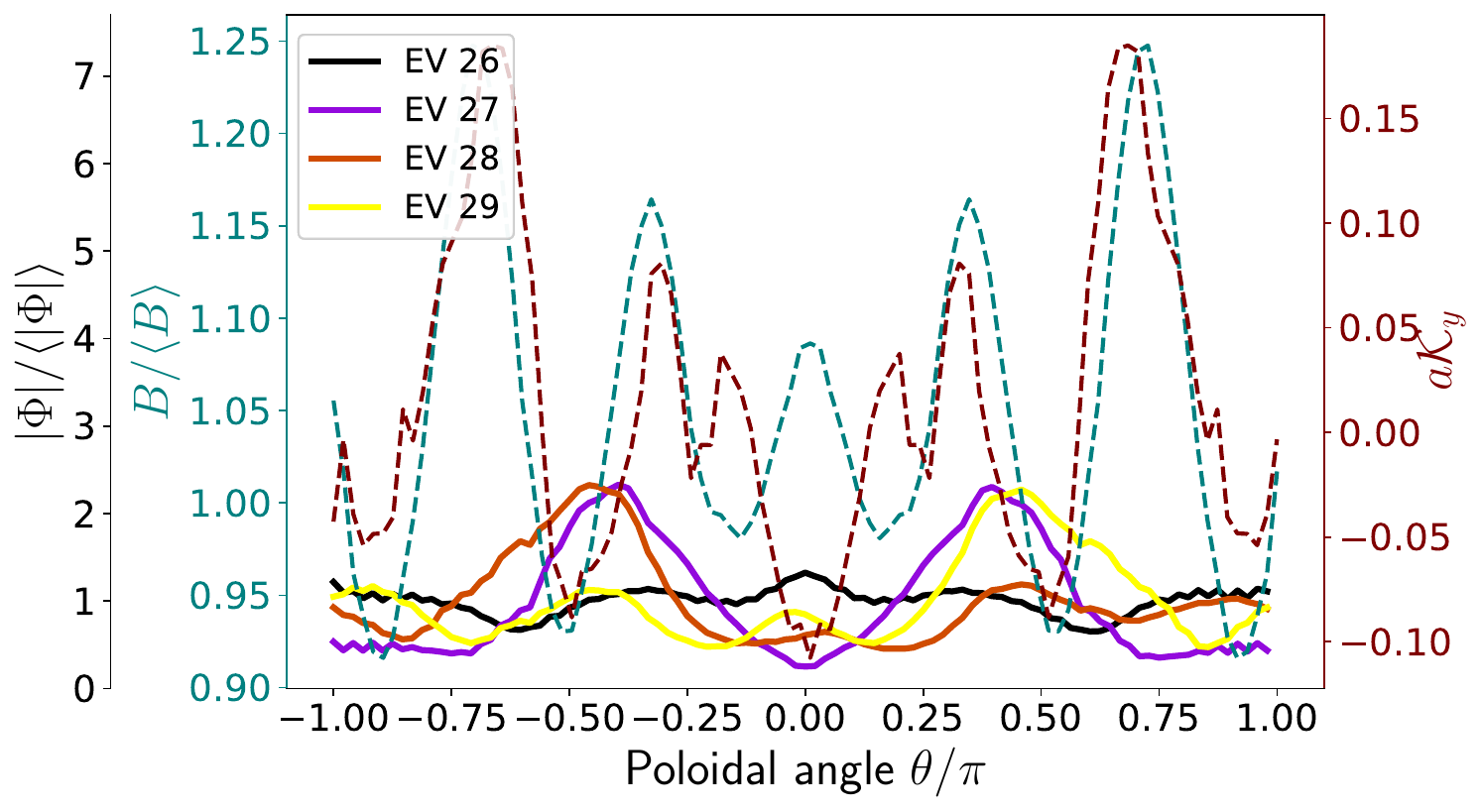}
        \caption{}
        \label{fig:app-UI-example}
    \end{subfigure}
    \caption{Typical eigenmode structures of the electrostatic potential $\phi$ along the field line at $\tilde{\nu}=0$. Shown are eigenmodes belonging to (a) the first cluster of modes, being identified as iTEMs, (b) the second cluster of modes, being identified as TEMs, and (c) the last cluster of modes, being identified as UIs.}
    \label{fig:app=eigenmode-examples}
\end{figure*}

The first cluster comprises eigenmodes $1\textrm{-}6, 11\textrm{-}16, 18, 24, 25 $, the second eigenmodes $7\textrm{-}10, 17,$ $19\textrm{-}21$ and the last eigenmodes $26\textrm{-}29$. Meanwhile the cross-phases of eigenmodes $22,23$ cannot be matched to any of the identified cluster, as these cross-phases exist between the typical cross-phases of the first and second cluster. It is noteworthy that the first cluster almost completely corresponds to all modes found propagating in the ion diamagnetic direction, and are hence expected to correspond to iTEMs. This suspicion is confirmed when looking at the localisation of the electrostatic potential in the magnetic geometry in Fig.~\ref{fig:app-iTEM-example}, showing similar structures to what has been observed in the collisionless initial value simulations. Likewise, the last cluster modes is easily identified as instances of the UI, Fig.~\ref{fig:app-UI-example}, showing no strong (correlated) localisation in any of the geometric features along the field line. Lastly, the second cluster of eigenmodes are shown in Fig.~\ref{fig:app-TEM-example}, which show similar behaviour to the iTEMs being localised in the trapping wells, but are slightly narrower. Combined with the negative frequency of these eigenmodes, these therefore correspond to TEMs. The remaining eigenmodes $22,23$ could not be classified as any of the aforementioned instabilities by combining signatures of mode frequency, cross-phases and localisation. \par

\begin{figure*}
    \centering
    \begin{subfigure}{.45\linewidth}
        \centering
        \includegraphics[width = \linewidth]{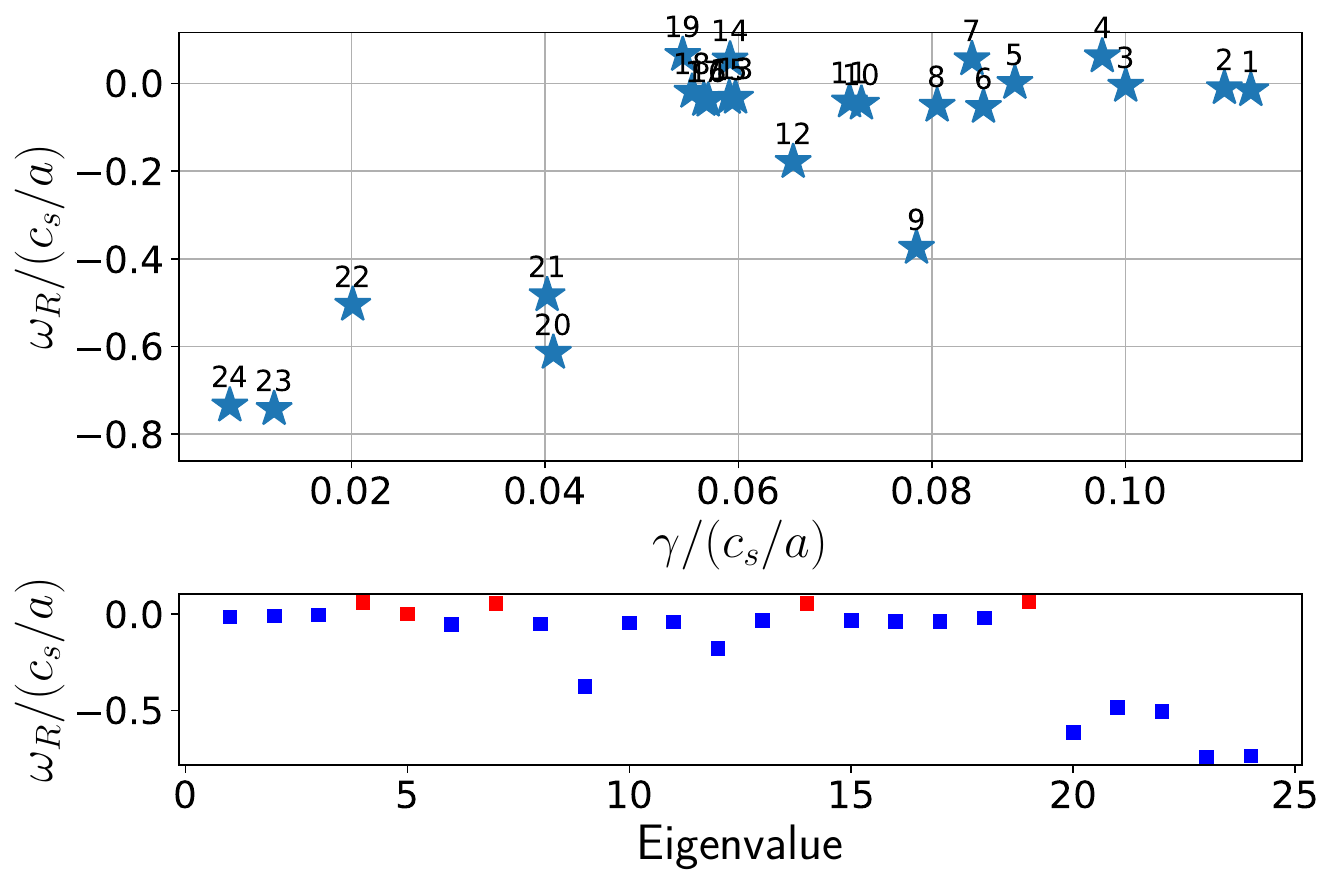}
        \caption{}
    \end{subfigure}
    \begin{subfigure}{.45\linewidth}
        \centering
        \includegraphics[width = \linewidth]{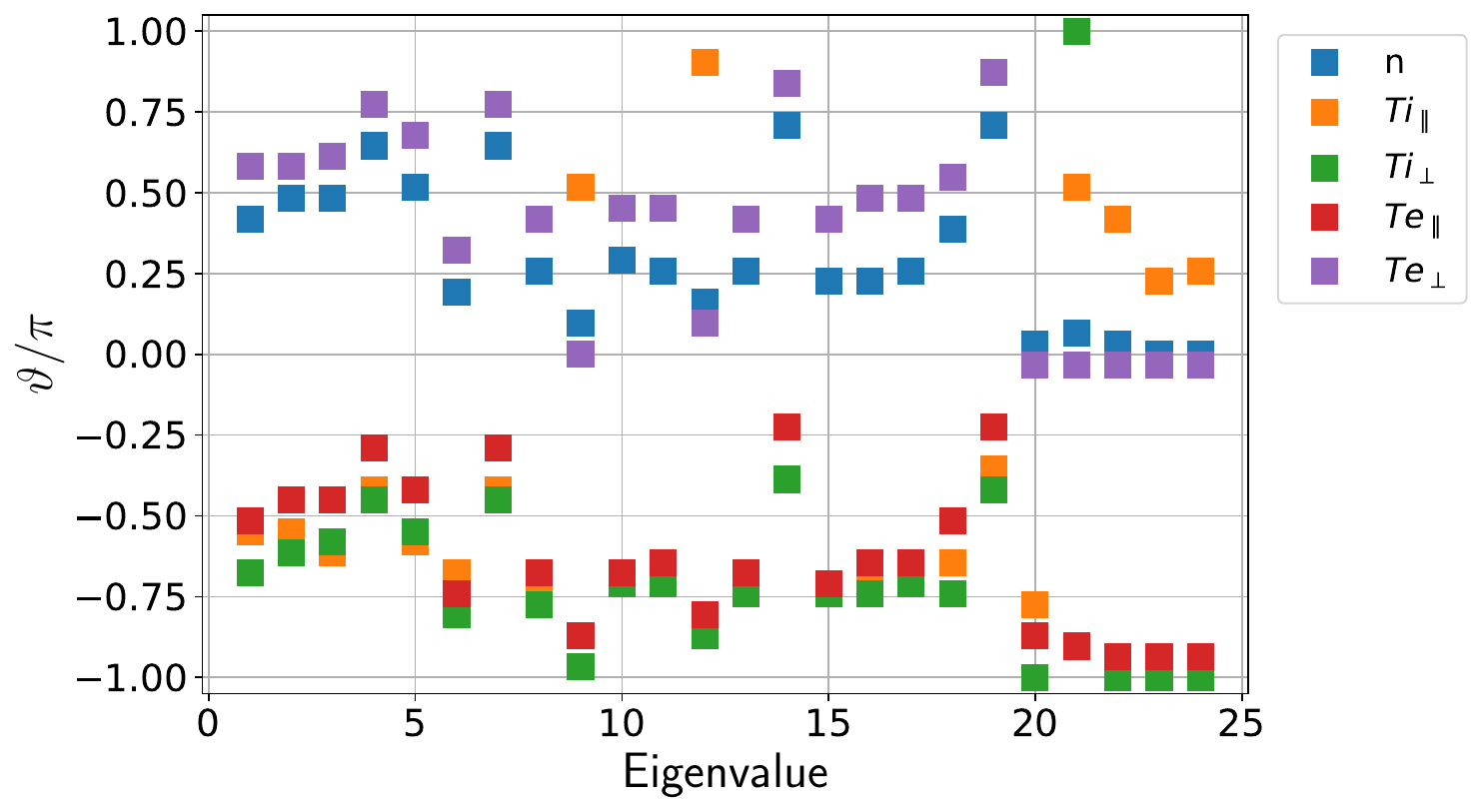}
        \caption{}
    \end{subfigure}
    \caption{(a) Eigenfrequency spectrum obtained by the eigenvalue at $k_y \rho = 0.5$ and $\tilde{\nu}=10^{-5}$, and (b) the cross-phases between the electrostatic potential and plasma perturbations. In the bottom plot of (a) red/blue colours represent modes propagating in the ion/electron diamagnetic direction respectively. Once more three distinct clusters can be identified in the cross-phases: EV $1\textrm{-}5, 7, 14, 19$ corresponding to iTEM, EV $ 6, 8, 10, 11, 13, 15\textrm{-}18 $ corresponding to TEM and EV $9, 12, 20\textrm{-}24$ corresponding to UI. }
    \label{fig:app-EV-nu1}
\end{figure*}

\begin{figure*}
    \centering
    \begin{subfigure}{.45\linewidth}
        \centering
        \includegraphics[width = \linewidth]{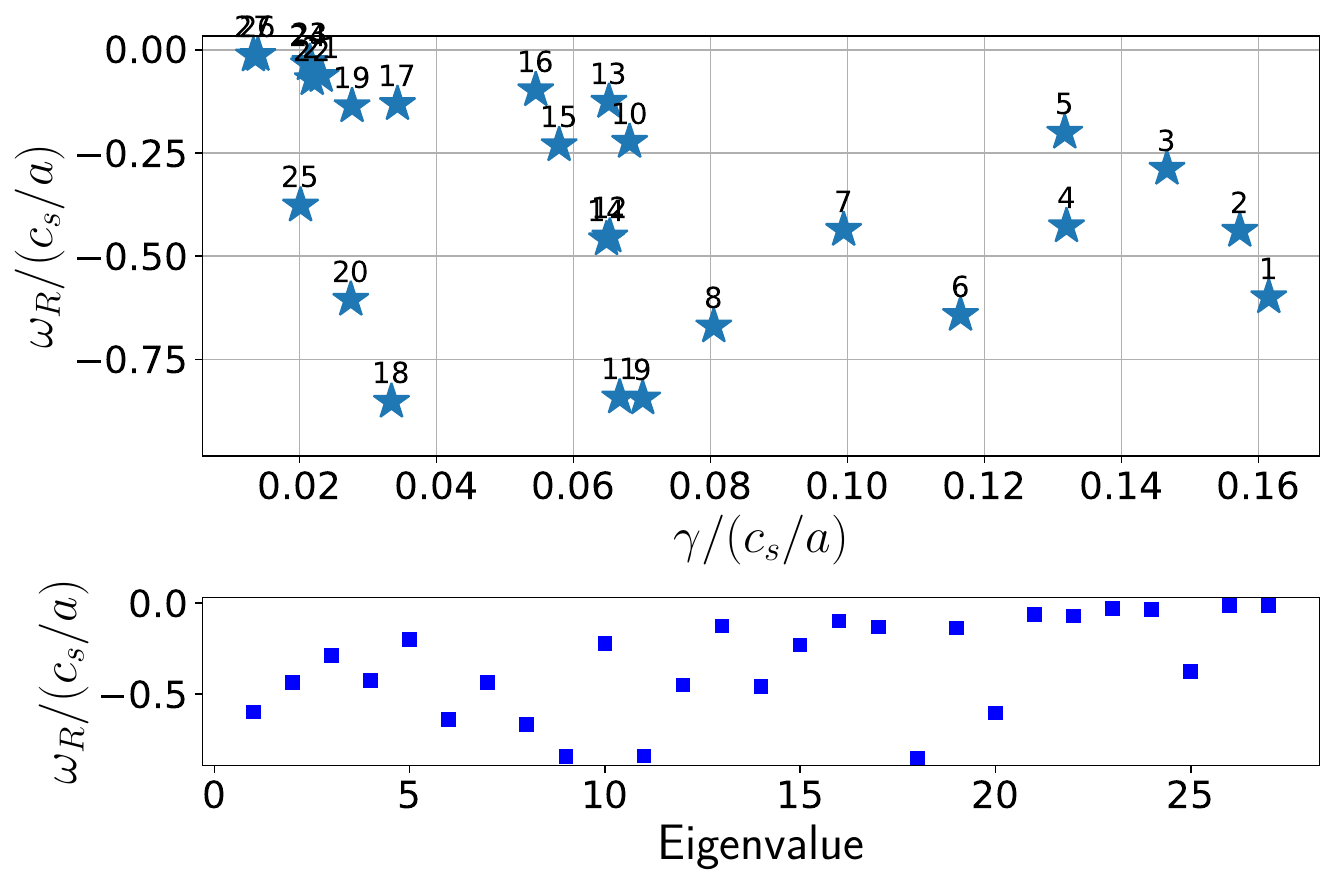}
        \caption{}
    \end{subfigure}
    \begin{subfigure}{.45\linewidth}
        \centering
        \includegraphics[width = \linewidth]{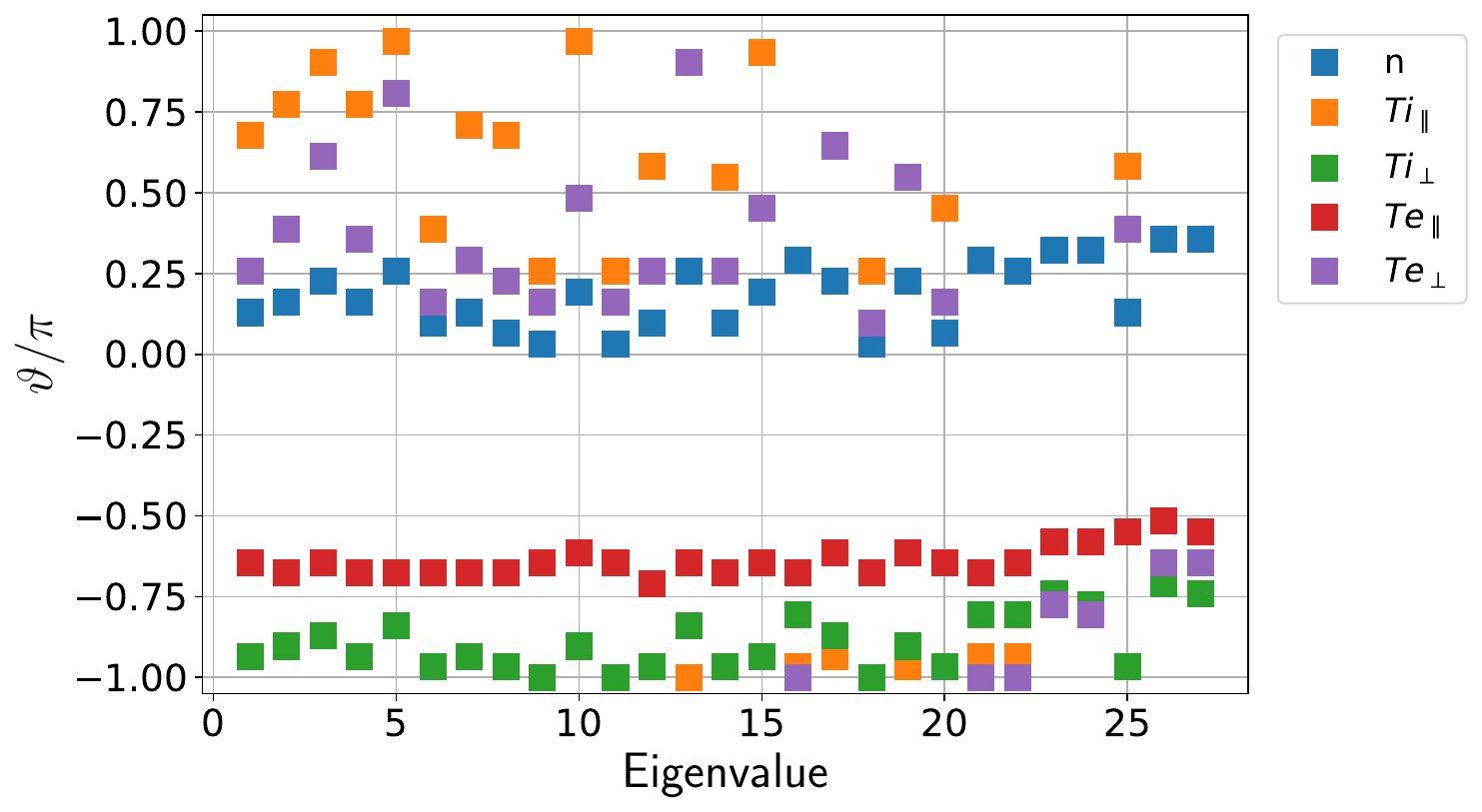}
        \caption{}
    \end{subfigure}
    \caption{(a) Eigenfrequency spectrum obtained by the eigenvalue at $k_y \rho = 0.5$ and $\tilde{\nu}=10^{-3}$, and (b) the cross-phases between the electrostatic potential and plasma perturbations. In the bottom plot of (a) red/blue colours represent modes propagating in the ion/electron diamagnetic direction respectively. In the cross-phases, an increase in the scatter for $\delta Ti_\parallel, \delta Te_\perp$ can be observed, but the increased scatter does not justify the introduction of separate clusters as the differences in other cross-phases remain marginal. Consequently all modes belong to a singular UI cluster.}
    \label{fig:app-EV-nu2}
\end{figure*}

The EV analysis for the the two collisional cases was carried out in the same way, with the $\tilde{\nu} = 10^{-5}$ and $\tilde{\nu} =10^{-3}$ cases shown in Figs.~\ref{fig:app-EV-nu1}, \ref{fig:app-EV-nu2} respectively. In the case of $\tilde{\nu}=10^{-5}$ it can be seen that the number of modes propagating in the ion direction direction has decreased significantly, however, as a result of frequency downshift caused by finite collisionality, some of the modes close to the frequency transition threshold could also be iTEMs, as also suggested by the similarity in cross-phases and mode-structures (not shown). Once more three clusters can be identified in the cross-phases, with cross-phase angles in very similar ranges to those identified in the collisionless case, which respectively identify the iTEM, TEM and UI. In the case of $\tilde{\nu} = 10^{-3}$, most of the frequencies are far enough in the negative $\omega_R$ of the spectrum to quell any doubts of the iTEM, and only one large cluster can be identified in the cross-phases (although scatter has somewhat increased). Most of the eigenmodes within the cluster are easily identified as UI by the mode structures (also not shown). Some mode structures, however, displayed odd behaviour of relative localisation in one half of the parallel domain, but the mode structures were much broader than any geometrical features like trapping wells, hence ruling out the possibility of TEMs, and therefore remained unidentified. 

\end{document}